\documentclass[prd,superscriptaddress,nofootinbib,notitlepage]{revtex4-1}
\usepackage{epsfig,amsmath,amssymb,slashed}
\usepackage{graphicx}
\usepackage{color}
\usepackage{siunitx}
\usepackage{color}
\usepackage{bm}
\usepackage{subfig}
\usepackage{tikz}
\usepackage[most]{tcolorbox}
\usepackage{cancel}
\usepackage[normalem]{ulem}
\tcbuselibrary{listings, breakable}

\def\wT{{\widehat T}}
\def\wj{{\widehat j}}
\def\wJ{{\widehat J}}

\def\wP{{\widehat P}}
\def\wA{{\widehat A}}
\def\wB{{\widehat B}}
\def\wQ{{\widehat Q}}
\def\wD{{\widehat{\cal D}}}
\def\wE{{\widehat{\cal E}}}

\def\wpsi{{\widehat{\psi}}}
 
\def\wrho{{\widehat{\rho}}}
\def\wPi{{\widehat\Pi}}

\def\wa{\widehat a}
\def\wad{\widehat a^{\dagger}}
\def\wb{\widehat b}
\def\wbd{\widehat b^{\dagger}}
\def\wA{\widehat A}
\def\wAd{\widehat A^{\dagger}}
\def\wB{\widehat B}
\def\wBd{\widehat B^{\dagger}}

\def\wW{\widehat W}

\def\wO{\widehat O}
\def\wQ{\widehat Q}
\def\wX{\widehat X}
\def\wY{\widehat Y}
\def\wJ{\widehat J}

\def\wphi{\widehat \phi}
\def\wpsi{\widehat \psi}

\def\wPi{\widehat \Pi}

\def\wmT{\widehat {\mathcal{T}}}

\def\wPi{\widehat{\Pi}}
\def\wpsi{\widehat \psi}

\newcommand{\Tr}{{\rm Tr}}  
\newcommand{\e}{{\rm e}}
\newcommand{\kk}{{\rm k}}
\newcommand{\pp}{{\rm p}}
\newcommand{\qq}{{\rm q}}
\newcommand{\di}{{\rm d}}
\newcommand{\ii}{{\rm i}}

\newcommand{\betav}{\boldsymbol{\beta}}

\newcommand{\p}{{\mathbf{p}}}
\newcommand{\q}{{\mathbf{q}}}
\newcommand{\kb}{{\mathbf{k}}}
\newcommand{\x}{{\mathbf{x}}}
\newcommand{\y}{{\mathbf{y}}}

\newcommand{\be}{\begin{equation}}
\newcommand{\ee}{\end{equation}}

\begin{document}


\title{Dissipative corrections to the particle momentum spectrum of a decoupling fluid} 

\author{Francesco Becattini}
\affiliation{Universit\`a degli studi di Firenze and INFN Sezione di Firenze,\\
Via G. Sansone 1, I-50019 Sesto Fiorentino (Florence), Italy}

\author{Daniele Roselli}
\affiliation{Universit\`a degli studi di Firenze and INFN Sezione di Firenze,\\
Via G. Sansone 1, I-50019 Sesto Fiorentino (Florence), Italy}

\author{Xin-Li Sheng}
\affiliation{Universit\`a degli studi di Firenze and INFN Sezione di Firenze,\\
Via G. Sansone 1, I-50019 Sesto Fiorentino (Florence), Italy}
\affiliation{Institut f\"{u}r Theoretische Physik, Johann Wolfgang Goethe-Universit\"{a}t,
Max-von-Laue-Str.~1, D-60438 Frankfurt am Main, Germany}
\affiliation{ExtreMe Matter Institute EMMI,
GSI Helmholtzzentrum f\"{u}r Schwerionenforschung GmbH,
Planckstrasse 1, 64291 Darmstadt, Germany}

\begin{abstract}
We present an \emph{ab initio} calculation within quantum statistical field theory and linear response theory,
of the dissipative correction to the momentum spectrum of scalar particles emitted at decoupling (freeze-out) from a 
relativistic fluid assuming the initial state to be in local thermodynamic equilibrium. 
We obtain an expansion of the Wigner function of the interacting quantum field in terms of the gradients of the 
classical thermo-hydrodynamic fields - four-temperature vector and reduced chemical potential - evaluated on the 
initial local-equilibrium hypersurface, rather than on the decoupling (freeze-out) hypersurface as usual 
in kinetic theory. The gradient expansion includes an unexpected zeroth order term depending on the differences between 
thermo-hydrodynamic fields at the decoupling and the initial hypersurface. This term encodes a memory of the initial 
state which is related to the long-distance persistence of the correlation function between Wigner operator and stress-energy
tensor and charged current that is discussed in detail. We address the phenomenological implications of these corrections 
for the momentum spectra measured in relativistic nuclear collisions.
\end{abstract}

\maketitle

\section*{Introduction}

Pinning down the momentum spectrum of particles emitted from an expanding gas, or a fluid in general, when they 
cease to interact is an important problem in several fields of physics, e.g. in cosmology or relativistic heavy
ion physics. Two stages can be identified in the fluid-particle conversion process. The first stage is the end of the 
hydrodynamic approximation, when the system can no longer be approximated by a fluid, i.e. a system close to local 
thermodynamic equilibrium; this stage will henceforth called {\em decoupling}. After decoupling, the system is
best seen as weakly interacting particles whose collisions drive it out of local equilibrium. The second stage 
occurs when these particles finally cease to interact; this stage is generally called {\em freeze-out}. 

If decoupling and freeze-out are very close, as a first approximation, the residual interaction after decoupling is 
neglected and one just considers free particles at local thermodynamic equilibrium. The momentum distribution is obtained 
as an integral of the J\"uttner distribution (in the relativistic regime) the so-called Cooper-Frye formula \cite{Cooper:1974mv}:
\begin{equation}\label{cfrye}
    \varepsilon(k)\frac{\di N_k}{\di^3 \kk} = \int_{\Sigma_D}\di\Sigma_\mu(x) k^\mu \; 
    \frac{1}{\e^{\beta(x) \cdot p -\zeta(x)}\pm1} \;,
\end{equation}	
where $\Sigma_D$ is the decoupling hypersurface, $\beta(x)$ is the four-temperature vector at the point $x$, $\zeta(x)$
is the reduced chemical potential $\mu/T$ at the point $x$; the sign + applies to fermions, the - to bosons. Yet, it is 
generally accepted that this distribution should include other terms: dissipative as well as quantum corrections. It is 
commonly believed that dissipative terms are, at the leading order, proportional to the gradients of the thermo-hydrodynamic 
fields (temperature, four-velocity, chemical potential) and quantum corrections proportional to the gradients squared \cite{Kovtun:2012rj,Denicol:2012cn,Jaiswal:2014isa}. In heavy ion physics, where the decoupling stage is also called 
{\em particlization} and quickly follows the transition from the Quark Gluon Plasma to hadron gas, these corrections have 
been addressed in several studies \cite{Teaney:2003kp,Dusling:2007gi,Denicol:2009am,Pratt:2010jt,Luzum:2010ad,Teaney:2013gca,McNelis:2021acu} and some models of them have been implemented in numerical codes \cite{ShenEtAl2014,McNelis:2019auj}. 

While the theoretical formulae for the dissipative corrections of local currents, such as the conserved charged current 
$j^\mu(x)$ and the stress-energy tensor $T^{\mu\nu}(x)$ are well established in a quantum-statistical framework \cite{Zub2,Becattini:2019dxo,Jeon:1995zm,Harutyunyan:2021rmb} (leading to relativistic extension to the so-called 
Kubo formulae \cite{Kubo1991}), those of momentum-dependent operators such as the Wigner operator are not yet fully established. 
Most calculations of the dissipative corrections to the spectrum were carried out in classical relativistic kinetic theory \cite{Hidaka:2016yjf,Zhang:2019xya,Romatschke:2017ejr,Weickgenannt:2022qvh,Bhadury:2020puc,Yang:2020hri,Wang:2025mfz} and
fewer in quantum statistical field theory \cite{Buzzegoli:2025zud,Li:2025pef} with some assumptions. 

In this work, we will present an {\em ab initio} calculation of the momentum spectrum at the decoupling stage of scalar particles within
a full quantum-relativistic statistical framework, by using a new approximation method developed in a recent work of ours 
\cite{Sheng:2025cjk}. By using this method, we obtain an expansion of the Wigner function in the gradients of the {\em initial} 
thermo-hydrodynamic fields, instead of the gradients of the same fields at the decoupling stage, as it is usually obtained in 
e.g. classical relativistic kinetic theory. We will find a somewhat surprising result, namely an off-equilibrium correction of 
the Wigner function on the zero-order gradients i.e. on the thermo-hydrodynamic fields themselves, whose meaning will be the 
subject of a detailed discussion.

The paper is organized as follows: in Section \ref{Wigner} we summarize the main definitions concerning the covariant Wigner 
function and its generalization to the case of interacting fields. In Section \ref{density} we deal with the basic notions of local thermodynamic equilibrium and dissipation in the framework of quantum-relativistic statistical mechanics and introduce a new method to calculate expectation values. In Section \ref{neqwigner} we develop the calculation of the off-equilibrium part of the Wigner 
function in the linear response theory approximation. In Section \ref{formfactors} we present a major part of our study, that is 
a detailed analysis of the dynamical correlators between the conserved currents and the field Fourier transforms. In 
Section \ref{hydrolimit} we work out the off-equilibrium correction to the Wigner function in the hydrodynamic limit, obtaining
a gradient expansion. In Section \ref{discussion} we discuss the obtained results and their relation with those of classical
relativistic kinetic theory; moreover, we delve into the features of the correlation function between Wigner operator and 
conserved currents. Finally in Section \ref{sec:spectrum} we determine the form of the momentum spectrum of particles emitted
from the decoupling fluid and its leading order off-equilibrium correction.

\subsection*{Notations}
Throughout this paper we use the mostly minus signature convention for the flat metric $g^{\mu\nu}=\mbox{diag}\left(+,-,-,-\right)$. 
We adopt the natural system, $\hbar=k_B=c=1$. The spacial part of a four-vector $k^\mu$ is denoted with the "bold", ${\bf{k}}$, 
the scalar product is denoted with a dot for both four-vectors $k\cdotp p=k_\mu p^\mu$ and three-vectors ${\bf k}\cdot{\bf p}=k_jp^j$. 
Einstein index conventions is assumed, contracted indexes are summed all over the possible values, $\mu=0,1,2,3$; $j=1,2,3$. The Heaviside $\theta$ function is defined as: $\theta(x) = 1$ if $x > 0$ and $0$ if $ x \le 0$.


\section{The covariant Wigner function and the momentum spectrum}
\label{Wigner}

The covariant Wigner function is an essential tool in quantum statistical field theory; all mean values 
of both local densities and momentum-dependent observables can be expressed as integrals thereof. For a free complex 
scalar field $\wphi$ the covariant Wigner operator is defined as the Fourier transform of the two points operator function:
\begin{equation}\label{ScalarField:WigDef}
    \wW\left(x,k\right)=\frac{2}{\left(2\pi\right)^4}\int\di^4 s \; \e^{-\ii s\cdot k}:
    \wphi^\dagger\left(x+\frac{s}{2}\right)\wphi\left(x-\frac{s}{2}\right):  \; ,
\end{equation}
where the semi-colon stands for normal ordering. The Wigner function is defined as the quantum expectation value 
of the Wigner operator \eqref{ScalarField:WigDef}:
\begin{equation}\label{meanwig}
    W(x,k)\equiv\Tr\left(\wrho\;\wW\left(x,k\right)\right)\; ,
\end{equation}
where $\wrho$ is the density matrix for a state, either pure or mixed. Henceforth, we will denote the expectation
values $\Tr(\wrho\,\widehat X)$ as $\langle \widehat X \rangle$.

According to its definition \eqref{ScalarField:WigDef}, the Wigner function is real but not positive definite. 
In general, even for free fields, the momentum $k$ is neither on-shell nor time-like. However, one can split the Wigner 
operator \eqref{ScalarField:WigDef} into three terms according to the signs of $k^2$ and $k^0$:
\begin{equation}\label{widecomp}
    \widehat W\left(x,k\right)\equiv \widehat W^+\left(x,k\right)+ \widehat W^-\left(x,k\right)+
    \widehat W_S\left(x,k\right)\; ,
\end{equation}
where:
\begin{equation*}
    \widehat W^\pm\left(x,k\right)\equiv \widehat W\left(x,k\right)\theta\left(k^2\right)
    \theta\left(\pm k^0\right),\quad \widehat W_S\equiv \widehat W\left(x,k\right)
    \theta\left(-k^2\right)\; .
\end{equation*}
The first two operators are associated with the particle/antiparticle terms of the free field expansion, 
while $\widehat W_S$ is a pure quantum term \cite{DeGroot:1980dk}.

We start considering free scalar fields satisfying the Klein--Gordon equation,
\begin{equation}\label{free kg}
    \left(\Box + m^2\right)\wphi(x) = 0 \, ,
\end{equation}
where $\Box \equiv \partial_\mu \partial^\mu$. The field operator can be expanded in plane-wave modes as
\begin{equation}\label{ScalarField:Expansion}
    \wphi(x)
    = \frac{1}{(2\pi)^{3/2}} \int \frac{\di^3\pp}{2\varepsilon_p}
    \left(
    \wa(p)\, e^{-\ii p \cdot x}
    + \wbd(p)\, e^{\ii p \cdot x}
    \right) \, ,
\end{equation}
where the annihilation and creation operators satisfy the canonical commutation relations
\begin{equation}\label{CommRelations}
    \left[\wa(p), \wad(p')\right]
    = \left[\wb(p), \wbd(p')\right]
    = 2\varepsilon_p\, \delta^3\!\left(\bf{p}-\bf{p}'\right) \, ,
\end{equation}
with $\varepsilon_p = \sqrt{m^2 + \bf{p}^2}$ denoting the on-shell energy. Under these assumptions, it can be shown that the particle momentum spectrum can be expressed as an integral over an arbitrary space-like hypersurface $\Sigma$:
\begin{equation}\label{spectrum}
    \frac{\di N_k}{\di^3 \kk} = \frac{1}{2\varepsilon_{\bf k}} \langle \wad(k) \wa(k) \rangle = 
    \int \di k^0 \int_\Sigma\di\Sigma_\mu k^\mu \; W^+\left(x,k\right) \;.
\end{equation}	
The integration can be taken over an arbitrary hypersurface because the Wigner function fulfills the equation
$k^\mu \partial_\mu W^\pm_S = 0$ for free fields. Besides, after integration, the variable $k$ becomes on-shell \cite{Becattini:2020sww} and the integral over $k^0$ removes a $\delta(\varepsilon_{\bf k} - k^0)$ distribution. 
A formula like the \eqref{spectrum} can be obtained for anti-particles as well. 

The use of the free fields in the equation \eqref{spectrum} with the local equilibrium density operator (see Section
\ref{density}) leads, in the first approximation, to the Cooper-Frye formula \eqref{cfrye}. Nevertheless, in 
principle, the quantum fields are not free and one should use interacting fields to express the momentum
spectrum. There is another very good reason to use interacting fields when dealing with formulae involving the Wigner
function; even though the Wigner operator depends on a given space-time point $x$, being a Fourier transform of
a two-point operator (see eq. \eqref{ScalarField:WigDef}) it is not in fact a local operator like the field operator 
or conserved currents. The Fourier transform involves space-time points $x\pm s/2$  which can be far from each other 
such that the approximation of free field at either points may not be a good one. 
\begin{figure}[h!]
    \centering
    \includegraphics[width=0.6\linewidth]{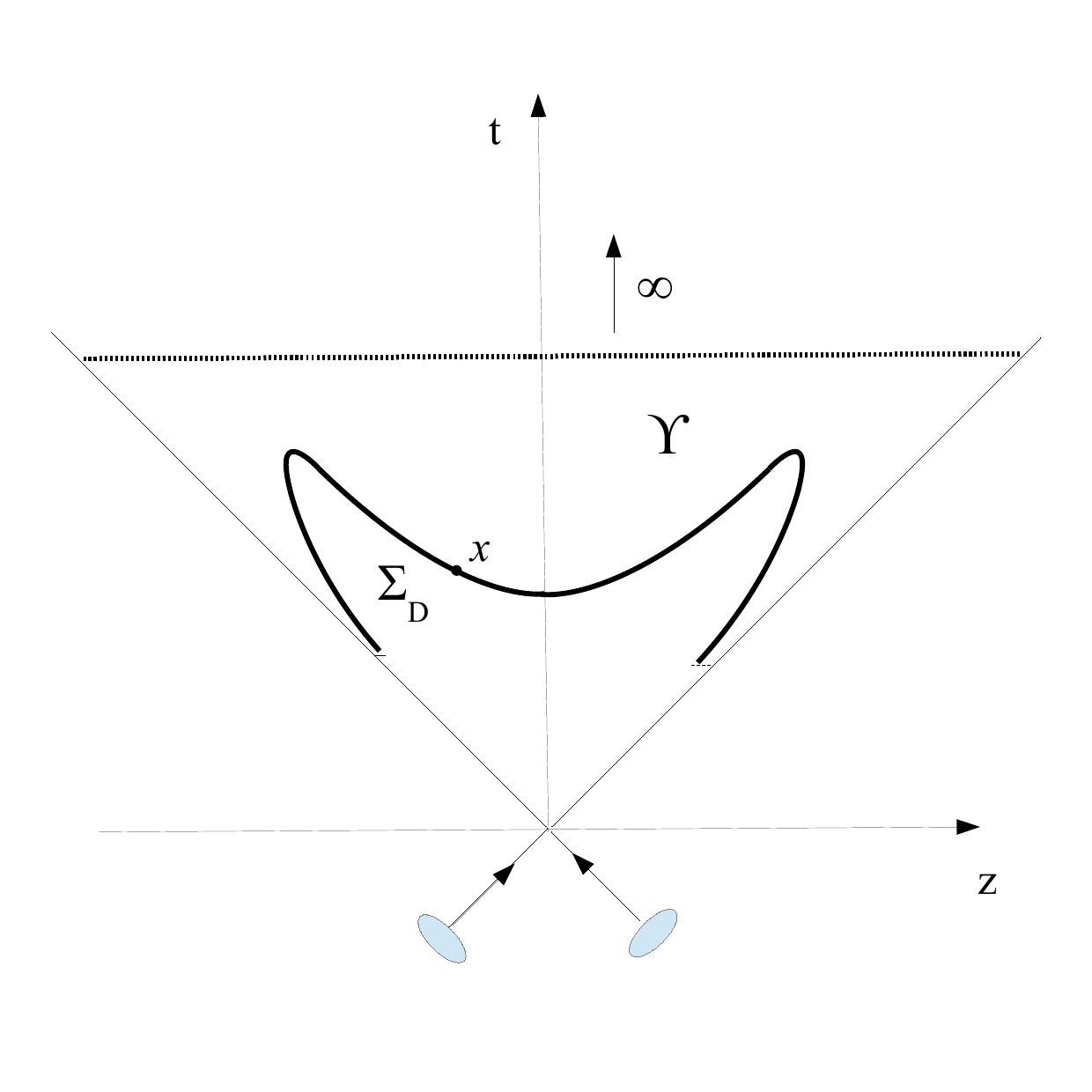}
    \caption{A typical shape of a decoupling hypersurface $\Sigma_D$ in a relativistic nuclear collision
    in a space-time diagram. The fluid decouples at $\Sigma_{\rm D}$ and the produced particles interact
    through collisions in the region $\Upsilon$ until all interaction cease and the spectra freeze out. The particles are eventually observed in the asymptotic future $t\to\infty$.}
    \label{postdecoupling}
\end{figure}

It is indeed possible to express the momentum spectrum of the finally produced particles with the Wigner function 
by replacing the free field with the interacting field in the equation \eqref{ScalarField:WigDef} and taking a 
suitable limit for $t \to +\infty$:
\begin{equation}\label{Interacting spectrum}
    \frac{\di N_k}{\di^3 \kk} = \lim_{t \to +\infty} 
    \int \di k^0 \int_{\Sigma(t)} \di\Sigma_\mu k^\mu \; W^+\left(x,k\right) = \frac{1}{2\varepsilon} \langle \wad_{\rm out}(k)
    \wa_{\rm out}(k) \rangle\;.
\end{equation}
Note that the integration hypersurface in this case is no longer arbitrary but it should be asymptotic in time.
The equation \eqref{Interacting spectrum} builds on the relation between the interacting field and the, asymptotic free, out-field 
stipulated by the Yang--Feldman equation \cite{PhysRev.79.972}:
\be\label{yangfel}
  \wphi(x) = \wphi_{\rm out}(x) - \int \di^4 y\;\Delta_{\rm adv}(x-y)\,\wJ(y)\;,
\ee
where $\wphi_{\rm out}$ fulfills the free Klein-Gordon equation \eqref{free kg}, whereas $\wphi(x)$ the interacting one with source $\wJ(x)$:
\begin{equation}\label{Interacting Klein Gordon}
    \left(\Box+m^2\right)\wphi(x)=\wJ(x)\;.
\end{equation}
 Note that $\Delta_{\rm adv}(x-y)$, being the advanced propagator, vanishes for $y^0 < x^0$. Obviously the \eqref{yangfel} implies 
the usual LSZ condition:
$$
 \lim_{t \to +\infty} \wphi(x) = \wphi_{\rm out}(x)\;.
$$

The main issue concerning the equation \eqref{Interacting spectrum} is that we are practically unable to reckon all 
scattering processes necessary to calculate the limit for $t \to +\infty$. What can be more easily done is to determine 
the Wigner function at the decoupling stage, i.e. when the fluid ceases to exist but not all interaction processes have ended. 
So, if $\Sigma_D$ is the decoupling hypersurface and using the Gauss theorem, one can recast the \eqref{Interacting spectrum}
as:
\begin{equation}\label{Interacting spectrum2}
    \frac{\di N_k}{\di^3 \kk} = \int \di k^0 \int_{\Sigma_D} \di\Sigma_\mu k^\mu \; W^+\left(x,k\right) + 
     \int \di k^0 \int_{\Upsilon} \di^4 x \; k^\mu \partial_\mu W^+\left(x,k\right)\;,
\end{equation}
where $\Upsilon$ is the space-time region encompassed by the hypersurfaces $\Sigma_D$ and $\Sigma(t \to +\infty)$
(see figure \ref{postdecoupling}).
The first term on the right hand side supposedly provides the dominant part of the spectrum while the second term is
a correction induced by the scattering processes in the dilute, post-decoupling phase, which can be estimated by using 
relativistic kinetic theory. In this work, we will focus only on the first term and its calculation including the
effect of interactions, i.e. by replacing the free fields with interacting fields. For this purpose, we will write
down a formal decomposition of the interacting field in plane waves which is the most suitable to deal with the 
problem at hand, i.e. the determination of the first term on the right hand side of the equation \eqref{Interacting spectrum2}.

A general interacting field $\wphi(x)$ can be Fourier-transformed and the inverse formula reads:
$$
 \wphi(x) = \frac{1}{(2\pi)^4}\int \di^4 p \; \e^{-\ii p \cdot x} \wphi_F(p)\;.
$$
We can split the integration for $p^0 \ge 0$ and $p^0 < 0$ and, after some simple redefinition of the integration
variables, turn the above formula into:
$$
 \wphi(x) = \frac{1}{(2\pi)^4}\int \di^4 p \; \theta(p^0) \left( \e^{-\ii p \cdot x} \wphi_F(p)
  + \e^{\ii p \cdot x} \wphi_F(-p) \right)\;.
$$
It is convenient to replace the integration variable $p^0>0$ with $M^2\equiv p^2=(p^0)^2-|\p|^2$ which can be either 
positive or negative, so to obtain:
$$
 \wphi(x) = \frac{1}{(2\pi)^4}\int \di^3 \pp \int^{+\infty}_{-|\p|^2} \di M^2 \; \frac{1}{2\sqrt{M^2+|\p|^2}} 
 \left( \e^{-\ii p \cdot x} \wphi_F(p) + \e^{\ii p \cdot x} \wphi_F(-p) \right)\;,
$$
where $p^0 = \sqrt{M^2+|\p|^2}$ in the four-vector $p$ is understood. 
The final step is extract some suitable factors from the Fourier transform operator $\wphi_F$ so as
to put the field expansion in a form which has a straightforward free limit:
\be\label{intfield}
 \wphi(x) = \frac{1}{(2\pi)^{3/2}}\int \di^3 \pp \int^{+\infty}_{-|\p|^2} \di M^2 \; \frac{1}{2\sqrt{M^2+|\p|^2}} 
 \frac{\varrho(p)}{2\pi} \left( \e^{-\ii p \cdot x} \wA(M^2,\p) + \e^{\ii p \cdot x} \wBd(M^2,\p) \right)\;,
\ee
where 
\be\label{invfourier}
 \wA(p) \equiv \frac{1}{(2\pi)^{3/2}\varrho(p)} \theta(p^0) \wphi_F(p)\;, \qquad\qquad \wBd(p) \equiv \frac{1}{(2\pi)^{3/2}\varrho(p)} \theta(p^0) \wphi_F(-p)\;, 
\ee
with $\varrho(p)$ being taken as the {\em spectral function} without loss of generality:
\be\label{specfun}
 \varrho (p) \equiv \int \di^4 x \; \e^{\ii p \cdot x} \left\langle \left[\wphi(x),\wphi^\dagger(0)\right]\right\rangle\;.
\ee
The spectral function depends on the quantum state, namely the density operator $\wrho$ chosen to calculate the expectation
value in the equation \eqref{specfun}. On the other hand, the field operator and its Fourier transform $\wphi_F(p)$ are 
state-independent, so the extraction of the factor $\varrho$ makes the operators $\widehat{A}$ and $\widehat{B}$ apparently 
state-dependent. The choice of the density operator for the calculation of $\varrho(p)$ is arbitrary and it is thus a matter
of convenience; we will find out later on the natural most convenient choice.  If the free field spectral function:
\begin{equation}\label{freespectral}
    \varrho_{\rm free}(p) = 2\pi \; {\rm sign}(p^0) \delta(p^2-m^2) = 2\pi \; {\rm sign}(p^0) \delta(M^2-m^2)\;,
\end{equation}
is plugged in the \eqref{intfield}, the free field expansion in plane waves \eqref{ScalarField:Expansion} is recovered with 
$\wA(p)=\wa(p)$ and $\wB(p)=\wb(p)$. In spite of the similarity between the free field expansion and the \eqref{intfield}, 
it should be kept in mind that in the interacting case the operators $\wA(p)$ and $\wB(p)$ do not fulfill the commutation relations \eqref{CommRelations}, nor they commute with operators of other fields; if, for instance, the field $\wphi$ described the charged 
pion field, the operators $\wA(p)$ and $\wB(p)$ would not commute with those describing the kaon, proton and other particles' field.
Nevertheless, some basic relations for the interacting-field creation and annihilation operators hold. A major one is the following:
\be\label{abtransl}
  \e^{\ii y \cdot \wP} \wA(p) \e^{-\ii y \cdot \wP} = \e^{-\ii p \cdot y} \wA(p)\;, \qquad
   \e^{\ii y \cdot \wP} \wBd(p) \e^{-\ii y \cdot \wP} = \e^{\ii p \cdot y} \wBd(p)\;,
\ee
where $\wP^\mu$ are the generators of space-time translation, i.e the full Hamiltonian and the total momentum. The equation
\eqref{abtransl} ensues from the known transformation of the field under space-time translation:
\be\label{heisenberg}
  \e^{\ii y \cdot \wP} \wphi(x) \e^{-\ii y \cdot \wP} =  \wphi(x+y) \;,
\ee
and the expansion \eqref{intfield}. Equation \eqref{heisenberg} encodes the complete dynamics of the field, since time translation 
corresponds to integrating the Heisenberg equation of motion. Similarly, for an internal U(1) transformation group with generator
$\wQ$ we have:
$$
 \e^{-\ii \varphi \wQ} \wphi(x) \e^{\ii \varphi \wQ} = \e^{\ii \varphi} \wphi(x) 
$$
whence:
\be\label{abtransl2}
\e^{- \ii \varphi \wQ} \wA(p) \e^{\ii \varphi \wQ} = \e^{\ii \varphi} \wA(p)\;, \qquad
  \e^{- \ii \varphi \wQ} \wBd(p) \e^{\ii \varphi \wQ} = \e^{\ii \varphi} \wBd(p)\;.
\ee
The transformation rules \eqref{abtransl} and \eqref{abtransl2} can be straightforwardly extended to complex values of $y$ and
$q$, which is very useful to determine thermodynamic equilibrium values of combinations of those operators.

We are now in a position to work out the Wigner operator with the interacting fields. By using the expansion
\eqref{intfield} into the equation \eqref{ScalarField:WigDef} and suitably extending normal ordering to the operators 
$\wA$ and $\wB$, we obtain:
\begin{equation}\label{ScalarField:WignerFuncExpansion}
\begin{split}
\wW(x,k)&= \frac{2}{(2\pi)^5} \int\di^3\pp\;\di^3\pp'\int ^{+\infty}_{-{\bf p}^2}
\int^{+\infty}_{-{\bf p}'^2}\frac{\di M^2}{2\varepsilon(M,\p)}\frac{\di M'^2}{2\varepsilon(M',\p')}
\varrho(p)\varrho(p')\\
&\times \left\{ \left[\e^{\ii x\cdot\left(p-p'\right)}
\delta^4\left(k-\frac{p+p'}{2}\right)\wAd(p)\wA(p')+ \e^{-\ii x\cdot\left(p-p'\right)}
\delta^4\left(k+\frac{p+p'}{2}\right) \wBd(p)\wB(p')\right]\right.\\
&\left.+ \left[\e^{\ii x\cdot\left(p+p'\right)} \delta^4\left(k-\frac{p-p'}{2}\right)
\wAd(p)\wBd(p')+\e^{-\ii x\cdot\left(p+p'\right)} \delta^4\left(k+\frac{p-p'}{2}\right)
\wB(p)\wA(p')\right]\right\}\;,
\end{split}
\end{equation}
where $\varepsilon(M,\p)=\sqrt{M^2+|\p|^2}=p^0$. Since $p^0+p'^0 \ge 0$ term involving $\wBd(p)\wB(p')$ in the 
equation \eqref{ScalarField:WignerFuncExpansion} does not contributes to $\wW^+(x,k)$, while the combinations 
$\wAd(p)\wBd(p')$ and $\wB(p)\wA(p')$ are not eliminated by the $\theta(k^2)$ unlike in the free field case.

A useful form of the Wigner operator $\wW^+$ (the "+" stands for particles, see definitions \eqref{widecomp}) is obtained by changing 
the integration variables in the equation \eqref{ScalarField:WignerFuncExpansion}. We first extend the integration to a 
four-vector $p$ and enforce the on-shell condition with a delta distribution:
\begin{equation*}
    \int\di^3\pp\int^{+\infty}_{-{\bf p}^2}\frac{\di M^2}{2\sqrt{M^2+{\bf p}^2}}=\int\di^4 p
    \int^{+\infty}_{-{\bf p}^2}\di M^2\delta\left(p^2-M^2\right)\theta(p^0)\;,
\end{equation*}
and likewise for the integral over $\di^3\pp'$. Then we change integration variables as follows:
\begin{equation*}
    P\equiv\frac{p+p'}{2},\quad q\equiv p-p' \implies p=P+q/2 \equiv P_+,\quad p'=P-q/2 \equiv P_-\;,
\end{equation*}
so to obtain:
\begin{align*}
  & \di^4 p \, \di^4 p' \; \delta\left(p^2-M^2\right)\theta(p^0) \delta\left(p'^2-M'^2\right)\theta(p^{\prime\ 0}) \\
   =\, & \di^4 P \, \di^4 q \; \delta\left(P^2+\frac{q^2}{4}+P\cdot q-M^2\right)\delta\left(P^2+\frac{q^2}{4}
   -P\cdot q-M'^2\right)\theta(P^0_+)\theta(P^0_-)\;.
\end{align*}
Now the integrations over $\di^4 P$ or $\di^4 q$ in the equation \eqref{ScalarField:WignerFuncExpansion}
are straightforward and the Wigner operator becomes:
\begin{equation*}
    \begin{split}
\wW^+\left(x,k\right)&=\frac{2}{(2\pi)^5}\int\di^4 q\int^{+\infty}_{-{\bf k}^2_+}\di M^2\int^{+\infty}_{-{\bf k}^2_-} \di M^{\prime 2}\Big\{\varrho(k_+)\varrho(k_-) \, \wAd(k_+)\wA(k_-) 
\, \e^{\ii x\cdot q}\\
&\qquad\qquad\times\delta\left(k^2+\frac{q^2}{4}+k\cdot q-M^2\right)\delta\left(k^2+\frac{q^2}{4}-k\cdot q-M^{\prime2}\right)
\theta(k^0_+)\theta(k^0_-)\Big\} \\
&+ \frac{2^5}{(2\pi)^5}\int\di^4 P \int^{+\infty}_{-{\bf P}^2_+}\di M^2\int^{+\infty}_{-{\bf P}^2_-}
\di M^{\prime 2}\Big\{\varrho(P+k)\varrho(P-k) \, \left[\e^{2 \ii x\cdot P}
\wAd(P+k)\wBd(P-k)+ {\rm h. c.}\right]\\
&\qquad\qquad\times \delta\left(P^2+k^2+2 k\cdot P-M^2\right)\delta\left(P^2+k^2-2 k\cdot P-M'^2\right)
\theta(P^0+k^0)\theta(P^0-k^0)\Big\}\;,
\end{split}
\end{equation*}
where we defined:
\be\label{defpm}
 k_\pm = k \pm q/2\;, \qquad \implies \qquad k_+-k_-=q\;, \qquad k = \frac{k_+ +k_-}{2}\;.
\ee
It is convenient to rename the integration variable in the second integral $q\equiv 2P$
and obtain: 
\begin{equation*}
    \begin{split}
\wW^+\left(x,k\right)&=\frac{2}{(2\pi)^5}\int\di^4 q\int^{+\infty}_{-{\bf k}^2_+}\di M^2\int^{+\infty}_{-{\bf k}^2_-}
\di M^{\prime 2}\Big\{\varrho(k_+)\varrho(k_-) \, \wAd(k_+)\wA(k_-) 
\, \e^{\ii x\cdot q}\\
&\qquad\qquad\times\delta\left(k^2+\frac{q^2}{4}+k\cdot q-M^2\right)\delta\left(k^2+\frac{q^2}{4}-k\cdot q-M^{\prime2}\right)
\theta(k^0_+)\theta(k^0_-)\Big\} \\
&+ \frac{2}{(2\pi)^5}\int\di^4 q \int^{+\infty}_{-{\bf k}^2_+}\di M^2\int^{+\infty}_{-{\bf k}^2_-}
\di M^{\prime 2}\Big\{\varrho(k_+)\varrho(-k_-) \, \left[\e^{\ii x\cdot q}
\wAd(k_+)\wBd(-k_-)+ {\rm h. c.}\right]\\
&\qquad\qquad\times \delta\left(k^2+\frac{q^2}{4}+k\cdot q-M^2\right)\delta\left(k^2+\frac{q^2}{4}-k\cdot q-M^{\prime2}\right)
\theta(k^0_+)\theta(-k^0_-)\Big\}\;.
\end{split}
\end{equation*}
Finally, we can use the \eqref{defpm} to rewrite the arguments of the delta distributions and integrate in $M^2,M'^2$:
\begin{align}\label{wigner2}
&\wW^+\left(x,k\right)=\frac{2}{(2\pi)^5}\int\di^4 q\int^{+\infty}_{-{\bf k}^2_+}\di M^2\int^{+\infty}_{-{\bf k}^2_-} 
\di M^{\prime 2} \delta\left(k_+^2-M^2 \right)\delta(k_-^2-M^{\prime 2}) \nonumber \\
&\times\Bigg\{\varrho(k_+)\varrho(k_-)  \, \e^{\ii x\cdot q} \wAd(k_+)\wA(k_-)\theta(k^0_+)\theta(k^0_-) 
+\varrho(k_+)\varrho(-k_-)\left[\e^{\ii x\cdot q} \wAd(k_+)\wBd(-k_-)+ {\rm h. c.}\right]\theta(k^0_+)\theta(-k^0_-)\Bigg\} 
\nonumber \\
& = \frac{2}{(2\pi)^5}\int\di^4 q \;\e^{\ii x\cdot q} \Big\{\varrho(k_+)\varrho(k_-)\wAd(k_+)\wA(k_-)\theta(k^0_+)\theta(k^0_-) \nonumber
\\
&\qquad\quad\qquad\qquad\qquad+\varrho(k_+)\varrho(-k_-)\left[\wAd(k_+)\wBd(-k_-)+ {\rm h. c.}\right]\theta(k^0_+)\theta(-k^0_-)\Big\}\,.
\end{align}
Before ending this Section, we derive the form of the expectation values of combinations of $\wA(p)$
and $\wB(p)$ operators at global thermodynamic equilibrium, with a density operator:
\be\label{gedens}
\wrho_{\rm GE}= \frac{1}{Z}\exp\left[-\beta \cdot \wP + \zeta \wQ\right]\;,
\ee
with $\beta$ and $\zeta$ constant, which will be very useful for the rest of this work. 
From equation \eqref{invfourier} and the definition of Fourier transform of the field we obtain:
\begin{align*}
& \langle \wAd(p)\wA(p')\rangle_{\rm GE} = \frac{1}{(2\pi)^3} \frac{\theta(p^0)\theta(p'^0)}{\varrho(p)\varrho(p')} \langle 
\wphi^\dagger_F(p)\wphi_F(p')\rangle_{\rm GE} = \frac{1}{(2\pi)^3} \frac{\theta(p^0)\theta(p'^0)}{\varrho(p)\varrho(p')}
 \int \di^4 x \, \di^4 x' \e^{-\ii p \cdot x} \e^{\ii p' \cdot x'} \langle \wphi^\dagger(x)\wphi(x')
 \rangle_{\rm GE} \\
& = \frac{1}{(2\pi)^3} \frac{\theta(p^0)\theta(p'^0)}{\varrho(p)\varrho(p')}
 \int \di^4 x \, \di^4 x' \e^{-\ii p \cdot x} \e^{\ii p' \cdot x'} \langle \wphi^\dagger(0)\wphi(x'-x)\rangle_{\rm GE}\\
 &
 = \frac{1}{(2\pi)^3} \frac{\theta(p^0)\theta(p'^0)}{\varrho(p)\varrho(p')}
 \int \di^4 x \, \di^4 y \; \e^{-\ii (p-p') \cdot x} \e^{\ii p' \cdot y} \langle \wphi^\dagger(0)\wphi(y) \rangle_{\rm GE}=\frac{2\pi\theta(p^0)\theta(p'^0)}{\varrho(p)\varrho(p')} \delta^4(p-p') \int \di^4 y \; \e^{\ii p' \cdot y} \langle \wphi^\dagger(0)\wphi(y) \rangle_{\rm GE}\;,
 \end{align*}
where we have taken advantage of translational invariance of the density operator and changed the integration
variable from $x'-x=y$ . Now, by using the definition of the lesser Wightman function at global equilibrium:
$$
 {\cal G}^<_{\rm GE}(q) \equiv \int \di^4 y \; \e^{\ii q \cdot y} \langle \wphi^\dagger(0)\wphi(y) \rangle_{\rm GE}\;,
$$
and its known relation with the spectral function \cite{LeBellac_1996}:
$$
 {\cal G}^<_{\rm GE}(q) = \frac{1}{\e^{\beta \cdot q - \zeta}-1}\, \varrho_{\rm GE}(q) = n_{\rm B}(q)\, \varrho_{\rm GE}(q)\;,
$$
we get the equation:
\begin{align}\label{tevaa}
\langle \wAd(p)\wA(p')\rangle_{\rm GE} & = \frac{2\pi}{\varrho^2(p)} \theta(p^0)\theta(p'^0)
\delta^4(p-p') n_B(p) \varrho_{\rm GE}(p) \nonumber \\
& = ({\rm if}\; \varrho(p) = \varrho_{\rm GE}(p)) \;\; 
\frac{2\pi}{\varrho(p)} \theta(p^0)\theta(p'^0) 2p^0 \delta(p^2-p'^2) \delta^3(\p-\p') n_B(p)\;.
\end{align}
Thus, in order to obtain the last simple form of the expectation values in \eqref{tevaa}, the spectral function 
in the field expansion \eqref{intfield} must be the one calculated with the density operator \eqref{gedens}; 
only in this case a cancellation between the spectral function in the numerator and denominator occurs. 

Similarly, we can obtain:
\begin{align} \label{tevab}
\langle \wAd(p)\wBd(p')\rangle_{\rm GE} &= \frac{2\pi}{\varrho(p')} \theta(p^0)\theta(p'^0) 
\delta^4(p+p') n_B(p) = 0 \;,
\end{align}
whence, from the hermiticity of the density operator:
\be\label{tevba}
\langle \wB(p)\wA(p')\rangle_{\rm GE} = \langle \wAd(p')\wBd(p)\rangle_{\rm GE}^*= 0\;.
\ee
%

\section{Density operator, gradient expansion and dissipation}
\label{density}

If we are to calculate the Wigner function, according to the equation \eqref{meanwig}, we need to know the quantum state of the 
system, that is the density operator. In the Heisenberg representation, this state is stationary and is thus fixed by the initial
conditions of the problem. 

In many applications of non-equilibrium thermal quantum field theory, the initial state is defined in the infinite past as 
that corresponding to global thermodynamic equilibrium with an unperturbed Hamiltonian. 
One of the typical problems is the calculation of the response of some quantities (currents etc.) to a perturbation of the 
Hamiltonian at the time $t=0$, like e.g. turning on an electric or a magnetic field etc. This scheme is also used to calculate 
intrinsic transport coefficients such as shear viscosity; in this case, the perturbation is a modification of the Minkowskian 
metric tensor \cite{Jeon:1994if}. 

However, in many other problems (such as the evolution of the Quark Gluon Plasma in relativistic nuclear collisions), there
is no external perturbation and one needs to describe the dynamics without changing the Hamiltonian. In this regard, the 
non-equilibrium stationary density operator approach known as Zubarev's \cite{Zub2} and its reformulation 
\cite{Becattini:2019dxo} is a well suited method. This formalism was employed, for instance, to derive the Kubo formula of shear viscosity 
\cite{Hosoya:1983id} and it is especially fit for relativistic heavy ion collisions, where the initial state is assumed
to be local thermodynamic equilibrium over a space-like hypersurface $\Sigma_0$. The corresponding density operator reads:
\begin{equation}\label{densop}
    \wrho=\wrho_{\rm LE}(\tau_0)=\frac{1}{Z}\exp\left[-\int_{\Sigma_{0}}\di\Sigma_\mu(y)
    \left(\wT^{\mu\nu}(y) \beta_\nu(y)-\wj^\mu(y)\zeta(y)\right)\right]\;,
\end{equation}
where $\wT^{\mu\nu}$ is the stress-energy operator; $\wj^\mu$ is a conserved charge current; $\beta(y)$ is the four-temperature 
vector field and $\zeta(y) = \mu/T$ is the reduced chemical potential and $\tau_0$ can be associated with the time at which the local equilibrium is attained \cite{Becattini:2019dxo}. The hypersurface $\Sigma_0$ plays the role of an initial 
Cauchy hypersurface for the evolution of quantum fields and it is the starting point for hydrodynamic simulations. 

\begin{figure}[h!]
    \centering
    \includegraphics[width=0.5\linewidth]{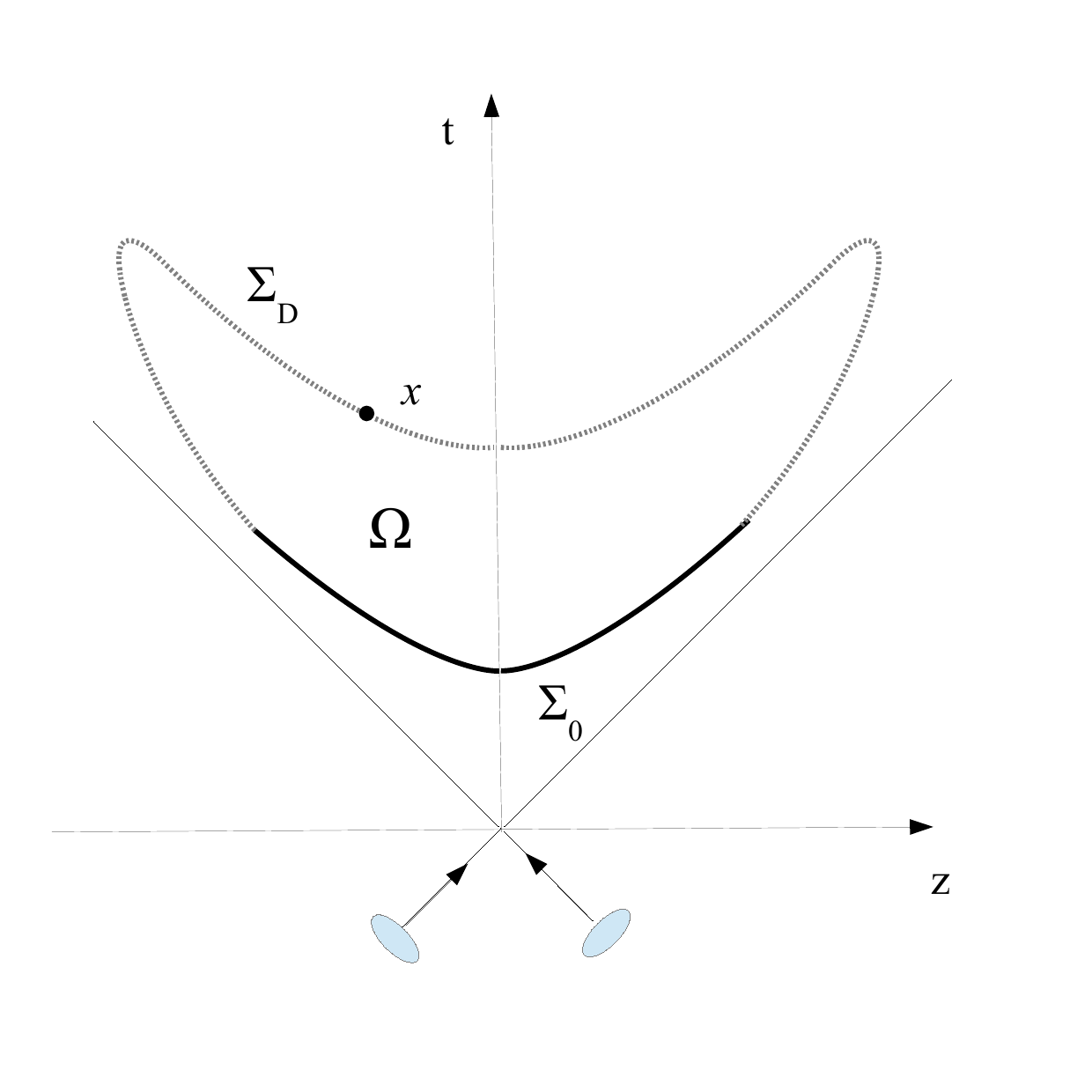}
    \caption{Schematic illustration of a nuclear collision at high energy. The initial local equilibrium hypersurface 
    is $\Sigma_0$ (solid line) and the decoupling hypersurface $\Sigma_{\rm D}$ (finely dotted line); the Quark Gluon 
    Plasma as a fluid lives in the encompassed region $\Omega$. A point $x$ on $\Sigma_{\rm D}$ is the most suitable place 
    where an approximate expression of the spectrum can be obtained from the Wigner operator.}
    \label{fig:Gauss Theorem Region}
\end{figure}

Suppose we want to calculate the mean value of an $x$-dependent operator, where $x$ lies in the future of $\Sigma_0$ (see figure 
\ref{fig:Gauss Theorem Region}). In principle, the density operator $\wrho$ in the equation \eqref{densop} is a {\em functional}
of the fields at the hypersurface $\Sigma_0$, that is it depends on the functional form of the fields $\beta$, $\zeta$ and $n$
where $n$ is the unit four-vector perpendicular to the hypersurface. Therefore, for the Wigner function we can write:
$$
  W^+(x,k) = \Tr (\wrho\,\wW^+(x,k)) = W^+[\beta_0,\zeta_0,n_0](x,k)\;,
$$
where the squared brackets stand for the functional dependence and the subscript $0$ denotes that the fields are evaluated on 
the initial hypersurface $\Sigma_0$. The functional dependence can be transformed into a dependence on an infinite number of 
arguments provided that the functions $\beta_0,\ \zeta_0,\ n_0$ are infinitely differentiable, so that they can be replaced by the 
values of their derivatives at some point $x_0$ on the hypersurface, which can be a function of $x$ and $k$. In symbols:
$$
 \Tr (\wrho\,\wW^+(x,k)) = W^+(x,k,\beta_0(x_0),\partial \beta_0(x_0),\partial^2 \beta_0(x_0),\ldots,\zeta(x_0),\partial \zeta(x_0),
 \ldots,n_0(x_0),\partial n_0(x_0),\ldots)\;,
$$
where the dependence of $x^0$ on $x,k$ is understood.
The function on the right hand side can be now expanded in a power series of the arguments, that is the gradients of the fields:
\be\label{gradexp}
 \Tr (\wrho\,\wW^+(x,k)) = w_0(x,k,\beta_0(x_0(x,k)) + w_1(x,k,\beta_0(x_0(x,k)) \partial\beta_0(x_0(x,k)) + \ldots
\ee
where the dots include terms which are powers of gradients of any order.
This expansion, truncated at some order, provides a good approximation of the actual value of the Wigner function 
at the point $x$ and for the four-momentum $k$ once truncated if the functions $\beta_0,\zeta_0,n_0$ are slowly varying 
compared to the intrinsic length scales of the problem. 
An example where this kind of expansion provides a good approximation is the free field Wigner function. In this case, the 
exact solution is known since the free Wigner function fulfills the equation:
$$
  k \cdot \partial W^+(x,k) = 0\;,
$$
and the free-streaming solution is:
\be\label{fssol}
 W^+\left(x,k\right) = W^+_0\left(x^0_0(k),\,\x-\frac{\kb}{k^0}\left(x^0-x^0_0\right),\,k\right)\;,
\ee
where $x_0$ is the intersection point of the characteristic line drawn from the point $x$ and the hypersurface $\Sigma_0$
and $W_0^+$ the Wigner function at $\Sigma_0$. The $W^+_0$ is simply the local equilibrium solution, whose gradient expansion 
as a function of the gradients calculated in the arguments of the right hand side of \eqref{fssol} can be obtained by the methods 
presented in ref. \cite{Sheng:2025cjk}. Thereby, an expansion of the kind \eqref{gradexp} is obtained.

If, on the other hand, the system behaves like a coupled fluid between $\Sigma_0$ and $x$, a different gradient expansion of 
the Wigner function is believed to provide a better approximation with respect to the \eqref{gradexp} once truncated 
at some fixed order. This is a series in gradients of the fields calculated in the same point $x$ where the Wigner function is 
evaluated:
\be\label{gradexp2}
 \Tr (\wrho\,\wW^+(x,k)) = w'_0(x,k,\beta(x)) + w'_1(x,k,\beta(x)) \partial \beta(x) + \ldots\;.
\ee
At least in theory, the \eqref{gradexp2} can be obtained from the \eqref{gradexp} itself by taking into account that the 
fields $\beta$ and $\zeta$ evolve according to the deterministic equations of relativistic hydrodynamics, so that $\beta(x),\zeta(x)$ are 
themselves functionals of $\beta_0(x)$, $\zeta_0(x)$ and $n_0(x)$. An expansion of the kind \eqref{gradexp2} can be obtained
by transforming the density operator \eqref{densop} by means of the Gauss' theorem \cite{Becattini:2019dxo} choosing a
hypersurface $\Sigma$ (a natural choice for $\Sigma$ in the context of HIC would be the decoupling hypersurface $\Sigma_{\rm D}$) passing through the point $x$:
\begin{equation}\label{gauss}
    \wrho=\frac{1}{Z}\exp\left[-\int_{\Sigma}\di\Sigma_\mu \; \left(\wT^{\mu\nu}\beta_\nu
     -\zeta \wj^\mu \right)+\int_\Omega\di^4y \; 
     \left(\wT^{\mu\nu}\partial_\mu\beta_\nu -\wj^\mu \partial_\mu\zeta  \right)\right]
     = \frac{\exp\left[\wE+\wD \right]}{\Tr\left(\exp\left[\wE +\wD \right]\right)}      \;,
\end{equation}
where $\Omega$ is the region (see fig.\ref{fig:Gauss Theorem Region}) enclosed by $\Sigma_0$ and $\Sigma$
and:
\begin{subequations}\label{eandd}
    \begin{align}
\wE&=-\int_{\Sigma}\di\Sigma_\mu \; \left(\wT^{\mu\nu}\beta_\nu-\zeta \wj^\mu \right)\;,\\
\wD&=\int_\Omega\di^4 y \; \left(\wT^{\mu\nu} \partial_\mu\beta_\nu - \wj^\mu \partial_\mu\zeta 
\right)\;.
\end{align}
\end{subequations}
The operator $\wE$ corresponds to the local equilibrium at the hypersurface $\Sigma$ defining the "simultaneity" space of $x$, 
that is at the present time, where the Wigner operator is assumed to be computed. Conversely, the second term $\wD$ corresponds to the dissipation \cite{Becattini:2019dxo}. 
The interpretation of the term operator $\wD$ as source of the dissipative terms of the observables follows from the entropy
production rate equation \cite{VANWEERT1982133,Becattini:2019dxo}:
\begin{equation}\label{Entropy production}
    \partial_\mu s^\mu(x)= \left(T^{\mu\nu}(x)-T^{\mu\nu}_{\rm LE}(x)\right)\partial_\mu\beta_\nu(x)- 
    \left(j^\mu(x)-j^\mu_{\rm LE}\right) \partial_\mu\zeta(x)\;,
\end{equation}
where $T^{\mu\nu}(x)$ and $j^\mu(x)$ are the actual mean values calculated with the density operator \eqref{densop}
or \eqref{gauss}, whereas those with the subscript LE are calculated with the density operator of local equilibrium
at the present time:
\begin{equation}\label{leqrho}
   \wrho_{\rm LE} = \frac{\e^\wE}{\Tr\left( \e^\wE \right)}\;.  
\end{equation}
If the system is not too far from equilibrium, the second term $\wD$ proportional to the gradients of the thermo-hydrodynamic
fields $\beta$ and $\zeta$ is supposedly "smaller" than $\wE$ and the \eqref{gauss} can be expanded in $\wD$ when calculating
the mean value of an operator $\wO$. The leading term of the expansion calculated with the density operator \eqref{leqrho},
corresponds to the local equilibrium value at the present time $\wO_{\rm LE}$ whereas the other terms are the {\em dissipative} 
corrections:
\begin{equation}\label{leq+dis}
    \langle\wO\rangle = \Tr (\wrho \, \wO ) = \Tr (\wrho_{\rm LE} \wO) +
    \Delta O_{\rm diss} = O_{\rm LE} + \Delta O_{\rm diss}\;.
\end{equation}
For a local operator $\wO(x)$, at the leading order of the expansion in $\wD$, the dissipative term turns out to be:
\begin{equation}\label{traditional}
    \begin{split}
 \Delta O(x)_{\rm diss} &= \int_{\Omega} \di^4 y \int^1_0\di z \; 
 \langle\wO\left(x\right), \e^{z\wE} \left(\wT^{\mu\nu}(y)\partial\beta_\nu(y)-
 \wj^\mu\partial\zeta(y)\right)\e^{-z\wE}\rangle_{c, \mathrm{LE}}\\
 & \equiv \int_{\Omega} \di^4 y \; \left( C_{OT}(x,y)\partial_\mu\beta_\nu(y) - C_{Oj}(x,y)\partial_\mu \zeta \right)
 \;,
\end{split}
\end{equation}
where the subscript $c$ stands for correlation, that is:
$$
   \langle \wX, \wY\rangle_c= \langle \wX \wY \rangle - \langle \wX \rangle \langle \wY \rangle\;.
$$
Tipically, the correlation functions $C_{OT}$ and $C_{Oj}$ are peaked around $y \sim x$ with a width
governed by microscopic scales of the theory (mass, temperature, interaction length) which is much smaller 
than the length of variation of the thermo-hydrodynamic fields $\beta$ and $\zeta$; this is the so-called hydrodynamic 
limit. With this assumption, the slowly varying gradients of $\beta$ and $\zeta$ can be evaluated around the point 
$y \sim x$ yielding:
\be\label{traditional2}
 \Delta O(x)_{\rm diss} \simeq \partial_\mu\beta_\nu(x) 
  \int_{\Omega} \di^4 y \; C_{OT}(x,y) - \partial_\mu\zeta(x) \int_{\Omega} \di^4 y \; C_{Oj}(x,y) \;.
\ee
In turn, the local equilibrium expectation value $\langle \wO(x) \rangle_{\rm LE}$ can be expanded from the global 
equilibrium value. Following ref. \cite{Sheng:2025cjk}, one can write, being $x$ the point where the local operator is
evaluated:
\begin{equation}\label{deltabeta}
    \beta_\nu(y) = \beta_\nu(x) + (\beta_\nu(y) - \beta_\nu(x)) = \beta_\nu(x) + \Delta\beta_\nu(y,x)\;,
    \qquad  \zeta(y) = \zeta(x) + (\zeta(y)-\zeta(x)) = \zeta(x) + \Delta\zeta(y,x)\;,
\end{equation}
and the local equilibrium density operator can be rewritten as:
\begin{equation}\label{densop2}
    \wrho =\frac{1}{Z}\exp\left[-\beta(x) \cdot \wP + \zeta(x) \wQ - \int_{\Sigma}\di\Sigma_\mu(y)
    \left(\wT^{\mu\nu}(y) \Delta\beta_\nu(y,x)-\wj^\mu(y)\Delta\zeta(y,x)\right)\right]\;,
\end{equation}
where, owing to the continuity equations:
\begin{equation}\label{conservation}
    \partial_\mu\wT^{\mu\nu}(x)=0\;,\qquad \partial_\mu\wj^\mu(x)=0\;,
\end{equation}
we have:
$$
 \wP^\nu = \int_\Sigma \di \Sigma_\mu \; \wT^{\mu\nu}\;,  \qquad \wQ = \int_\Sigma \di \Sigma_\mu \; \wj^{\mu}\;,
$$
for any arbitrary space-like hypersurface $\Sigma$.
The basic idea is that the mean value of $\wO(x)$ is mostly determined by the values of the thermo-hydrodynamic fields 
$\beta$ and $\zeta$ around $x$ if they are slowly varying. Hence, if $\Delta \beta$ and $\Delta \zeta$ are not large we 
can expand the density operator around the global equilibrium configuration:
\be\label{geq}
 \wrho_{\rm GE} = \frac{1}{Z_{\rm GE}}\exp\left[-\beta(x) \cdot \wP + \zeta(x) \wQ \right]\;
= \frac{\exp\left[\wE_{\rm GE}\right]}{\Tr(\exp[\wE_{\rm GE}])}\;.
\ee
We thus have, for a local operator $\wO(x)$:
\begin{equation}\label{lrtO}
    \langle\wO(x)\rangle_{\rm LE} \simeq \langle\wO(x)\rangle_{\rm GE}+ \Delta O(x)_{\rm LE}\;,
\end{equation}
where:
\begin{equation*}
    \langle\wO(x)\rangle_{\rm GE}= \frac{\Tr \left(\e^{\wE_{\rm GE}} \wO(x) \right)}
    {\Tr\left( \e^{\wE_{\rm GE}}\right)}\;,
\end{equation*}
is the {\em global equilibrium} value calculated at the four-temperature $\beta(x)$ and reduced chemical potential 
$\zeta(x)$, while $\Delta O(x)$ is the correction to the global equilibrium value which, at the leading order
in $\Delta\beta$, and $\Delta \zeta$ reads:
\begin{equation}\label{deltaO}
    \Delta O(x)_{\rm LE}=-\int_{\Sigma}\di\Sigma_\mu(y) \; \int^1_0\di z \; \langle\wO(x),
    \e^{z\wE_{\rm GE}}\left(\wT^{\mu\nu}(y)\Delta\beta_\nu(y,x)-\wj^\mu\Delta\zeta(y,x)\right)\e^{-z\wE_{\rm GE}}
    \rangle_{c,\mathrm{GE}} \;.
\end{equation}
The equation \eqref{deltaO} can be shown to generate an expansion in the gradients of the fields $\beta,\zeta,n$ evaluated
in the point $x$ like \eqref{gradexp2} (with some provisoes, see ref. \cite{Sheng:2025cjk}). 

Plugging the gradient expansion of the \eqref{deltaO} and the equation \eqref{traditional2} into the equation
\eqref{leq+dis}, one obtains the leading terms of the general expansion of the observable $\langle O \rangle$ into gradients
at the point $x$, of the kind \eqref{gradexp2}. This approach, based on the splitting in eq. \eqref{gauss}, proved to be 
very fruitful for the determination of the constitutive equations of conserved currents \cite{Hosoya:1983id} 
and corresponding Kubo formulae, as well as for the calculation of the spin polarization at local thermodynamic equilibrium
\cite{Becattini:2021suc,Liu:2021uhn}. Yet, this method has a crucial requirement: for the dissipative term it relies on the assumption
of narrow-peaked correlation functions $C(x-y)$ around $y \sim x$, see transition from equation \eqref{traditional} to
\eqref{traditional2}. As we will see, this assumption may not hold (and, most likely it does not) for the correlator between 
the Wigner operator and the conserved currents $\wT^{\mu\nu}$ and $\wj^\mu$. In this case, one should keep the dissipative
term in its integral form \eqref{traditional}, like in the original Kubo formula, expand the gradients of $\beta$ in a 
Taylor series about $x$ and evaluating the integral coefficients. This is of course possible, but it is much more complicated
and it will make the truncation of the gradient expansion at low orders a not so good approximation. 

In this paper, for the problem of the decoupling fluid, we thus propose a new method, which will eventually lead to an 
expansion of the form \eqref{gradexp}, that is in the gradients of the initial fields. Instead of choosing a hypersurface 
$\Sigma$ passing through the point $x$, we use the original form \eqref{densop} of the density operator and we evaluate 
$\wO(x)$ by first decomposing $\beta(y)$ and $\zeta(y)$ like in the equation \eqref{deltabeta} and then writing the \eqref{densop} 
as:
\begin{equation}\label{densop3}
    \wrho =\frac{1}{Z}\exp\left[-\beta(x) \cdot \wP + \zeta(x) \wQ - \int_{\Sigma_0}\di\Sigma_\mu(y)
    \left(\wT^{\mu\nu}(y) \Delta\beta_\nu(y,x)-\wj^\mu(y)\Delta\zeta(y,x)\right)\right]\;,
\end{equation}
that is without applying the Gauss theorem in the first place. If $\Delta\beta(y,x)$ and $\Delta\zeta(y,x)$ are sufficiently
small, we can expand at the linear order and obtain:
\begin{equation}\label{linresp}
    \langle\wO(x)\rangle \simeq \langle\wO(x)\rangle_{\rm GE}+ \Delta O(x)\;,
\end{equation}
where:
\begin{equation}\label{ddO}
    \Delta O(x) = -\int_{\Sigma_0}\di\Sigma_\mu(y) \; \int^1_0\di z \; \langle\wO(x),
    \e^{z\wE_{\rm GE}}\left(\wT^{\mu\nu}(y)\Delta\beta_\nu(y,x)-\wj^\mu\Delta\zeta(y,x)\right)\e^{-z\wE_{\rm GE}}
    \rangle_{c,\mathrm{GE}} \;,
\end{equation}
will be henceforth denoted as the off-equilibrium correction of the observable $\wO(x)$. The equation \eqref{ddO}
looks very similar to the equation \eqref{deltaO} except for the crucial difference that the integration is now
over the hypersurface $\Sigma_0$ where local equilibrium is established and not over $\Sigma$ passing
through $x$. In fact, the term $\Delta O(x)$ in the equation \eqref{ddO} includes both the 
local equilibrium correction to the global equilibrium term quoted in eq. \eqref{deltaO} and the dissipative term
in eq. \eqref{traditional}. This can be readily shown by applying the Gauss theorem to the equation \eqref{ddO},
yielding:
\begin{equation}\label{gauss2}
    \begin{split}
         \Delta O(x) & =-\int_{\Sigma}\di\Sigma_\mu(y) \; \int^1_0\di z \; \langle\wO(x),
    \e^{z\wE_{\rm GE}}\left(\wT^{\mu\nu}(y)\Delta\beta_\nu(y,x)-\wj^\mu\Delta\zeta(y,x)\right)\e^{-z\wE_{\rm GE}}
    \rangle_{c,\mathrm{GE}}  \\
    & + \int_{\Omega}\di^4 y \; \int^1_0\di z \; \langle\wO(x),
    \e^{z\wE_{\rm GE}}\left(\wT^{\mu\nu}(y)\partial_\mu \beta_\nu(y)
    -\wj^\mu \partial_\mu \zeta(y)\right)\e^{-z\wE_{\rm GE}} \rangle_{c,\mathrm{GE}}\;.
    \end{split}
\end{equation}
The first term on the right hand side of \eqref{gauss2} is indeed the leading order correction $\Delta O(x)_{\rm LE}$ in 
equation \eqref{deltaO}, while the second term is the dissipative correction, at the leading order, $\Delta O(x)_{\rm diss}$ 
in the equation \eqref{traditional} evaluated at global equilibrium, i.e. at the leading order of the expansion of
the local equilibrium density operator at the present time. 

Since in the equation \eqref{ddO} the integration is not over a 4D domain but over the initial hypersurface, the point 
$y$ can be far from $x$ and one may wonder about the impossibility of evaluating the correlation functions for $y \sim x$ 
as usual. However, as has been mentioned, the behaviour of the correlation function of the Wigner operator and the conserved currents is significantly different from the naive expectation.

\section{Off-equilibrium scalar Wigner function}
\label{neqwigner}

We are now going to work out the Wigner function with the method described at the end of the previous Section. According to 
the equation \eqref{lrtO}, we have, with $\wW^+$ given by the equation \eqref{wigner2}:
\begin{equation*}
    \langle \wW^+(x,k)\rangle \simeq \langle\wW^+(x,k)\rangle_{\rm GE}+ \Delta W^+(x,k)\;,
\end{equation*}
where $\Delta W^+(x,k)$ is obtained by replacing $\wO(x)$ with $\wW^+(x,k)$ in the equation 
\eqref{ddO}. The main term $\langle\wW^+(x,k)\rangle_{\rm GE}$ is readily found by using the equation \eqref{wigner2} and the
relations \eqref{tevaa}, \eqref{tevab} and \eqref{tevba}:
\begin{equation}\label{leading order W}
    \langle\wW^+\left(x,k\right)\rangle_{\rm GE}=\frac{2}{\left(2\pi\right)^4}\,
    n_{\rm B}(\beta(x)\cdot k) \, \varrho(k)\;,
\end{equation}
where $\varrho(k)$ is the spectral function calculated with $\wrho$ in \eqref{gedens} with $\beta=\beta(x)$
and $\zeta=\zeta(x)$. Plugging the above expression into the \eqref{Interacting spectrum}, 
integrating in $x$ over the hypersurface of decoupling one obtains the well known relativistic formula for the momentum 
spectrum of particles emitted from a decoupling fluid at local equilibrium, known as Cooper-Frye:
\begin{equation*}
  \frac{\di N_k}{\di^3 \kk} = \frac{2}{(2\pi)^4}\int_0^{+\infty} \di k^0 \; \varrho(k) 
  \int_{\Sigma_{\rm D}} \di\Sigma \cdot k \; n_B (\beta(x) \cdot k)\;,
\end{equation*}
where $\varrho(k)$ encodes the interaction corrections at the decoupling. Note that $k$ is off-shell and $k^2=m^2$ only in 
the limit of free fields for which $\varrho(k)\mapsto\varrho_{\rm free}(k)\propto\delta(k^2-m^2)$.

The calculation of the linear response correction $\Delta W^+(x,k)$ proceeds just like in the ref. \cite{Sheng:2025cjk} 
integrating over the hypersurface $\Sigma_0$ instead of $\Sigma_{\rm D}$ and replacing the free field expansion with
the interacting one in eq. \eqref{intfield}. By using the eq. \eqref{wigner2} and $\wW(x,k)$ replacing $\wO(x)$ in the
eq. \eqref{ddO}:
\begin{align}\label{deltawig1}
 &\Delta W^+(x,k) = - \frac{2}{(2\pi)^5} \int_{\Sigma_{0}} \di \Sigma_{\mu}(y)  
  \int \di^{4}q\;
   \nonumber\\
&\times \Bigg\{\varrho(k_+)\varrho(k_-)\theta(k_+^0)\theta(k_-^0) \e^{\ii x\cdot q}\int_0^1 \di z \, \left\langle 
\wAd (k_+) \wA (k_-), \e^{z\wE_{\rm GE}} \left(
\wT^{\mu\nu}(y) \Delta\beta_\nu(y,x) - \wj^\mu(y) \Delta\zeta(y,x) \right) \e^{-z\wE_{\rm GE}} 
   \right\rangle_{c,{\rm GE}}  \nonumber\\
& + \varrho(k_+)\varrho(-k_-) \theta(k^0_+)\theta(-k^0_-)  \int_0^1 \di z  \left[ \e^{\ii x\cdot q} 
\left\langle \wAd(k_+)\wBd(-k_-),
\e^{z\wE_{\rm GE}} \left(\wT^{\mu\nu}(y) \Delta\beta_\nu(y,x) - \wj^\mu(y) \Delta\zeta(y,x) \right) 
\e^{-z\wE_{\rm GE}} \right\rangle_{c,{\rm GE}} \right. \nonumber \\
& + \left. \left. \e^{-\ii x\cdot q} \left\langle \wB(-k_-)\wA(k_+), \e^{z\wE_{\rm GE}} \left(\wT^{\mu\nu}(y) 
\Delta\beta_\nu(y,x) - \wj^\mu(y) \Delta\zeta(y,x) \right) \e^{-z\wE_{\rm GE}} \right\rangle_{c,{\rm GE}} \right] 
\right\}\;,
\end{align}
with $\wE_{\rm GE}$ as in the equation \eqref{geq}. The equation \eqref{deltawig1} can be further worked out by
extracting exponential factors. Since:
\begin{equation}\label{transtranf}
    \wT^{\mu\nu}(y)=\e^{\ii\wP\cdot y}\,\wT^{\mu\nu}(0)\,\e^{-\ii\wP\cdot y}\;,
    \qquad \wj^{\,\mu}(y)=\e^{\ii\wP\cdot y}\,\wj^{\,\mu}(0)\,\e^{-\ii\wP\cdot y}\;,
\end{equation}
where $\wP$ is the total four-momentum (including interacting terms), and taking into account that the operator 
$\wE_{\rm GE}=-\beta(x)\cdot\wP+\zeta(x)\wQ$ commutes with the generator of translations and the main term of the density 
operator $\e^{\wE_{\rm GE}}/\Tr(\e^{\wE_{\rm GE}})$, we obtain for the $\wAd\wA$ term:
\begin{eqnarray*}
&& \left\langle \wAd (k_+) \wA (k_-),\; \e^{z\wE_{\rm GE}} \left(
   \wT^{\mu\nu}(y) \Delta\beta_\nu(y,x) - \wj^\mu(y) \Delta\zeta(y,x) \right) \e^{-z\wE_{\rm GE}} 
   \right\rangle_{c,{\rm GE}} \\
&& =\left\langle 
\e^{-z\wE_{\rm GE}-iy\cdot\wP}
\wAd (k_+) \wA (k_-)\e^{z\wE_{\rm GE}+iy\cdot\wP} ,\;
   \left( \wT^{\mu\nu}(0) \Delta\beta_\nu(y,x) - \wj^\mu(0) \Delta\zeta(y,x) \right) \right\rangle_{c,{\rm GE}}\;;
\end{eqnarray*}
for the terms involving $\wAd\wBd$ and $\wB\wA$ a similar expression can be derived. By using the transformation properties \eqref{abtransl} and \eqref{abtransl2} and taking into account that $[\wP^\mu,\wQ]=0$, it can be readily shown that the 
quadratic combinations of $\wA$ and $\wB$ operators fulfill the following relations:
\begin{equation*}
 \begin{split}
\e^{-z\wE_{\rm GE}-iy\cdot\wP}\wAd(k_+)\wA(k_-)\e^{z\wE_{\rm GE}+iy\cdot\wP} &=\e^{-\ii y\cdot q }\e^{z\beta(x)\cdot q}\wAd(k_+)\wA(k_-)\;,\\
\e^{-z\wE_{\rm GE}-iy\cdot\wP}\wAd(k_+)\wBd(-k_-)\e^{z\wE_{\rm GE}+iy\cdot\wP} &=\e^{-\ii y\cdot q }\e^{z\beta(x)\cdot q}\wAd(k_+)\wBd(-k_-)\;,\\
\e^{-z\wE_{\rm GE}-iy\cdot\wP}\wB(-k_-)\wA(k_+)\e^{z\wE_{\rm GE}+iy\cdot\wP} &=\e^{\ii y\cdot q }\e^{-z\beta(x)\cdot q}\wB(-k_-)\wA(k_+)\;,
 \end{split}
\end{equation*}
where we used the equality $k_+-k_-=q$ in the equation \eqref{defpm}. Replacing the above expression in 
\eqref{deltawig1} and integrating in $z$ we obtain:
\begin{eqnarray}\label{deltawig2}
&&\Delta W^+(x,k) =\frac{2}{(2\pi)^5} \int_{\Sigma_{0}} \di \Sigma_{\mu}(y) 
\int \di^{4}q\Bigg\{ \varrho(k_+)\varrho(k_-)\frac{1 -\e^{\beta(x)\cdot q}}{\beta(x) \cdot q}
\e^{\ii q\cdot (x-y)}\; \theta(k_+^0)\theta(k_-^0)\nonumber \\
& \times& \left( \left\langle \wAd (k_+) \wA(k_-),\; \wT^{\mu\nu}(0) 
\right\rangle_{c,{\rm GE}} \!\!\! \Delta\beta_\nu(y,x) - \left\langle \wAd (k_+) \wA(k_-),\; \wj^{\mu}(0) 
\right\rangle_{c,{\rm GE}} \!\!\! \Delta\zeta(y,x) \right) \\
&+& \varrho(k_+) \varrho(-k_-) \theta(k_+^0)\theta(-k_-^0)\frac{1-\e^{\beta(x)\cdot q}}{\beta(x)\cdot q}
\nonumber\\
&\times&\left(\e^{\ii q\cdot\left(x-y\right)} \left\langle \wAd (k_+) \wBd(-k_-),\; \wT^{\mu\nu}(0) 
\right\rangle_{c,{\rm GE}} \!\!\! \Delta\beta_\nu(y,x) - \left\langle \wAd (k_+) \wBd(-k_-),\; \wj^{\mu}(0) 
\right\rangle_{c,{\rm GE}} \!\!\! \Delta\zeta(y,x) +{\rm c.c}\right)\Bigg\}\;,\nonumber
\end{eqnarray}
where we took advantage of the relation:
\begin{equation*}
    \begin{split}
        \left\langle \wAd (k_+) \wBd(-k_-),\; \wT^{\mu\nu}(0) 
\right\rangle^*_{c,{\rm GE}}&=
\e^{-\beta\cdot q}
\, \left\langle \wB (-k_-) \wA(k_+),\; \wT^{\mu\nu}(0) 
\right\rangle_{c,{\rm GE}}\;,\\
\left\langle \wAd (k_+) \wBd(-k_-),\; \wj^{\mu}(0) 
\right\rangle^*_{c,{\rm GE}}&=
\e^{-\beta\cdot q} \, \left\langle \wB (-k_-) \wA(k_+),\; \wj^{\mu}(0) 
\right\rangle_{c,{\rm GE}}\;.
    \end{split}
\end{equation*}
again ensuing from the equations \eqref{abtransl} and \eqref{abtransl2}.

\section{Thermal-Gravitational form factors}
\label{formfactors}

The non-equilibrium correction \eqref{deltawig2} depends on the correlators:
\be\label{thff}
\Theta^{\mu\nu}(k,q,\beta) \equiv \langle\wAd(k_+)\wA(k_-),\;\wT^{\mu\nu}(0)\rangle_{c,{\rm GE}} \;,
\ee
as well as 
\be\label{thff2}
\langle\wAd(k_+)\wBd(-k_-),\;\wT^{\mu\nu}(0) \rangle_{c,{\rm GE}}\;, 
\qquad \qquad \langle\wB(-k_-)\wA(k_+),\;\wT^{\mu\nu}(0)\rangle_{c,{\rm GE}}\;,
\ee
and those involving $\wj^\mu(0)$ alike. In essence, the \eqref{thff} and the \eqref{thff2} are extensions of the gravitational 
(and charged) form factors at finite temperature and chemical potential and we will refer to them as thermo-gravitational 
and thermo-charged correlators respectively. They are in general unknown and depend on the specific dynamical 
quantum field theory. 

We first consider the thermo-gravitational correlators in eq. \eqref{thff}. In general, this is a symmetric tensor 
which is a function of the arguments in \eqref{thff}, hence $k,q,\beta$ and $\zeta$, the pseudo-vector 
$a^\mu=\epsilon^{\mu\rho\sigma\tau}k_\rho q_\sigma \beta_\tau$, as well as the metric tensor $g_{\mu\nu}$. The most general 
combination of vectors and tensors one can write to express a symmetric tensor reads: 
\begin{equation}\label{inizialeT}
    \begin{split}
 \Theta^{\mu\nu}(k,q,\beta)
        &=\Theta_1(S)\,k^\mu k^\nu+\Theta_2(S)\,q^\mu q^\nu
        + \Theta_3(S)\,\beta^\mu \beta^\nu \\
        &+\Theta_4(S)\left(k^\mu q^\nu + k^\nu q^\mu\right)
        + \Theta_5(S)\left(k^\mu \beta^\nu + k^\nu \beta^\mu\right) \\
        &+ \Theta_6(S)\left(q^\mu \beta^\nu + q^\nu \beta^\mu\right)
        + \Theta_7(S)\,g^{\mu\nu} \\
        &  + \Theta_8(S)\left(a^\mu k^\nu + a^\nu k^\mu\right)
         + \Theta_{9}(S)\left(a^\mu q^\nu + a^\nu q^\mu\right)\\
         &+\Theta_{10}(S)\left(a^\mu\beta^\nu+a^\nu\beta^\mu\right)\;,
\end{split}
\end{equation}
where the $\Theta_1,\ldots,\Theta_{10}$ are scalar coefficients depending on all the possible scalars (denoted 
collectively by the letter $S$):
\begin{equation}\label{scalars}
    S=\left\{k^2\,,\; q^2\,,\; \beta^2\,,\; k\cdot\beta\,,\;k\cdot q\,,\; q\cdot\beta,\;\zeta\right\}\,,
\end{equation}
that can be formed out of $\beta(x), q, k ,a$ and the reduced chemical potential $\zeta(x)$\footnote{The tensor $\Theta^{\mu\nu}$ also depends on $\zeta$ through the form factors $\Theta_i(S)$. However we do not write it explicitly in order not to overly complicate the notation.}. 
Note that since $a^\mu$ is orthogonal to $k$, $q$, and $\beta$, no pseudo-scalar can be formed, hence all the $\Theta_i$ are pure Lorentz scalars.
It can be shown (see Appendix~\ref{SECTION: aa is not independent}) that the 
tensor $a^\mu a^\nu$ can be written as a suitable combinations of the others,
so the corresponding term has been omitted. The thermo-gravitational correlators fulfills other important relations 
dictated by the transformation properties of the statistical operator under conjugation, time-reversal and parity (see Appendix \ref{SECTION: symmetries}):
\begin{equation}\label{adjoint}
 \Theta^{\mu\nu} \left(k,q,\beta\right)^*=
 \e^{-\beta(x)\cdot q}\Theta^{\mu\nu} \left(k,-q,\beta\right)\;,
\end{equation}
and:
\begin{subequations}\label{trevpar}
    \begin{align}
        \Theta^{\mu\nu} \left(k,q,\beta\right)&=\e^{-\beta(x)\cdot q}\theta^\mu_\alpha\theta^\nu_\beta 
        \Theta^{\alpha\beta} \left(\Tilde{k},-\Tilde{q},\Tilde{\beta}\right),\\
        \Theta^{\mu\nu} \left(k,q,\beta\right)&=\theta^\mu_\alpha\theta^\nu_\beta \Theta^{\alpha\beta} \left(\Tilde{k},\Tilde{q},\Tilde{\beta}\right)\;,
    \end{align}
\end{subequations}
where $\theta^\mu_\alpha=\mbox{diag}\left(1,-1,-1,-1\right)$ is the transformation matrix associated to parity, while 
$\widetilde{V}$ indicates the time-reversal or parity transformed of a four-momentum or a four-temperature:
$$
\widetilde{V}=\left(V^0,-{\bf V}\right)\;.
$$
Now, taking into account that:
\begin{equation*}
    \theta^\mu_\alpha\theta^\nu_\beta V^{\alpha}_1 V^{\beta}_2=V^\mu_1V^\nu_2,\quad \theta^\mu_\alpha\theta^\nu_\beta 
    a^{\alpha} V^{\beta}=-a^\mu V^\nu\;,
\end{equation*}
and combining \eqref{adjoint} with the relations \eqref{trevpar}, it turns out that:
\begin{subequations}\label{G q to meno q}
    \begin{align}
        \Theta_i\left(k,q,\beta\right)&\in\mathbb{R}\;,\;\;\;i=1\ldots10\;,\\
        \Theta_i\left(k,q,\beta\right)&=\Theta_i\left(k,-q,\beta\right)\e^{-\beta(x)\cdot q} \;\;\; i=1\ldots7\;,\\
      \Theta_i\left(k,q,\beta\right)&=-\Theta_i\left(k,q,\beta\right)=0 \;\;\; i=8,\ 9,\ 10\;.
    \end{align}
\end{subequations}
The decomposition of the correlator in eq. \eqref{thff} then comes down to:
\begin{equation}\label{inizialeT2}
    \begin{split}
 \Theta^{\mu\nu}(k,q,\beta)
        &=\Theta_1(S)\,k^\mu k^\nu+\Theta_2(S)\,q^\mu q^\nu
        +\Theta_3(S)\,\beta^\mu \beta^\nu \\
        &+\Theta_4(S)\left(k^\mu q^\nu + k^\nu q^\mu\right)
        + \Theta_5(S)\left(k^\mu \beta^\nu + k^\nu \beta^\mu\right) \\
        &+ \Theta_6(S)\left(q^\mu \beta^\nu + q^\nu \beta^\mu\right)
        + \Theta_7(S)\,g^{\mu\nu}\;,
\end{split}
\end{equation}
and this holds for the correlators in eq. \eqref{thff2} too.

Albeit being in general unknown, the coefficients $\Theta_{j}$ are effectively constrained by general conservation laws.
Since the stress-energy tensor is a conserved current \eqref{conservation}, its integral over any space-like hypersurface $\Sigma$ 
yields the four-momentum operator, so that:
\begin{equation}\label{corrint}
    \int_{\Sigma}\di\Sigma_\mu(y)\langle\wAd(k_+)\wA(k_-),\;\wT^{\mu\nu}(y)\rangle_{c,{\rm GE}}=
\langle\wAd(k_+)\wA(k_-),\;\wP^\nu\rangle_{c,{\rm GE}}\;,
\end{equation}
and similarly for the other combinations of $\wA$ and $\wB$. Choosing $\Sigma$ as the hyperplane $t=0$ and
using the \eqref{transtranf}, \eqref{abtransl} and \eqref{abtransl2} one gets:
\begin{eqnarray}\label{corrint2}
&& \int_{\Sigma}\di\Sigma_\mu
(y)
\langle\wAd(k_+)\wA(k_-),\wT^{\mu\nu}(y)\rangle_{c,{\rm GE}} 
=\int \di^3 \y \; \langle\wAd(k_+)\wA(k_-),\;\wT^{0\nu}(0,{\bf y})\rangle_{c,{\rm GE}}  \nonumber \\
&& =\langle\wAd(k_+)\wA(k_-),\;\wT^{0\nu}(0)\rangle_{c,{\rm GE}} \int_{t=0} \di^3 {\rm y} \; \e^{\ii {\bf q}\cdot {\bf y}} = \langle\wAd(k_+)\wA(k_-),\;\wT^{0\nu}(0)\rangle_{c,{\rm GE}} \left(2\pi\right)^3 \delta^3({\bf q})\;,
\end{eqnarray}
for $q = k_+ - k_-$. The correlator on the right hand side of the equation \eqref{corrint} can be related to the derivative 
with respect to $\beta_\nu(x)$ of the density operator \eqref{gedens} with an important caveat: according to the equation 
\eqref{invfourier} the operators $\wA(p)$, $\wAd(p')$, $\wB(p)$ and $\wBd(p')$ are $\beta$-dependent if the spectral function in the field 
expansion is $\varrho_{\rm GE}(p)$ calculated with $\beta = \beta(x)$. In order to make them $\beta$-independent, it suffices to multiply them by the spectral function. We can then write:
\begin{equation*}
       \varrho(k_+)\varrho(k_-) \langle\wAd(k_+)\wA(k_-),\wP^\nu\rangle_{c,{\rm GE}} = - 
\frac{\partial}{\partial \beta_\nu(x)} \left(\varrho(k_+)\varrho(k_-) \langle\wAd(k_+)\wA(k_-)\rangle_{\rm GE}
\right)\; .
\end{equation*}
From the above equation, with $\varrho(p) = \varrho_{\rm GE}(p)$ with $\beta=\beta(x)$, by using the \eqref{tevaa} 
the \eqref{corrint} and the \eqref{corrint2}, the following relation is obtained:
\be
  \varrho(k^0_+,\kb)\varrho(k^0_-,\kb)\langle\wAd(k^0_+,\kb)\wA(k^0_-,\kb),\wT^{0\nu}(0)\rangle_{c,{\rm GE}} 
  (2\pi)^3 \delta^3({\bf q}) \nonumber \\
 = -(2\pi) \delta^3({\bf q}) \delta(k^0_+-k^0_-) \theta(k^0_+)\theta(k^0_-)
 \frac{\partial}{\partial \beta_\nu(x)} \left( n_B(k) \varrho(k) \right)\;,
\ee
which implies:
\be\label{correl1}
 \langle\wAd(k^0_+,\kb)\wA(k^0_-,\kb),\wT^{0\nu}(0)\rangle_{c,{\rm GE}} = \Theta^{0\nu}(k,q^0,\q=0,\beta) =
 \frac{\theta(k^0)}{(2\pi)^2} \frac{1}{\varrho^2(k)} \delta(q^0)
 \left[-\frac{\partial}{\partial \beta_\nu(x)} \left(n_B(k)\varrho(k)\right)\right]\;,
\ee
which vanishes for $q^0 \ne 0$ and for any value of $k$ and $\beta(x)$. Hence all terms on the right hand side of 
\eqref{inizialeT2} with $\mu=0$ and  $\nu \ne 0$ must be vanishing for $q^0 \ne 0$. This requirement constrains the coefficients
$\Theta_i$ to be proportional to Dirac $\delta$ distributions such that they reduce to a $\delta(q^0)$ for $\q =0$. 
Since these coefficients must be Lorentz scalars, we can write them in general as a sum over all possible 
delta distributions of the scalars $S$ in eq. \eqref{scalars} multiplied by tensors $\Gamma^{\mu\nu}_j(k,q,\beta)$. 
In formulae:
\begin{equation}\label{Theta deltas}
    \Theta^{\mu\nu}(k,q,\beta) = \sum_j \delta(F_{j}(S)) \Gamma^{\mu\nu}_{j}(k,q,\beta) \,,
\end{equation}
where $F_{j}(S)$ is a scalar functions such that:
\begin{equation*}
    \delta(F_{j}(S))\Big|_{\q =0} \propto \delta(q^0)\;,
\end{equation*}
so as to fulfill the equation \eqref{correl1}. The above condition requires the functions $F_j(S)$ to vanish for $q=0$ 
and that they do not have zeroes with $q^0 \ne 0$ and $\q =0$, for any value of $k$ and $\beta$. Furthermore, its 
derivative with respect to $q^0$ should not vanish in $q^0=0$:
\be\label{firsderivcond}
 \frac{\partial F_j(S)}{\partial q^0}\Big|_{q=0} =  \frac{\partial F_j(S)}{\partial (q\cdot k)}\Bigg|_{q=0} k^0 +
 \frac{\partial F_j(S)}{\partial (q\cdot \beta)}\Bigg|_{q=0} \beta^0 + 
 \lim_{q^0\to 0}\frac{\partial F_j(S)}{\partial q^2}\Bigg|_{\q=0} q^0 \ne 0\;,
\ee
which, for instance, rules out a term like $ F_j(S) = q^2$.
Nevertheless, in principle, there are infinite functions fulfilling those conditions, hence infinite terms on the right 
hand side of \eqref{Theta deltas}.
All the functions $F_j(S)$ can be expressed explicitly in terms of either $q\cdot k$ or $q \cdot \beta$ or $q^2$ so that, 
for instance:
$$
   \delta(F_{j}(S)) =  \delta(q \cdot k - f_j(S))\Bigg|\frac{\partial F_j(S)}{\partial (q\cdot k)}\Bigg|^{-1}_{q\cdot k=f_j(S)} \;,
$$
where $f_j(S)$ is a function of the remaining scalars ($S$ does not include $q\cdot k$ in the above example) 
that vanishes for $q=0$. This must be possible because if all the derivatives of $F_j$ with respect to the above scalars 
involving $q$ vanished for $q \to 0$, then the condition \eqref{firsderivcond} would be violated. Note that the pre-factors 
such as $|\partial F_j/\partial (q\cdot k)|$ are scalars and can be re-absorbed into a re-definition of the scalar coefficients 
$\Theta_i$ without loss of generality, so we can recast the equation \eqref{Theta deltas} as:
\begin{equation}\label{Theta deltas2}
     \Theta^{\mu\nu}(k,q\cdot\beta)=\sum_{j}\delta(s_j-f_j(S))\Gamma^{\mu\nu}_{j}(k,q,\beta)\;,  \qquad {\rm with}\;\; 
     s_j=k\cdot q,\, q\cdot\beta,\, q^2\;,
\end{equation}
with the condition \eqref{firsderivcond} now becoming:
$$
   \lim_{q \to 0} \frac{\partial}{\partial q^0} \left( s_j - f_j(S) \right) \ne 0\;.
$$
There are some peculiar limitations on these functions owing to the continuity equation of the stress-energy 
tensor \eqref{conservation}. Indeed, by using the \eqref{transtranf} and \eqref{tevaa} and the ciclicity of the trace, 
it can be readily shown that:
\begin{align*}
& 0=\frac{\partial}{\partial y^\mu}\langle\wAd(k_+)\wA(k_-),\wT^{\mu\nu}(y)\rangle_{c,{\rm GE}}
=\langle\wAd(k_+)\wA(k_-),\wT^{\mu\nu}(0)\rangle_{c,{\rm GE}}\frac{\partial}{\partial y^\mu}
\left(\e^{-\ii q\cdot y}\right) \\
& =-\ii q_\mu\langle\wAd(k_+)\wA(k_-),\wT^{\mu\nu}(0)\rangle_{c,{\rm GE}}
\e^{-\ii q\cdot y}\;,
\end{align*}
implying the transversality, or Ward identity, condition:
\begin{equation}\label{qTheta=0}
   q_\mu\Theta^{\mu\nu}\left(k,q,\beta\right)=0\;, \qquad \forall \, k,\ q,\ \beta \; .
\end{equation}
Plugging the equation \eqref{Theta deltas2} into the transversality condition leads to:
\begin{equation}\label{ward gamma}
     \delta(s_j-f_j(S))q_\mu\Gamma^{\mu\nu}_{j}(k,q,\beta)=0\;,\qquad\forall\; j\;.
\end{equation}
The equation \eqref{inizialeT2} implies that each tensor $\Gamma^{\mu\nu}_j(k,q,\beta)$ can be 
decomposed as:
\begin{equation}\label{iniziale2}
    \begin{split}
        \Gamma^{\mu\nu}_{j}(k,q,\beta)
        &=G^{j}_1(S)\,k^\mu k^\nu+G^{j}_2(S)\,q^\mu q^\nu
        + G^{j}_3(S)\,\beta^\mu \beta^\nu \\
        &+G^{j}_4(S)\left(k^\mu q^\nu + k^\nu q^\mu\right)
        +  G^{j}_5(S)\left(k^\mu \beta^\nu + k^\nu \beta^\mu\right) \\
        &+ G^{j}_6(S)\left(q^\mu \beta^\nu + q^\nu \beta^\mu\right)+  
        G^{j}_7(S)\,g^{\mu\nu}\;,
    \end{split}
\end{equation}
with suitable scalar coefficients $G^{j}_i(S)$. 
For each tensor $\Gamma^{\mu\nu}_j$ not all coefficients $G^j_i$ appearing in Eq.~\eqref{iniziale2} are independent. 
Indeed, the Ward identity \eqref{ward gamma} implies relations reducing the number of independent coefficients. 
These relations depend on the specific constraint imposed by the associated $\delta$ function; consequently, different
tensors $\Gamma^{\mu\nu}_j$ satisfy different sets of relations among the corresponding $G^j_i$.

To systematically implement these constraints, we introduce a fundamental assumption: each tensor $\Gamma^{\mu\nu}_j$ 
appearing in the decomposition \eqref{Theta deltas2} is assumed to be an analytic function of the momentum $q$ in a 
neighborhood of $q=0$ \footnote{This requirement is met in the free field case, where $\wT^{\mu\nu}(0)$ is known and the
correlator \eqref{thff} can be explicitly calculated:
$$
 \left\langle\widehat{a}^\dagger(k_+)\widehat{a}(k_-),\;\wT^{\mu\nu}(0)\right\rangle_{c,\rm GE}=\frac{2}{(2\pi)^3}n_{\rm B}(k_+)\left(1+n_{\rm B}(k_-)\right)\left[k^\mu k^\nu-\frac{1}{4}\left(q^\mu q^\nu-q^2g^{\mu\nu}\right)\right]\;,\quad k^2_\pm=m^2\;.
$$
For free fields $k_\pm$ are always on-shell due to \eqref{freespectral} and the only kinetic constraint is $\delta\left(k\cdot q\right)$ (see \cite{Sheng:2025cjk}).
}. More precisely each tensor $\Gamma^{\mu\nu}_j(k,q,\beta)$ is required to have a finite limit and
be infinitely differentiable at $q=0$. This assumption plays a crucial role in the following analysis.

Once the transversality conditions \eqref{qTheta=0} are enforced under this assumption it then follows that, in the limit 
$q \to 0$, the only non-vanishing contributions arise from the two terms on the right-hand side of Eq.~\eqref{Theta deltas2} 
for which $s_j = k \cdot q$ and $s_j = q \cdot \beta$, with $f_j(S)=0$. All remaining terms must vanish at $q=0$. 
This conclusion is very important for a twofold reason:
\begin{enumerate}
    \item {} among all possible terms of the series in the equation \eqref{Theta deltas}, the two delta distributions
    $\delta(q\cdot k)$ and $\delta(q\cdot \beta)$ must exist in order to ensure the validity of the equation \eqref{correl1}, see the derivation of the equation \eqref{Gamma1q0} below;
    \item {} as it will be shown in Section \ref{hydrolimit}, there is a one-to-one correspondence between the order of 
    the $q$ expansion and the order of the hydrodynamic expansion in the gradients of the thermo-hydrodynamic fields $\beta,\, \zeta$; the equation \eqref{qvanish} tells us that all terms in the series except the two mentioned do not
    contribute at the lowest orders of the gradient expansion.
\end{enumerate}

We begin by discussing the aforementioned two terms: $\Gamma^{\mu\nu}_k$, corresponding to $s=q\cdot k$ and $f_j(S)=0$ 
and $\Gamma^{\mu\nu}_\beta$, corresponding to $s=q\cdot \beta$; and $f_j(S)=0$. For the $k$ branch, being $q\cdot k=0$ enforced by 
the delta in eq. \eqref{Theta deltas2}, three constraints turn up from the transversality condition 
\eqref{qTheta=0}:
\begin{equation*}
    \begin{split}
        G^k_7(S)&=-q^2G^k_2(S)-q\cdot\beta\; G^k_6(S)\;,\\
        G^k_3(S)&=-\frac{q^2}{q\cdot\beta}G^k_6(S)\;,\\
        G^k_5(S)&=-\frac{q^2}{q\cdot\beta}G^k_4(S)\;.
    \end{split}
\end{equation*}
Plugging the above relations in \eqref{iniziale2} and absorbing the denominators into a redefinition of the 
scalar factors the following final expression is obtained:
\begin{equation}\label{Gammak}
\begin{split}
 \Gamma^{\mu\nu}_k (k,q,\beta)&=\Gamma^{k}_1(S)k^\mu k^\nu+\Gamma^{k}_2(S)
 (q^\mu q^\nu-q^2g^{\mu\nu}) +\Gamma^{k}_3(S)\left[q\cdot\beta\left(k^\mu q^\nu+k^\nu q^\mu\right)-q^2\left(k^\mu\beta^\nu+k^\nu\beta^\mu\right)\right]\\
&+ \Gamma^{k}_4(S)\left[q\cdot\beta\left(\beta^\mu q^\nu+\beta^\nu q^\mu\right)-
\left(q\cdot\beta\right)^2g^{\mu\nu}-q^2\beta^\mu\beta^\nu\right]\;,
\end{split}
\end{equation}
where we have renamed the four remaining unknown coefficients; they will be henceforth called {\em thermo-gravitational
form factors} in analogy with more familiar vacuum form factors. Because of the analyticity requirement, these
form factors must be analytic functions of $q$. A similar calculation can be carried out for $\Gamma^{\mu\nu}_\beta$, 
associated to $\delta(q\cdot\beta)$, and the following expression is obtained:
\begin{equation}\label{Gammabeta}
    \begin{split}
        \Gamma^{\mu\nu}_{\beta}(k,q,\beta)&=\Gamma^{\beta}_{1}(S)\,\beta^\mu\beta^\nu+
        \Gamma^{\beta}_{2}(S)\left(q^\mu q^\nu-q^2g^{\mu\nu}\right)\\
        &+\Gamma^{\beta}_{3}(S)\left[q\cdot k\left(q^\mu\beta^\nu+q^\nu\beta^\mu\right)-
        q^2\left(k^\mu\beta^\nu+k^\nu\beta^\mu\right)\right]\\
        &+\Gamma^{\beta}_{4}(S)\left[q\cdot k\left(k^\mu q^\nu+k^\nu q^\mu\right)-
        q^2k^\mu k^\nu-(q\cdot k)^2g^{\mu\nu}\right]\,,
    \end{split}
\end{equation}
implying four more form factors besides those in \eqref{Gammak}. It can be seen that the tensors $\Gamma_{k,\beta}$ may
have a finite limit for $q \to 0$. Comparing the expressions \eqref{Gammak} and \eqref{Gammabeta} with 
the gravitational form factors in the vacuum for scalar particles (e.g a pion) \cite{Donoghue:1991qv,Pagels:1966zza,Polyakov:2018zvc} 
there appear six additional form factors: $\Gamma^k_3,\Gamma^k_4$ and all the $\Gamma^\beta_i$, ought to the 
additional four-temperature vector. Clearly, the form factors $\Gamma^k_{1,2}$ which in literature are often denoted with $A,B$, 
are different from their expression in the vacuum and also depend on $\beta$ and $\zeta$ through the scalars \eqref{scalars}. 

A similar calculation can be carried out for all the remaining tensors $\Gamma_j$ in eq. \eqref{Theta deltas2}. In
the most general case $s_j = q\cdot k$, $s_j = q^2$ or $s_j = q \cdot \beta$ and $s_j = f_j(S)$ with $f_j(S) \ne 0$.
with the constraint $\lim_{q \to 0} f_j(S) = 0$, as required by Eq.~\eqref{firsderivcond}.
As an illustrative example, consider the case $s_j = q\cdot k$ with $f_j(S)\neq 0$. The Ward identity and the requirement
of analyticity turn the \eqref{iniziale2} into:
\begin{equation}\label{gammakj}
\begin{split}
    \Gamma^{\mu\nu}_{k,j}\left(k,q,\beta\right)
&= \Gamma^{k,j}_1(S)\left(q^\mu q^\nu - q^2 g^{\mu\nu}\right) + \Gamma^{k,j}_2(S)\left[
f_j(S)\left(k^\mu q^\nu + k^\nu q^\mu\right)- q^2 k^\mu k^\nu - f_j^2(S) g^{\mu\nu} \right] \\
&\quad + \Gamma^{k,j}_3(S)\left[f_j(S)(q\cdot\beta)\left(k^\mu \beta^\nu + k^\nu \beta^\mu\right)
- f_j^2(S)\beta^\mu \beta^\nu - (q\cdot\beta)^2 k^\mu k^\nu \right] \\
&\quad + \Gamma^{k,j}_4(S)\left[ (q\cdot\beta)\left(q^\mu \beta^\nu + q^\nu \beta^\mu\right)
- q^2 \beta^\mu \beta^\nu - (q\cdot\beta)^2 g^{\mu\nu} \right] \; .
    \end{split}
\end{equation}
Since $f_j(S)$ ought to vanish for $q=0$ and must be such that the conditions \eqref{firsderivcond} are fulfilled, 
it must be at least linear in $q$, hence all terms in the equation \eqref{gammakj} are at least quadratic in $q$. As 
a result, $\Gamma^{\mu\nu}_{k,j}$ vanishes at $q=0$ together with its first derivative with respect to $q$:
\begin{equation}
\label{qvanish}
\Gamma^{\mu\nu}_{k,j}(k,0,\beta)=0\;, \qquad\qquad
\frac{\partial}{\partial q^\lambda}\Gamma^{\mu\nu}_{k,j}\!\left(k,q,\beta\right)\Big|_{q=0}=0 \, .
\end{equation}
The analysis for all remaining possible combinations with $s_j = q\cdot k$ or $s_j = q^2$ and $f_j(S)\neq 0$ leads
to the same result and for all of them the tensors $\Gamma^{\mu\nu}_j$ fulfill the condition~\eqref{qvanish}.

Now the equation \eqref{Theta deltas2}, taking into account the \eqref{qvanish}, can be matched to the equation 
\eqref{correl1} to obtain a major result on the thermal-gravitational form factors:
\begin{equation}\label{comparison}
    \frac{1}{k^0}\Gamma^{0\nu}_k\left(k,q=0,\beta\right) +\frac{1}{\beta^0}
    \Gamma^{0\nu}_\beta\left(k,q=0,\beta\right)=-\frac{\theta(k^0)}{\left(2\pi\right)^2\varrho^2(k)}
    \frac{\partial}{\partial\beta^\nu}\left(n_{\rm B}(k)\varrho(k)\right)\;.
\end{equation}
Since $\varrho$ is a scalar function, it can only depend on $k^2,\;k\cdot\beta\;,\beta^2$ hence:
\begin{equation*}
    \frac{\partial}{\partial\beta^\nu}\left[n_{\rm B}(k)\varrho(k)\right] = - n_{\rm B}(k)[1+n_{\rm B}(k)] \varrho(k)
    k^\nu + n_{\rm B}(k) \frac{\partial \varrho(k)}{\partial (k \cdot \beta)} k^\nu + 2 n_{\rm B}(k) 
    \frac{\partial \varrho(k)}{\partial \beta^2} \beta^\nu\;.
\end{equation*}
By using the equations \eqref{Gammak} and \eqref{Gammabeta} in the eq. \eqref{comparison} an explicit expression 
of the form factors $\Gamma_1^k(q=0)$ and $\Gamma_1^\beta(q=0)$ is thus obtained:
\begin{subequations}\label{Gamma1q0}
\begin{align}
    \Gamma_1^k\left(k,0,\beta\right)&=\theta(k^0) \frac{1}{(2\pi)^2} \frac{n_{\rm B}(k)}{\varrho(k)} \left[ 1+n_{\rm B}(k) 
    - \frac{\partial \log \varrho(k)}{\partial (k \cdot \beta)} \right]\;, \\
  \Gamma_1^\beta\left(k,0,\beta\right)&=- \theta(k^0) \frac{2}{(2\pi)^2} \frac{n_{\rm B}(k)}{\varrho(k)} 
  \frac{\partial \log \varrho(k)}{\partial \beta^2} \;.
\end{align}
\end{subequations}
For $q=0$, as it is apparent from the eqs. \eqref{Gammak} and \eqref{Gammabeta} and the \eqref{qvanish},
$\Gamma_1^k$ and $\Gamma_1^\beta$ are the only relevant form factors. Therefore, for $q=0$ the thermo-gravitational 
form factors are entirely determined by the spectral function without additional unknown dynamical coefficient.
In the limit of a free field $\varrho$ is independent from $\beta$ hence $\Gamma_1^\beta$ vanishes and only 
$\Gamma_1^k$ is left. As it was discussed above, the equations \eqref{Gamma1q0} confirms that the two terms in 
equation \eqref{Theta deltas2} with $s=q\cdot k,\, q\cdot \beta$ and $f_j(S)=0$ must exist with non-vanishing 
form factors to ensure that the equation \eqref{correl1} is fulfilled. All other terms in the expansion 
\eqref{Theta deltas2} are possible terms, meaning that they may in principle exist, but they are not necessary 
for the purpose of energy-momentum conservation.

The value of the derivatives of the correlators in $q=0$ is also well constrained by complex conjugation, parity and 
time-reversal transformations (see Appendix \ref{SECTION: symmetries}). It turns out that the tensors $\Gamma_{k,\beta}$
in \eqref{Gammak} and \eqref{Gammabeta} fulfill the following relations:
\begin{subequations}\label{Gamma q = Gamma -q exp}
    \begin{align}
        \Gamma^{\mu\nu}_k\left(k,q,\beta\right)&=\e^{-\beta(x)\cdot q}\Gamma^{\mu\nu}_k\left(k,-q,\beta\right)\;,\\
        \Gamma^{\mu\nu}_\beta\left(k,q,\beta\right)&=\Gamma^{\mu\nu}_\beta\left(k,-q,\beta\right)\;,
    \end{align}
\end{subequations}
From the \eqref{Gamma q = Gamma -q exp} it follows:
\begin{subequations}\label{qderiv Gamma}
    \begin{align}
        \frac{\partial}{\partial q^\lambda}\Gamma^{\mu\nu}_k\left(k,q,\beta\right)\Big|_{q=0}&=-\frac{1}{2}\beta_\lambda(x)\Gamma^{\mu\nu}_k\left(k,0,\beta\right)\;,\\
         \frac{\partial}{\partial q^\lambda}\Gamma^{\mu\nu}_\beta\left(k,q,\beta\right)\Big|_{q=0}&=- \frac{\partial}{\partial q^\lambda}\Gamma^{\mu\nu}_\beta\left(k,q,\beta\right)\Big|_{q=0}=0\;.
    \end{align}
\end{subequations}

Much in the same way as for the correlator in eq. \eqref{correl1}, for the correlator involving $\wAd\wBd$, the 
equation \eqref{tevab} implies:
\be\label{correl2}
\langle\wAd(k^0_+,\kb)\wBd(-k_-^0,-\kb),\;\wT^{0\nu}(0)\rangle_{c,{\rm GE}} = 0\;.
\ee
For this correlator, in principle the same tensor decomposition in eq. \eqref{inizialeT2} can be written,
and from the equation \eqref{correl2}, the conclusion trivially follows:
$$
 \langle\wAd(k^0_+,\kb)\wBd(-k_-^0,-\kb),\;\wT^{0\nu}(0)\rangle_{c,{\rm GE}} = 0
\quad \forall\, k,\,\beta \implies \Theta_{i\wAd\wBd} \equiv 0\;.
$$
so, the correlator $\langle\wAd(k_+)\wBd(-k_-),\;\wT^{\mu\nu}(0)\rangle_{c,{\rm GE}}$ and its complex conjugate vanish and do not 
play any role.

\subsection{Charged thermal form factors}

A similar calculation can be carried out for the correlators involving the charged current. The correlator can be 
expanded in terms of the independent vectors $k,\,q$ and $\beta$:
\begin{equation}\label{Y}
        \langle\wAd(k_+)\wA(k_-),\;\wj^\mu(0)\rangle_{c,\rm GE} \equiv Y^\mu\left(k,q,\beta\right)=Y_1(S)\,k^\mu+Y_2(S)\, q^\mu+Y_3(S)\,\beta^\mu\,.
\end{equation}
Again a term proportional to the pseudo-vector $a^\mu$ is forbidden by parity and time-reversal while:
\begin{subequations}\label{Ypar}
    \begin{align}
        Y^\mu\left(k,q,\beta\right)&=\e^{-\beta(x)\cdot q}\theta^\mu_\alpha\, Y^\alpha\left(\widetilde{k},-\widetilde{q},\widetilde{\beta}\right)\,,\\
        Y^\mu\left(k,q,\beta\right)&=\theta^\mu_\alpha\, Y^\alpha\left(\widetilde{k},\widetilde{q},\widetilde{\beta}\right)\,.
    \end{align}
\end{subequations}
Since the charged current is a conserved according to the eq. \eqref{conservation}, for an arbitrary space-like hypersurface 
$\Sigma$ we have a globally conserved charge operator:
\begin{equation}\label{integral current}
    \int_{\Sigma}\di\Sigma_\mu(y)\,\wj^\mu(y)=\wQ\;.
\end{equation}
Hence, integrating over the hyperplane at $t=0$, we obtain: 
\begin{equation*}
    \int_{\Sigma}\di\Sigma_\mu(y)\langle\wAd(k_+)\wA(k_-),\;\wj^\mu(y)\rangle_{c,\rm GE}=\langle\wAd(k_+)\wA(k_-),\;\wQ\rangle_{c,\rm GE}\,,
\end{equation*}
which in turn implies, in view of the form of the density operator \eqref{geq} (see analogous derivation in the
equations \eqref{corrint2}-\eqref{correl1} above):
\begin{equation}\label{Y0}
    \langle\wAd\left(k^0_+,{\bf k}\right)\wA\left(k^0_-,{\bf k}\right),\;\wj^{\,0}(0)\rangle_{c,\rm GE}=\frac{\theta(k^0)}{(2\pi)^2}\frac{1}{\varrho^2(k)}\delta(q^0) \frac{\partial}{\partial\zeta}\left(n_{\rm B}(k)\varrho(k)\right)\,.
\end{equation}
From the conservation of the four-current \eqref{conservation} the Ward identity ensues:
\begin{equation*}
    q_\mu Y^\mu\left(k,q,\beta\right)=0\;,\qquad\forall\,k,\, q,\, \beta\;,
\end{equation*}
implying that the correlator \eqref{Y} can be expanded much the same way as we have seen in eq. \eqref{Theta deltas}:
\begin{equation}\label{Ydeltas}
    Y^\mu\left(k,q,\beta\right)=\sum_{j}\delta\left(s_j-f_j(S)\right)\Upsilon^\mu_{j}\left(k,q,\beta\right)\,,
\end{equation}
where $S$ are all the scalars \eqref{scalars} while $s_j=k\cdot q,\, q\cdot\beta\,,q^2 $. The four-vectors $\Upsilon^\mu_{j}$ 
are assumed to be analytic functions of $q$ so that they are finite and infinitely differentiable in $q=0$.

Again, according to the condition \eqref{Y0}, the only two terms in the expansion \eqref{Ydeltas} which are non-vanishing 
for $q=0$ are those associated with the $q\cdot k=0$ and $q\cdot\beta=0$ delta distributions:
\begin{subequations}\label{Upsilon k e Upsilon beta}
    \begin{align}
        \Upsilon^\mu_k\left(k,q,\beta\right) &=\Upsilon^k_1(S)k^\mu+\Upsilon^k_2(S)
    \left[\left(q\cdot\beta\right) q^\mu-q^2\beta^\mu\right]\;,\\
    \Upsilon^\mu_\beta\left(k,q,\beta\right) &=\Upsilon^\beta_1(S)\beta^\mu+\Upsilon^\beta_2(S)
    \left[\left(q\cdot k\right) q^\mu-q^2k^\mu\right]\;,
    \end{align}
\end{subequations}
whereas all remaining $\Upsilon^\mu_{j}$, along with their first order derivative in $q$ vanish for $q=0$. 
Combining the \eqref{Upsilon k e Upsilon beta} and the \eqref{Y0}, we thus obtain:
\begin{equation}\label{Upsilon 1 in q=0}
 \Upsilon^k_{1}\left(k,q=0,\beta\right)+\Upsilon^{\beta}_1\left(k,q=0,\beta\right) =\frac{\theta(k^0)}{(2\pi)^2\varrho^2(k)}\frac{\partial}{\partial\zeta}\left(n_{\rm B}(k)\varrho(k)\right)\,.
\end{equation}
Note that, in principle, for the four-current terms we are not able to determine separately the two coefficients in $q=0$ like in the case of the stress energy tensor \eqref{Gamma1q0}, but only their sum. However, in the free-limit, $\varrho$ turns out to be independent from $\zeta$ and thus we can conclude that:
\begin{subequations}\label{Upsilonkbeta q0}
    \begin{align}
        \Upsilon^k_1\left(k,0,\beta\right)&=\frac{\theta(k^0)}{(2\pi)^2\varrho(k)}n_{\rm B}(k)\left(1+n_{\rm B}(k)\right)+\Lambda\left(k,\beta\right)\;,\\
        \Upsilon^\beta_1\left(k,0,\beta\right)&=\frac{\theta(k^0)}{(2\pi)^2}n_{\rm B}(k)\frac{\partial\log\varrho}{\partial\zeta}-\Lambda\left(k,\beta\right)\;,
    \end{align}
\end{subequations}
with $\Lambda\left(k,\beta\right)$ scalar function which must be vanishing for free fields so that, for free fields, $\Upsilon^\beta_1\left(k,0,\beta\right)\mapsto0$.

Finally, due to the relations \eqref{Ypar}, using the results from appendix \ref{SECTION: symmetries},   the vectors $\Upsilon_{k/\beta}$ fulfill:
\begin{subequations}\label{Upsiloj q = Upsilon -q exp}
    \begin{align}
        \Upsilon^{\mu}_k\left(k,q,\beta\right)&=\e^{-\beta(x)\cdot q}\,\Upsilon^{\mu}_k\left(k,-q,\beta\right)\,,\\
        \Upsilon^{\mu}_\beta\left(k,q,\beta\right)&=\Upsilon^{\mu}_\beta\left(k,-q,\beta\right)\;,
    \end{align}
\end{subequations}
impying:
\begin{subequations}\label{qderiv Upsilon}
    \begin{align}
        \partial^q_\lambda\,\Upsilon^{\mu}_k\left(k,q,\beta\right)\Big|_{q=0}&=-\frac{1}{2}\beta_\lambda(x)\Upsilon^{\mu}_\beta\left(k,0,\beta\right)\;,\\
        \partial^q_\lambda\,\Upsilon^{\mu}_\beta\left(k,q,\beta\right)\Big|_{q=0}&=-\partial^q_\lambda\,\Upsilon^{\mu}_\beta\left(k,q,\beta\right)\Big|_{q=0}=0\;.
    \end{align}
\end{subequations}
%

\section{Hydrodynamic limit and gradient expansion}
\label{hydrolimit}

We can use the results of the foregoing Section to further develop the off-equilibrium correction of the Wigner
function. With the definition \eqref{thff}, taking into account the vanishing of the correlators involving 
$\wAd\wBd$ and plugging the \eqref{Theta deltas2} into the equation \eqref{deltawig2} a new expression is obtained:
\begin{align}\label{deltawig3}
\Delta W^+(x,k) &=\frac{2}{(2\pi)^5} \sum_{j} \int_{\Sigma_{0}} \di \Sigma_{\mu}(y) 
\int \di^{4}q\;  
\varrho(k_+)\varrho(k_-)\theta(k_+^0)\theta(k_-^0) \e^{\ii q\cdot (x-y)} \frac{1 -\e^{\beta(x)\cdot q}}{\beta(x) \cdot q} \nonumber\\
&\times
 \delta(s_j - f_j(S)) \left[ \Gamma_{j}^{\mu\nu}(k,q,\beta)\Delta\beta_\nu(y,x)-\Upsilon_{j}^\mu(k,q,\beta)\Delta\zeta(y,x) \right]\;.
\end{align}
The standard approach to evaluate this double integral is to introduce a suitable approximation of the integration 
hypersurface \cite{Fu:2021pok,Liu:2021uhn,Becattini:2021suc,Sheng:2024pbw,Zhang:2024mhs} and to integrate first 
in $\di \Sigma_\mu(y)$, assuming that the hydrodynamic fields are slowly varying, so that they can be evaluated
for $y \simeq x$. This method, however, relies on strong geometric assumptions and we recently showed for the local
equilibrium calculation \cite{Sheng:2025cjk} that a much better approximation is obtained by reversing the order of 
integrations, i.e. first integrating in $\di^4 q$, thereafter in $\di \Sigma_\mu(y)$. There is another very good reason 
why the traditional method cannot be adopted: we cannot evaluate the hydrodynamic fields around $y \simeq x$ because $x$ 
lies on $\Sigma_{\rm D}$ and not on $\Sigma_0$, so we cannot expand $\Delta\beta$ and $\Delta\zeta$ around $y=x$. 
Therefore, we deal with the integral \eqref{deltawig3} using a similar way as in our previous work \cite{Sheng:2025cjk}.

The formula \eqref{deltawig3} can be rewritten in a way which makes it apparent the effect of the hydrodynamic limit:
\be\label{deltawig4}
\begin{split}
    \Delta W^+(x,k) =\frac{2}{(2\pi)^5} \sum_{j} \int \di^4 q\;
    \delta (s_j - f_j(S))& \left[ G^{\mu\nu}_{j}\left(k,q,\beta\right) F^{(\beta)}_{\mu\nu}\left(x,q\right)-H_{j}^\mu\left(k,q,\beta\right)F^{(\zeta)}_{\mu}\left(x,q\right)\right]\;, 
\end{split}
\ee
where:
\begin{subequations}\label{Gfunct}
\begin{align}
    &   G_{j}^{\mu\nu}\left(k,q,\beta\right) \equiv \theta(k_+^0)\theta(k_-^0) \frac{1 - \e^{\beta(x)\cdot q}}{\beta(x) \cdot q }
 \varrho(k_+)\varrho(k_-)\Gamma_{j}^{\mu\nu}\left(k,q,\beta\right)\;, \\
&  H_{j}^{\mu}\left(k,q,\beta\right) \equiv \theta(k_+^0)\theta(k_-^0) \frac{1 -\e^{\beta(x)\cdot q}}{\beta(x) \cdot q}
 \varrho(k_+)\varrho(k_-) \Upsilon_{j}^{\mu}\left(k,q,\beta\right)\;,
\end{align}
\end{subequations}
and:
\begin{subequations}\label{Ffunct}
    \begin{align} F^{(\beta)}_{\mu\nu}\left(x,q\right) &= \int_{\Sigma_{0}} \!\!\! \di \Sigma_{\mu}(y) \; 
  \e^{iq\cdot(x-y)} \Delta \beta_\nu(y,x)\;, \\
 F^{(\zeta)}_{\mu}\left(x,q\right) &= \int_{\Sigma_{0}} \!\!\! \di \Sigma_{\mu}(y) \; 
  \e^{\ii q\cdot(x-y)} \Delta\zeta(y,x)\;.
    \end{align}
\end{subequations}
In the hydrodynamic limit, $\Delta\beta_{\nu}$, $\Delta\zeta$ and the normal vector $n_\mu$ to the hypersurface
$\Sigma_0$ are slowly varying functions in space and time, implying that the $F^{(\beta)}_{\mu\nu},\, F^{(\zeta)}_\mu$, which are Fourier 
transform in the variable $q$ integrated in the variable $y$, are functions peaked around $q^{\mu}=0$. This makes 
it possible to obtain a good approximation of the \eqref{deltawig4} by expanding the functions $G^{\mu\nu}_{j}(q)$ 
and $H_{j}^\mu(q)$ around $q^\mu=0$. 

First, we study one of the two cases where the form factors present non-vanishing contributions at the lowest 
order in the $q$ expansion, namely $s=q \cdot k$ and $f_j(S)=0$. This case arose in the calculation of the corrections
to the Wigner function at local equilibrium that we carried it out in detail in ref. \cite{Sheng:2025cjk}. 
Plugging the $q$ expansion in the \eqref{deltawig3} \footnote{The two functions $G_{j}$ and $H_{j}$ feature a $\theta(k^0_+)\theta(k^0_-)$ function which is differentiable for any $k^0 > 0$, being trivially constant equal to 1 
around $q^0=0$; for $k^0=0$ both the functions vanish because $\theta(q^0/2)\theta(-q^0/2) = 0$},
taking into account the definitions of the functions in \eqref{Ffunct} and integrating over $q$, the contribution
to the Wigner function from this term (denoted by $j \to k$) reads:
\begin{equation}\label{deltawiggen}
\begin{split}
    \Delta W_k^+(x,k) &= \frac{2}{(2\pi)^5} \sum_{N=0}^{\infty}\frac{1}{N!}\int_{\Sigma_0} \di \Sigma_{\mu}(y) 
\; I_{N}^{\nu_{1}\nu_{2}\cdots\nu_{N}}(y-x)  \\
& \times \left( \Delta\beta_\nu(y,x) \left[\partial_{\nu_{1}}^{q}\partial_{\nu_{2}}^{q}
\cdots\partial_{\nu_{N}}^{q} G_k^{\mu\nu}\left(k,q,\beta\right)\right]\Big|_{q^{\mu}=0} - \Delta\zeta(y,x) 
\left[\partial_{\nu_{1}}^{q}\partial_{\nu_{2}}^{q}\cdots\partial_{\nu_{N}}^{q}
H_k^{\mu}\left(k,q,\beta\right)\right]\Big|_{q^{\mu}=0} \right)\;,
\end{split}
\end{equation}
where: 
\begin{eqnarray}\label{integrale y-x}
&& I_{N}^{\nu_{1}\nu_{2}\cdots\nu_{N}}(y-x)\equiv\int \di^{4}q\, \delta(q \cdot k)
\e^{-\ii q\cdot(y-x)}q^{\nu_{1}}q^{\nu_{2}}\cdots q^{\nu_{N}} \nonumber \\
=&& (-\ii)^N \partial_x^{\nu_1}\partial_x^{\nu_2}\cdots \partial_x^{\nu_N} 
\int \di^{4}q\,\delta(q\cdot k) \e^{-\ii q\cdot(y-x)} = (2\pi)^3
\frac{(-\ii)^N}{k^0} \partial_x^{\nu_1}\partial_x^{\nu_2}\cdots \partial_x^{\nu_N} \delta^3\left(
\y-\x - \frac{\kb}{k^0} (y^0-x^0) \right)\;,
\end{eqnarray}
keeping in mind that $k^0 > 0$ for the $W^+(x,k)$. The inverse Leibniz formula:
$$
  f g^{(N)} = \sum_{M=0}^N (-1)^M \binom{N}{M} (f^{(M)} g)^{(N-M)}\;,
$$
can be used to isolate the delta function so the \eqref{deltawiggen} becomes:
\begin{eqnarray}\label{deltawig5}
&& \Delta W_k^+(x,k) = \frac{2}{k^0\left(2\pi\right)^2}\sum_{N=0}^{\infty} \frac{(-i)^N}{N!}
\left[\partial^{q}_{\nu_{1}}\cdots\partial^{q}_{\nu_{N}}G^{\mu\nu}\left(k,q,\beta\right)\right]\Big|_{q=0}  
\sum_{M=0}^N \frac{N!(-1)^M}{M!(N-M)!}
 \nonumber \\
&& \times \partial^{\nu_{M+1}}_{x}\cdots\partial^{\nu_{N}}_{x}\int_{\Sigma_{\text{0}}} \di \Sigma_{\mu}(y)
\; \delta^3\left(\y-\x - \frac{\kb}{k^0} (y^0-x^0) \right)\partial^{\nu_{1}}_{x}\ldots\partial^{\nu_{M}}_{x}
\Delta\beta_\nu(y,x) \, +\; {\rm analogous \; term \; for \; 
 \Delta\zeta}\;.
\end{eqnarray}
The key observation is that the presence of the $\delta$--function restricts the support of the integral to those points lying at 
the intersection between the equilibrium hypersurface and:
\begin{equation}\label{worldline}
    {\bf y} = {\bf x} - \frac{\bf k}{k^{0}}(y^0 - x^0)\;,
\end{equation}
corresponding to the world-line of a free particle emitted from $x=(x^0,{\bf x})$ and propagating to $y=(y^0,{\bf y})$ with velocity 
${\bf k}/k^0$. It is important to note, however, that since $k^2 \ne m^2$ the particle is off-mass-shell hence it is a virtual 
particle in the language of Feynman diagrams. Indeed, if $\Sigma_0$ is space-like, at it supposedly is, there is at most one 
intersection between the world-line \eqref{worldline} and the hypersurface, that we will henceforth denote by $\bar y_k(x)$. 
Consequently, one can show (see Appendix~\ref{SECTION: Wigner computation}) that the \eqref{deltawig5} can be transformed into:
\be\label{deltawigk}
\begin{split}
    \Delta W_k^+(x,k) &=\sum_{N=0}^{\infty} \frac{2\theta_k(x)}{\left(2\pi\right)^2}\frac{\left(-\ii \right)^N}{N!} 
\left[D_y(\bar{y}_k(x))\right]^N\\
&\times\left\{ \frac{n_{\mu}(y)}{|k\cdot n(y)|} \left[ G_k^{\mu\nu}\left(k,q,\beta\right) 
\Delta\beta_{\nu}(y,x) - H_k^\mu\left(k,q,\beta\right) \Delta \zeta(y,x) \right] \right\} \Bigg|_{q=0,y=\bar y_k(x)}\;,
\end{split}
\ee
where $\theta_k(x)$ is a Heaviside-like function:
$$
  \theta_k(x) = \begin{cases} 1 \qquad {\rm if \; an \; intersection \; point}\; \bar y_k(x) \; {\rm exists}\;, 
  \\ 0 \qquad {\rm otherwise}\;, \end{cases}
$$
and where we defined:
\begin{equation}\label{Dgrande1}
D_y(\bar y_k(x)) \equiv \Delta^{\nu\rho}(\bar y_k(x))
\partial_{\rho}^{y}\partial_{\nu}^{q}\;,\qquad \Delta^{\nu\rho}(\bar y_k(x))\equiv g^{\nu\rho}-
\frac{n^\nu(\bar{y}_k(x))k^\rho}{\left|k\cdot n(\bar y_k(x))\right|}\; ,
\end{equation}
with $n$ normal vector to the hypersurface  $\Sigma_0$ and $\bar y_k(x)$ intersection between the worldline \eqref{worldline} and $\Sigma_{0}$. 
Note that the operator $\Delta^{\nu\rho}$ is independent on $y$ and therefore in the formula \eqref{deltawigk} the operator 
$D_y$ does not act on itself. The equation \eqref{deltawigk} includes all linear terms in the gradients of $\Delta\beta, \Delta\zeta$ 
to all orders as well as gradients of the normal vector $n$ to the hypersurface; the latter are obviously absent if the hypersurface 
$\Sigma_0$ is a hyperplane. The gradients are evaluated on the initial equilibrium hypersurface $\Sigma_0$, which makes the \eqref{deltawigk}
an expansion of the Wigner function of the kind \eqref{gradexp} discussed in Section \ref{density}. 

A crucial feature of \eqref{deltawigk} is that the order of the gradient of $\beta,\zeta,n$ is $N$, that is the order of the 
expansion in powers of $q$. Therefore, the $q$ expansion of the functions \eqref{Gfunct} corresponds, order by order, to the 
expansion in gradients of this particular contribution to the Wigner function or, otherwise stated, the gradient expansion in 
the thermo-hydrodynamic ($\beta,\zeta$) and geometric ($n$) fields is generated by the expansion in $q$ of the functions 
\eqref{Gfunct}. More specifically, in the \eqref{deltawigk}, the space-time gradients are coupled to derivatives in $q$ of the 
same order for $q=0$, according to the \eqref{Dgrande1}, so the vanishing of a $q$-gradient in $q=0$ implies the vanishing of 
the corresponding term in the space-time gradient expansion. 

The correspondence between the powers of $q$ and the gradients of the fields extends to all other contributions to 
the Wigner function. We now consider the second case where we have non-vanishing terms at the leading order of the 
$q$ expansion of the thermal form factors, namely $s= q \cdot \beta$ and $f_j=0$. In this case, the functions in 
\eqref{Gfunct} can be replaced by:
\begin{subequations}\label{Gfunct2}
    \begin{align}
          G_\beta^{\mu\nu}\left(k,q,\beta\right) &\equiv -\theta(k_+^0)\theta(k_-^0) 
 \varrho(k_+)\varrho(k_-)\Gamma_\beta^{\mu\nu}\left(k,q,\beta\right)\;,  \\
  H_\beta^{\mu}\left(k,q,\beta\right) &\equiv -\theta(k_+^0)\theta(k_-^0) 
 \varrho(k_+)\varrho(k_-) \Upsilon_\beta^{\mu}\left(k,q,\beta\right)\;,
    \end{align}
\end{subequations}
because of the constraint $q \cdot \beta = 0$. The procedure is the same as for the previous $s=q \cdot k$ case,
with the only difference that the equation \eqref{worldline} is replaced by:
\begin{equation}\label{worldline2}
    {\bf y} = {\bf x} - \frac{\betav}{\beta^{0}}(y^0 - x^0)\;,
\end{equation}
which is the world-line of a particle moving with the velocity of the fluid. The resulting contribution to the off-equilibrium correction of the Wigner function are therefore obtained by just replacing $k$ with $\beta$ in the foregoing expressions and 
with the re-definition of the functions $G_\beta$ and $H_\beta$ as in equations \eqref{Gfunct2}:
\be\label{deltawigbeta}
\begin{split}
    \Delta W_\beta^+(x,k) &=\sum_{N=0}^{\infty} \frac{2\theta_\beta(x)}{\left(2\pi\right)^2}\frac{\left(-\ii \right)^N}{N!}\\
    &\times
\left[D_y(\bar{y}_\beta(x))\right]^N \left\{ \frac{n_{\mu}(y)}{|\beta \cdot n(y)|} \left[ G_\beta^{\mu\nu}\left(k,q,\beta\right) 
\Delta\beta_{\nu}(y,x) - H_\beta^\mu\left(k,q,\beta\right) \Delta \zeta(y,x) \right) \right\} \Bigg|_{q=0,y=\bar y_\beta(x)}\;,
\end{split}
\ee
where $\theta_\beta(x)$ is a Heaviside-like function:
$$
  \theta_\beta(x) = \begin{cases} 1 \qquad {\rm if \; an \; intersection \; point}\; \bar y_\beta(x) \; {\rm exists}\;, 
  \\ 0 \qquad {\rm otherwise}\;, \end{cases}
$$
and where we defined:
\begin{equation}\label{Dgrande2}
D_y(\bar y_\beta(x)) \equiv \Delta^{\nu\rho}(\bar y_\beta(x))
\partial_{\rho}^{y}\partial_{\nu}^{q}\;,\qquad \Delta^{\nu\rho}(\bar y_\beta(x))\equiv g^{\nu\rho}-
\frac{n^\nu(\bar{y}_\beta(x))\beta^\rho}{\left|\beta\cdot n(\bar y_\beta(x))\right|}\; .
\end{equation}
The general case can be tackled by first solving the equation $s_j = f_j(S)$ with respect to $q^0$ so 
to turn the delta distribution in \eqref{deltawig3} into:
$$
 \delta(s_j-f_j(S)) = \frac{1}{\chi(\q)}\delta(q^0 - \varphi_j(\q)) \; ;
$$
the arguments $k,\beta$ in the functions $\chi,\varphi$ are understood. In order to fulfill the requirement discussed
in Section \ref{formfactors}, that the delta distributions must reduce to a $\delta(q^0)$ for $\q =0$, the function $\chi(\q)$ must be 
non-vanishing for $\q=0$ whereas $\varphi(\q)$ is ought to vanish for $\q=0$. Hence, the latter can be expanded as:
$$
  \varphi_j(\q) = \nabla_\q \varphi_j|_{\q=0} \cdot \q + R_j(\q) \equiv {\bf v}_j \cdot \q + R_j(\q)\;,
$$
where $R(\q)$ is at least quadratic in the components $q^i$. Therefore, the single contribution $\Delta W^+_{j}$
in eq. \eqref{deltawig3} can be rewritten as:
\begin{align*}
\Delta W_{j}^+(x,k) &=\frac{2}{(2\pi)^5} \int_{\Sigma_{0}} \di \Sigma_{\mu}(y) 
\int \di^{4}q\; \varrho(k_+)\varrho(k_-)\theta(k_+^0)\theta(k_-^0) \e^{\ii q\cdot (x-y)} 
\frac{1 -\e^{\beta(x)\cdot q}}{\beta(x) \cdot q} \nonumber\\
&\times \delta(q^0- {\bf v}_j\cdot \q - R_j(\q))) \frac{1}{\chi(\q)} 
\left[ \Gamma_{j}^{\mu\nu}(k,q,\beta)\Delta\beta_\nu(y,x)-\Upsilon_{j}^\mu(k,q,\beta)\Delta\zeta(y,x) \right]
\nonumber \\
& = \frac{2}{(2\pi)^5} \int_{\Sigma_{0}} \di \Sigma_{\mu}(y) 
\int \di^{3} \qq\; \varrho(k_+)\varrho(k_-)\theta(k_+^0)\theta(k_-^0) \e^{\ii \q \cdot( \y -\x - {\bf v}_j (y^0-x^0))} \e^{-\ii R(\q) (y^0-x^0)} \nonumber\\
& \times \frac{1}{\chi(\q)} \left\{ \frac{1 -\e^{\beta(x)\cdot q}}{\beta(x) \cdot q} \left[ \Gamma_{j}^{\mu\nu}(k,q,\beta)\Delta\beta_\nu(y,x)-\Upsilon_{j}^\mu(k,q,\beta)\Delta\zeta(y,x) \right] \right\}\Bigg|_{q^0={\bf v}_j\cdot \q + R_j(\q)}\;.
\end{align*}
To further proceed, two new vectors can be defined:
$$
q'=q - r = q - (R_j(\q),{\bf 0}) = (q^0-R_j(\q),\q)\;, \qquad \qquad v_j = (1,{\bf v_j})\;.
$$
The vector $r = (R_j(\q),{\bf 0})$ can be seen as a function of either $q$ or $q'$ (since $\q = \q'$) and it has no 
dependence on the time component, i.e. $\partial r/\partial q^0 = \partial r/\partial q'^0 = 0$. $\Delta W^+_j\left(x,k\right)$
can be thus rewritten as:
\begin{align*}
\Delta W_{j}^+(x,k) &= \frac{2}{(2\pi)^5} \int_{\Sigma_{0}} \di \Sigma_{\mu}(y) 
\int \di^4 q\; \delta (q' \cdot v_j) \varrho(k_+)\varrho(k_-)\theta(k_+^0)\theta(k_-^0) \e^{\ii q \cdot (x-y)} 
\nonumber\\
& \times \frac{1}{\chi(\q)} \left\{ \frac{1 -\e^{\beta(x)\cdot q}}{\beta(x) \cdot q} \left[ \Gamma_{j}^{\mu\nu}(k,q,\beta)\Delta\beta_\nu(y,x)-\Upsilon_{j}^\mu(k,q,\beta)\Delta\zeta(y,x) \right] \right\}\;. \nonumber \\
\end{align*}
The integration variable can be changed from $q$ to $q'$ by using the definition above, and the resulting
Jacobian determinant is just 1, so that we can recast the above expression as:
\begin{align*}
\Delta W_{j}^+(x,k) &= \frac{2}{(2\pi)^5} \int_{\Sigma_{0}} \di \Sigma_{\mu}(y) 
\int \di^4 q'\; \, \delta (q' \cdot v_j) \varrho(k_+)\varrho(k_-)\theta(k_+^0)\theta(k_-^0) 
 \e^{\ii q' \cdot (x-y)} \e^{\ii r(q') \cdot (x-y)} \nonumber\\
& \times \frac{1}{\chi(\q')} \left\{ \frac{1 -\e^{\beta(x)\cdot (q'+r(q'))}}{\beta(x) \cdot (q'+r(q'))} 
\left[ \Gamma_{j}^{\mu\nu}(k,q'+r(q'),\beta)\Delta\beta_\nu(y,x)-\Upsilon_{j}^\mu(k,q'+r(q'),\beta)\Delta\zeta(y,x) \right] \right\}\;.
\end{align*}
Then, expanding the exponential $\exp[\ii r(q')\cdot (x-y)]$:
\begin{align*}
\Delta W_{j}^+(x,k) &= \sum_{\ell=0}^\infty \frac{2}{(2\pi)^5} \int_{\Sigma_{0}} \di \Sigma_{\mu}(y) 
\int \di^4 q'\; \, \delta (q' \cdot v_j) \varrho(k_+)\varrho(k_-)\theta(k_+^0)\theta(k_-^0) 
 \e^{\ii q' \cdot (x-y)}(x^0-y^0)^\ell \frac{\ii^\ell R^\ell(\q)}{\ell! \chi(\q')} \nonumber\\
& \times  \left\{ \frac{1 -\e^{\beta(x)\cdot (q'+r(q'))}}{\beta(x) \cdot (q'+r(q'))} 
\left[ \Gamma_{j}^{\mu\nu}(k,q'+r(q'),\beta)\Delta\beta_\nu(y,x)-\Upsilon_{j}^\mu(k,q'+r(q'),\beta)\Delta\zeta(y,x) \right] \right\}\;,
\end{align*}
whence:
\begin{align}\label{deltawigjs3}
\Delta W_{j}^+(x,k) &= \sum_{\ell=0}^\infty \frac{2}{(2\pi)^5} 
\int \di^4 q'\; \, \delta (q' \cdot v_j) \left[ G^{\prime\,\mu\nu}_{j\,(\ell)}\left(k,q',\beta\right) F^{\prime\,(\beta)}_{\mu\nu\,(\ell)}\left(q';x\right)-
H_{j\,(\ell)}^{\prime\,\mu}\left(k,q',\beta\right)F^{\prime\,(\zeta)}_{\mu\nu\,(\ell)}\left(q';x\right) \right]\;. 
\end{align}
Each term of the series can be written in a form which is similar to the two foregoing cases $\Delta W_{k,\beta}^+(x,k)$,
by defining:
\begin{subequations}\label{Gfunctprime}
 \begin{align}
     &  G^{\prime\,\mu\nu}_{j\,(\ell)}\left(k,q',\beta\right)\equiv \varrho(k_+)\varrho(k_-)\theta(k_+^0)\theta(k_-^0) 
 \frac{\ii^\ell R^\ell(\q')}{\chi(\q')}  \frac{1 -\e^{\beta(x)\cdot (q'+r(q'))}}{\beta(x) \cdot (q'+r(q'))} 
 \Gamma_{j}^{\mu\nu}\left(k,q'+r(q'),\beta\right)\;, \\
&  H_{j\,(\ell)}^{\prime\,\mu}\left(k,q',\beta\right) \equiv \varrho(k_+)\varrho(k_-)\theta(k_+^0)\theta(k_-^0) 
 \frac{\ii^\ell R^\ell(\q')}{\chi(\q')} \frac{1 -\e^{\beta(x)\cdot \left(q'+r(q')\right)}}{\beta(x) \cdot (q'+r(q'))} 
  \Upsilon_{j}^\mu(k,q'+r(q'),\beta)\;,
 \end{align}
\end{subequations}
and:
\begin{subequations}\label{Ffunctprime}
\begin{align}
    F^{\prime\,(\beta)}_{\mu\nu\,(\ell)}\left(x,q\right) &= \int_{\Sigma_{0}} \!\!\! \di \Sigma_{\mu}(y) \; 
  \e^{\ii q'\cdot(x-y)} \Delta \beta_\nu(y,x)(x^0-y^0)^\ell\;, \\
 F^{\prime\,(\zeta)}_{\mu\,(\ell)}\left(x,q\right)&= \int_{\Sigma_{0}} \!\!\! \di \Sigma_{\mu}(y) \; 
  \e^{\ii q' \cdot(x-y)} \Delta\zeta(y,x)(x^0-y^0)^\ell \;.
\end{align}
\end{subequations}
The functions in eq. \eqref{Ffunctprime} are strongly peaked around $q'=0$ in the hydrodynamic limit, so those
in \eqref{Gfunctprime} can be expanded around $q'=0$, just like in the previous cases. Because of the distribution 
$\delta(q' \cdot v)$ in the integral \eqref{deltawigjs3}, all the conclusions previously achieved hold. The 
$\Delta W_{j}$ can be written in the form of an expansion, complicated as it may be, of the form \eqref{deltawigk}
and \eqref{deltawigbeta}, where the orders of the $q'$ power expansion correspond to the gradients of the $\beta,\zeta$ 
and $n$ fields evaluated at the intersection point between the hypersurface $\Sigma_0$ and the lines:
$$
   \y = \x - {\bf v}_j(k,\beta) (y^0-x^0)\;,
$$
where we have restored the previously understood dependence of $v_j$ on $k$ and $\beta$. 
For the functions $\Gamma_{j}\left(k,q,\beta\right)$ and $r(q)$ are at least quadratic in $q$, also the functions 
$G'_{j}(q')_{(\ell)}$ and $H'_{j}(q')_{(\ell)}$ turn out to be quadratic in $q'$, implying that there are no contributions 
from the zeroth order and first order gradients in the \eqref{deltawigjs3}.

In conclusion, stopping at the first order of the $q$ expansion amounts to stop at the first order gradient expansion and, 
in this case, the only non-vanishing contributions to the off-equilibrium part of the Wigner function stems from the $\Gamma_k$ 
and $\Gamma_\beta$ form factors. In formulae, the correlators in the equation \eqref{Theta deltas2} read:
$$
 \Theta^{\mu\nu}\left(k,q,\beta\right) = \delta(q \cdot k)\, \Gamma^{\mu\nu}_k\left(k,q,\beta\right) + \delta(q \cdot \beta)\, 
 \Gamma^{\mu\nu}_\beta\left(k,q,\beta\right) + \sum_{j} \delta(s_j - f_j(S)){\cal O}(q^2)\;,
$$
and the off-equilibrium correction, neglecting ${\cal O}(q^2)$ terms, turns out to be:
\begin{align}\label{deltawig1storder}
\Delta W^+(x,k) &\simeq \nonumber \\
& \frac{2}{\left(2\pi\right)^2}\sum_{N=0}^{1} 
\frac{\left(-\ii \right)^N}{N!}\Bigg\{\theta_k(x) 
\left[D_y(\bar{y}_k(x))\right]^N \left\{ \frac{n_{\mu}(y)}{|k\cdot n(y)|} \left[G_k^{\mu\nu}\left(k,q,\beta\right) 
\Delta\beta_{\nu}(y,x) - H_k^\mu\left(k,q,\beta\right) \Delta \zeta(y,x) \right]\right\}\Bigg|_{\substack{q=0 \\y=\bar y_k(x)}} \nonumber \\
&+\theta_\beta(x) 
\left[D_y(\bar{y}_\beta(x))\right]^N \left\{ \frac{n_{\mu}(y)}{|\beta \cdot n(y)|} \left[ G_\beta^{\mu\nu}\left(k,q,\beta\right) 
\Delta\beta_{\nu}(y,x) - H_\beta^\mu\left(k,q,\beta\right) \Delta \zeta(y,x) \right) \right\} \Bigg|_{\substack{q=0 \\y=\bar y_k(x)}} \Bigg\}\;.
\end{align}
It should be stressed, however, that if the hypersurface is curved there are contributions at
the lowest order gradients (zeroth and first order) of the thermo-hydrodynamic fields $\beta$ and $\zeta$ from all $N > 1$ terms, 
obtained by letting derivatives $\partial_y$ in the above corrections \eqref{deltawigk}, \eqref{deltawigbeta} and 
\eqref{deltawigjs3} to act on the field $n^\mu(y)$. Only if the curvature of the hypersurface is small such terms can be 
regarded as small corrections of the main terms with $N=0,1$ \cite{Zhang:2025vlk}. We will henceforth disregard those terms,
keeping in mind though that their potential relevance should be considered for moderately curved hypersurfaces.

Up to the second order in the $q-$derivatives the two functions $G^{\mu\nu}_{k}$ \eqref{Gfunct} and $G^{\mu\nu}_\beta$ 
\eqref{Gfunct2} can be expanded taking into account:
\begin{equation*}
    \begin{split}
        \frac{1-\e^{\beta(x)\cdot q}}{\beta(x)\cdot q}&=-1-\frac{1}{2}\beta_\tau (x)q^\tau-\frac{1}{6}\beta_{\tau_1}(x)
        \beta_{\tau_2}(x)q^{\tau_1}q^{\tau_2}+\mathcal{O}(q^3)\;,\\
        \Gamma^{\mu\nu}_{k/\beta}\left(k,q,\beta\right)&=\Gamma^{\mu\nu}_{k/\beta}\left(k,0,\beta\right)+q^\tau\,
        \frac{\partial}{\partial q^\tau}\Gamma^{\mu\nu}_{k/\beta}\left(k,q,\beta\right)\Bigg|_{q=0}+\frac{1}{2}q^{\tau_1}q^{\tau_2}\,
        \frac{\partial^2}{\partial q^{\tau_1}\partial q^{\tau_2}}\Gamma^{\mu\nu}_{k/\beta}\left(k,q,\beta\right)\Bigg|_{q=0}+\mathcal{O}(q^3)\;,\\
        \varrho(k_+)\varrho(k_-)&=\varrho^2(k)+\frac{1}{8}
        \frac{\partial^2}{\partial q^{\tau_1}\partial q^{\tau_2}}\varrho(k_+)\Bigg|_{q=0}-\frac{1}{4}
        \frac{\partial}{\partial q^{\tau_1}}\varrho(k_+)\Bigg|_{q=0}
        \frac{\partial}{\partial q^{\tau_2}}\varrho(k_-)\Bigg|_{q=0}+\mathcal{O}(q^4)\;,
    \end{split}
\end{equation*}
where $k_\pm=k\pm q/2$.
Similarly we can expand the tensors $\Upsilon^\mu_{k/\beta}$ and the functions $H^\mu_{k/\beta}$ so the functions $G^{\mu\nu}_{k/\beta}$, at lowest  order in $q$, are given by:
\eqref{Gamma1q0}:
\begin{equation*}
    \begin{split}
        G^{\mu\nu}_k\left(k,0,\beta\right)&=-\varrho^2(k)\theta(k^0)\,\Gamma^{\mu\nu}_k\left(k,0,\beta\right)=-\frac{\theta(k^0)}{\left(2\pi\right)^2}\varrho(k)n_{\rm B}(k) \left[ 1+n_{\rm B}(k) 
    - \frac{\partial \log \varrho(k)}{\partial (k \cdot \beta)} \right]k^\mu k^\nu\;,\\
    G^{\mu\nu}_\beta\left(k,0,\beta\right)&=-\varrho^2(k)\theta(k^0)\,\Gamma^{\mu\nu}_\beta\left(k,0,\beta\right)=\frac{2\theta(k^0}{(2\pi)^2} )\varrho(k)n_{\rm B}(k)
  \frac{\partial \log \varrho(k)}{\partial \beta^2}\beta^\mu\beta^\nu\, .
    \end{split}
\end{equation*}
In the same way for the functions $H^\mu_{k/\beta}$ we have:
\begin{equation*}
    \begin{split}
        H^\mu_k\left(k,0,\beta\right)&=-\frac{\theta(k^0)}{(2\pi)^4}\varrho^2(k)\Upsilon^k_1(0)\,k^\mu\;,\\
        H^\mu_\beta\left(k,0,\beta\right)&=-\frac{\theta(k^0)}{(2\pi)^4}\varrho^2(k)\Upsilon^\beta_1(0)\,\beta^\mu\;,
    \end{split}
\end{equation*}
with $\Upsilon^k_1\left(k,0,\beta\right)$ and $\Upsilon^\beta_1\left(k,0,\beta\right)$ given in \eqref{Upsilonkbeta q0}.

Plugging these expressions in \eqref{deltawig1storder} and taking the $N=0$ term we then obtain the off-equilibrium 
correction to the Wigner function at the leading order of the gradient expansion which reads:
\begin{align}\label{zerorder}
\Delta^{(0)}W\left(x,k\right)&=\frac{2\varrho(k)n_{\rm B}(k)}{(2\pi)^4}\Bigg\{
{\theta_k(x)}\left(1+n_{\rm B}(k)-\frac{\partial\log\varrho(k)}{\partial(k \cdot \beta)} \right)\Big[k\cdot\beta(x)-k\cdot\beta\left(\overline{y}_k(x)\right)\Big] \nonumber\\
&\qquad\qquad\qquad\quad-2\theta_\beta(x) \frac{\partial\log\varrho(k)}{\partial\beta^2}\Big[\beta^2(x)-\beta(x)\cdot \beta(\overline{y}_\beta(x))\Big]\Bigg\} \\
&+\frac{2\varrho^2(k)}{(2\pi)^2}\Bigg\{\theta_k(x)\Upsilon^k_1\left(k,0,\beta\right)\left[\zeta(\overline{y}_k(x))-\zeta(x)\right]+\theta_\beta(x)\Upsilon^\beta_1\left(k,0,\beta\right)\left[\zeta(\overline{y}_\beta(x))-\zeta(x)\right]\Bigg\}\,.\nonumber
\end{align}

At the first order in the gradient expansion the correction is given by the $N=1$ term in \eqref{deltawig1storder}:
\begin{align*}
\Delta^{(1)} W^+(x,k) &= -\frac{\ii\theta_k(x)}{(2\pi)^2}\Big\{ 
\partial^q_\sigma G^{\mu\nu}_k\left(k,q,\beta\right)\Big|_{q=0}\Delta^{\sigma\gamma}(\bar{y}_k(x))\partial^y_\gamma 
\left[ \frac{n_{\mu}(y)\Delta\beta_{\nu}(y,x)}{|k\cdot n(y)|} \right] \Bigg|_{y=\bar y_k(x)}\nonumber\\
&-\partial^q_\sigma H^{\mu}_k\left(k,q,\beta\right)\Big|_{q=0}\Delta^{\sigma\gamma}(\bar{y}_k(x))\partial^y_\gamma 
\left[ \frac{n_{\mu}(y)\Delta\zeta(y,x)}{|k\cdot n(y)|} \right] \Bigg|_{y=\bar y_k(x)}\Big\} \nonumber \\
& -\frac{\ii\theta_\beta(x)}{\left(2\pi\right)^2}
\Big\{ \partial^q_\sigma G^{\mu\nu}_\beta\left(k,q,\beta\right)\Big|_{q=0}\Delta^{\sigma\gamma}(\bar{y}_\beta(x)\partial^y_\gamma 
\left[ \frac{n_{\mu}(y)\Delta\beta_{\nu}(y,x)}{|k\cdot n(y)|} \right] \Bigg|_{y=\bar y_\beta(x)}\nonumber\\
&-\partial^q_\sigma H^{\mu}_\beta\left(k,q,\beta\right)\Big|_{q=0}\Delta^{\sigma\gamma}(\bar{y}_\beta(x))\partial^y_\gamma 
\left[ \frac{n_{\mu}(y)\Delta\zeta(y,x)}{|k\cdot n(y)|} \right] \Bigg|_{y=\bar y_\beta(x)}\Big\}\;.
\end{align*}
This contribution exactly vanishes. This can be shown by calculating the first order derivative with respect to $q$, 
of the functions $G^{\mu\nu}_{k/q}$ and $H^\mu_{k/q}$. For the $q\cdot k=0$ channel one has:
\begin{equation*}
    \begin{split}
        \partial^q_\lambda G^{\mu\nu}_k\left(k,q,\beta\right)\Big|_{q=0}&=\varrho^2(k)\left[-\frac{1}{2}\beta_\lambda(x) 
        \Gamma^{\mu\nu}_k\left(k,0,\beta\right)+\partial^q_\lambda\Gamma^{\mu\nu}_k\left(k,q,\beta\right)\Big|_{q=0}\right]=0\,,\\
        \partial^q_\lambda H^{\mu\nu}_k\left(k,q,\beta\right)\Big|_{q=0}&=\varrho^2(k)\left[-\frac{1}{2}\beta_\lambda(x) 
        H^{\mu\nu}_k\left(k,0,\beta\right)+\partial^q_\lambda H^{\mu\nu}_k\left(k,q,\beta\right)\Big|_{q=0}\right]=0\,,
    \end{split}
\end{equation*}
while for the $q\cdot\beta=0$ channel one has:
\begin{equation*}
    \begin{split}
        \partial^q_\lambda G^{\mu\nu}_k\left(k,q,\beta\right)\Big|_{q=0}&=\varrho^2(k)\,\partial^q_\lambda\Gamma^{\mu\nu}_k\left(k,q,\beta\right)\Big|_{q=0}=0\,,\\
        \partial^q_\lambda H^{\mu\nu}_k\left(k,q,\beta\right)\Big|_{q=0}&=\varrho^2(k)\,\partial^q_\lambda H^{\mu\nu}_k\left(k,q,\beta\right)\Big|_{q=0}=0\,,
    \end{split}
\end{equation*}
where both vanish due to the relations \eqref{Gamma q = Gamma -q exp} and \eqref{Upsiloj q = Upsilon -q exp}. Hence:
\begin{equation*}
    \Delta^{(1)}W\left(x,k\right)=0\;.
\end{equation*}
Altogether, the off-equilibrium correction of the Wigner function can be written as:
\begin{equation}
    \Delta W^+\left(x,k\right)=\Delta^{(0)}W\left(x,k\right)+\mathcal{O}(\partial^2)\;.
\end{equation}
The higher order corrections in the above equation are far more complicated than the the leading term 
\eqref{zerorder} and will depend on all the form factors and on the all possible delta channels $\delta(s_j -f_j(S))$
in the equation \eqref{deltawig4}.

\section{Discussion}
\label{discussion}

In the previous section we have obtained a gradient expansion of the off-equilibrium correction of the Wigner 
function where all the gradients are evaluated on the initial equilibrium hypersurface instead of the final one. 
In a sense, we have obtained a constitutive equation of the Wigner function - that is a relation with the geometric and 
the thermo-hydrodynamic fields - which is non-local in time. More precisely, we have found the leading order solution of 
the equation of motion of the Wigner 
function which parametrically depends on the initial conditions, i.e. the fields $\beta$ and $\zeta$ on $\Sigma_0$.
Surprisingly, according to the equation \eqref{zerorder}, the gradient expansion includes a zeroth order term, which depends on 
the finite difference between the thermo-hydrodynamic fields at the point $x$ and over a point lying on the initial hypersurface 
$\Sigma_0$. This term suggests that memory effects are present in the full quantum statistical approach to the calculation of the 
Wigner function, hence of the momentum spectrum, at the decoupling.

It might be argued that if we had started from the traditional decomposition of the density operator \eqref{gauss} we would have
obtained an expansion in gradients evaluated at the same point $x$ of the Wigner function. As has been mentioned in Section
\ref{density}, this result crucially depends on the shape of the correlation functions in the equation \eqref{traditional}, i.e. whether 
it has a maximum at $y \sim x$. In this Section, we will show that this does not necessarily occur for the Wigner function and so, 
even if we had used the decomposition into local equilibrium and dissipative terms, we would eventually get the same expansion in terms of
the initial gradients. 

In this Section, we will delve into the features of the off-equilibrium correction found for the Wigner function, including the
aforementioned non-locality. 

\subsection{Non-locality of the thermal correlation function}

The non-locality of the correlation function can be understood rewriting the equation \eqref{deltawig3} by 
using the definitions \eqref{Gfunct}, as:
\be\label{deltawig-special}
\Delta W^+(x,k) =\frac{2}{(2\pi)^5} 
\int_{\Sigma_0} \!\!\!\! \di \Sigma_{\mu}(y) \Delta \beta_\nu(y,x) 
\left( \int \di^{4}q\; \sum_{j} \delta(s_j - f_j(S)) \e^{\ii q\cdot (x-y)} G_{j}^{\mu\nu}\left(k,q,\beta\right) \right)\;,
\ee
where we omitted the the term in the chemical potential for the sake of simplicity. By comparison with e.g. eq. \eqref{ddO}, the 
equation \eqref{deltawig-special} identifies the correlation function of the Wigner operator and the stress-energy tensor operator:
\be\label{corrfunc}
  \begin{split}
      C^{\mu\nu}_{WT}(x-y,k) &= -(2\pi)^5 \int^1_0\di z \; \langle\wW^+(x,k),\e^{z\wE_{\rm GE}}\wT^{\mu\nu}(y)\e^{-z\wE_{\rm GE}}
\rangle_{c,\mathrm{GE}}\\
&= \sum_{j} \int \di^{4}q\; \delta(s_j - f_j(S)) \e^{\ii q\cdot (x-y)} 
G_{j}^{\mu\nu}(k,q,\beta)\;. 
  \end{split}
\ee
This correlation function appears, once $\wE_{\rm GE}$ is approximated by $\wE$, in the equation \eqref{traditional} as well;
a similar expression hold for the four-current term. 

\begin{figure}[h!]
    \centering
    \includegraphics[width=0.55\linewidth]{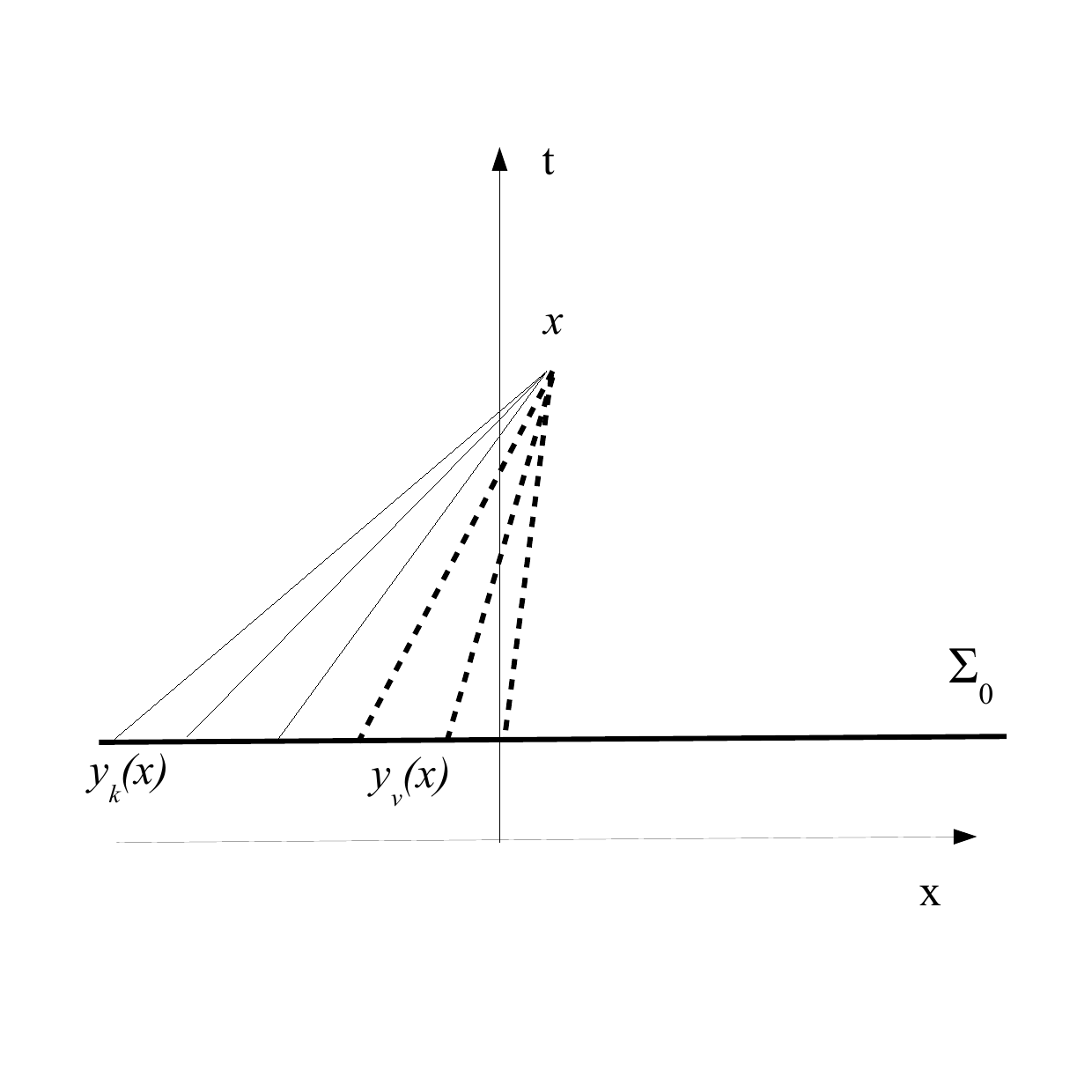}
    \caption{The correlation function between the Wigner operator $\wW^+(x,k)$ and stress-energy tensor (or vector current) operator 
    in a point $y$ features terms which are constant over the worldlines with tangent vector $k$ (solid lines) and terms which are 
    constant over worldlines with tangent vector $v(k,\beta)$ (dashed lines). The integration over $k$ eventually yields correlation functions
    which are strongly peaked around $y = x$.}
    \label{fig:correlfunc}
\end{figure}

It is common wisdom that such a function decays rapidly for macroscopic $(x-y)^2$, that is much larger than the typical microscopic 
scales, such as $1/m$, $\sqrt{\beta^2}$, interactions lengths and combinations thereof. In fact, because each term of the series in 
the eq. \eqref{corrfunc} can be rearranged so as to contain $\delta (q \cdot k)$, $\delta( q \cdot \beta)$, or 
$\delta (q \cdot v_j(k,\beta))$ in general - as has been shown in the Section \ref{hydrolimit} - this is not possibly the case. 
Because of these distributions, the various terms of the correlation function \eqref{corrfunc} turn out to be constant over the lines:
$$
  (x-y)^\mu = \tau \, k^\mu\;, \qquad \qquad (x-y)^\mu = \tau \, \beta^\mu\;,
$$
for $\delta (q \cdot k)$ and $\delta( q \cdot \beta)$ respectively, where $\tau$ is a real parameter. For $\delta(q \cdot v_j(k,\beta))$ 
also the correlation function is constant along the world-line $(x-y)^\mu = \tau v_j^\mu$, provided that the correlation function is 
re-defined by letting a factor $(x^0-y^0)^\ell$ with $\ell\ge1$ to be extracted from it, see Section \ref{hydrolimit}. For each of the contributing terms
labeled by $j$ in equation \eqref{corrfunc} thus the point $y=x$ is not an isolated maximum and the intersection points between the 
above lines and the hypersurface $\Sigma_0$ are responsible for the largest contribution to the integral \eqref{deltawig-special}. 
If the number of $j$ terms in the \eqref{corrfunc} is very large, it may happen that the sum yields a correlation function which 
is peaked around $y \sim x$ (see figure \ref{fig:correlfunc}), but the possibility of an infinite series of different delta distributions
in the equation \eqref{corrfunc} seems unlikely.

This non-locality feature is at odds with the familiar one when considering the correlation function of two local operators $\wO_1$ 
and $\wO_2$ depending on the quantum fields:
$$
C_{O_1O_2}(x-y) = \int^1_0\di z \; \langle\wO_1(x),\;\e^{z\wE_{\rm GE}}\,\wO_2(y)\,\e^{-z\wE_{\rm GE}}
\rangle_{c,\mathrm{GE}}\;,
$$
e.g. two components of conserved currents, which has a typical maximum for $x \sim y$ with a width driven by 
microscopic lengths. The reason for the different behaviour between an actual local operator and the Wigner operator as to 
their correlation with another local operator is that the Wigner operator is not truly local, being the Fourier transform of the 
product of field operators in two points, see eq. in eq. \eqref{ScalarField:WigDef}. Furthermore, the Wigner operator also depends 
on an additional argument $k$ besides space-time point $x$. 

It should be emphasized that if we had used the traditional method of decomposing the density operator with the Gauss theorem, 
separating the local equilibrium from the dissipative contribution like in eq. \eqref{gauss}, we would not get an expansion in terms 
of the gradients at the point $x$ anyway, in fact we would get the same expansion in terms of the initial gradients including the 
zeroth order term. An explicit calculation, starting from from equation \eqref{gauss} and proceeding to combine the local equilibrium 
with the dissipative corrections is reported in ref. \cite{roselliphd}. The method of expanding the density operator like in 
eq. \eqref{gauss} is tantamount to expand the off-equilibrium correction of the Wigner operator in eq. \eqref{deltawig-special} with 
the Gauss theorem; the resulting expression is:
\begin{align*}
 \Delta W^+(x,k)&= \int_{\Sigma_x} \!\!\! \di \Sigma_{\mu}(y) \; \Delta \beta_\nu(y,x) 
 C_{WT}^{\mu\nu}(x-y,k) + 
 \int_\Omega \di^4 y \; \Delta \beta_\nu(y,x) \frac{\partial}{\partial y^\mu} C_{WT}^{\mu\nu}(y-x,k) \\
&+ \int_\Omega \di^4 y \;  C_{WT}^{\mu\nu}(y-x,k) \frac{\partial}{\partial y^\mu}\Delta \beta_\nu(y,x)\;,
\end{align*}
where again the contribution proportional to $\Delta \zeta$ has been neglected for the sake of simplicity. Unless the correlation
function is narrow-peaked around $y \sim x$, we cannot neglect the second integral on the right hand side nor can we approximate 
the third integral by extracting the gradient of $\beta$ in $x$. Therefore, a consistent elaboration of the above equation taking 
into account the shape of the correlation function again leads to the eq. \eqref{deltawig-special}.

The long-distance persistence of the correlation function \eqref{corrfunc} makes it apparent that memory effects, which 
are manifest in the term \eqref{zerorder}, play an important role for the Wigner function in an expanding fluid. 
Nevertheless, it is reasonable to expect that if we generate an actual local operator by integrating the Wigner operator 
in the momentum variable $k$, a rapidly decaying correlation function in $(x-y)$ occurs and the typical behaviour is restored.
This happens because one integrates the correlation functions $C^{\mu\nu}_{WT}(x-y,k)$ and $C^\mu_Wj(x-y,k)$ over an 
infinite set of lines with fixed $k$, all of them converging in $x$ (see figure \ref{fig:correlfunc}). We show that this is a 
likely result by working out an example for the case $\delta(s_j - f_j(S))=\delta(q\cdot k)$ and the local operator:
$$
 \int \di^4 k \; \wW(x,k) = : \wphi^\dagger(x) \wphi(x) :\;,
$$
considering its correlation function with the stress-energy tensor $\wT^{\mu\nu}(y)$. A contributing term to the
this function is obtained by integrating the $\wW^+(x,k)$ part of the Wigner operator (the anti-particle and
space-like parts in eq. \eqref{widecomp} should also be included) in equation \eqref{corrfunc}, that is:
$$
-(2\pi)^5 \int^1_0\di z \; \langle : \wphi^\dagger(x) \wphi(x) :\,,\,\e^{z\wE_{\rm GE}}\wT^{\mu\nu}(y)\e^{-z\wE_{\rm GE}}
\rangle_{c,\mathrm{GE}} = \sum_{j} \int \di^4 k \int \di^{4}q\; \delta(s_j - f_j(S)) \e^{\ii q\cdot (x-y)} 
G_{j}^{\mu\nu}\left(k,q,\beta\right) \;.
$$
Let us now restrict to $\delta(q \cdot k)$, expand the function $G_k\left(k,q,\beta\right)$ and proceed like in the 
equation \eqref{deltawiggen} and following. We obtain:
\begin{align*}
& \int \di^4 k \int \di^{4}q\; \delta(q \cdot k) \, \e^{\ii q\cdot (x-y)} G_k^{\mu\nu}\left(k,q,\beta\right)
    \nonumber \\
& = \sum_{N=0}^{\infty}\frac{1}{N!} \partial_x^{\nu_1}\partial_x^{\nu_2}\cdots \partial_x^{\nu_N} 
    \int \di^4 k \left[\partial_{\nu_{1}}^{q}\partial_{\nu_{2}}^{q} \ldots\partial_{\nu_{N}}^{q} 
    G_k^{\mu\nu}\left(k,q,\beta\right)\right]\Big|_{q=0} \frac{(-\ii)^N}{k^0} \delta^3\left(\y-\x - \frac{\kb}{k^0} (y^0-x^0) \right)\nonumber \\
& = \sum_{N=0}^{\infty}\frac{(-\ii)^N}{N!} \partial_x^{\nu_1}\partial_x^{\nu_2}\cdots \partial_x^{\nu_N} 
    \int \di k^0 \frac{(k^0)^2}{\left(y^0-x^0\right)^3}\left\{\partial_{\nu_{1}}^{q}\partial_{\nu_{2}}^{q} \ldots\partial_{\nu_{N}}^{q} 
    G_k^{\mu\nu}\left[\left(k^0,\frac{k^0(\y -\x)}{y^0-x^0}\right),q,\beta\right]\right\}\Big|_{q=0}\;. 
\end{align*}
The general term of last expression is difficult to work out, however the term $N=0$ can be expanded based on the 
equation \eqref{Gamma1q0}. This term becomes:
\begin{align*}
   \int \di k^0 \frac{(k^0)^2}{\left(y^0-x^0\right)^3} \theta(k^0) \varrho(k) \frac{1}{(2\pi)^2} n_B(k\cdot\beta) 
   \left( 1 + n_B(k\cdot\beta) - \frac{\partial \log \varrho(k)} {\partial (k\cdot \beta)} \right) k^\mu k^\nu   \;,
\end{align*}
with $\kb = k^0(\y-\x)/(y^0-x^0)$. For a free field the integral can be solved analytically and we obtain,
by using the \eqref{freespectral}:
\be\label{decrease}
\frac{1}{4\pi} \theta((y-x)^2) {\rm sign}(y^0-x^0) \frac{m^3}{[(y-x)^2]^{5/2}} n_B( k \cdot \beta) 
\left(1+n_B(k \cdot \beta)\right) (y-x)^\mu (y-x)^\nu\;,
\ee
where $k = m \, {\rm sign}(y^0-x^0) (y-x)/\sqrt{(y-x)^2}$. Apparently, the function \eqref{decrease} decays for
large values of $(y-x)^2$ in all directions, as expected, with a rate dictated by the mass of the field, and it
diverges for $y \to x$. In general, in the interacting case, and taking into account all terms in the expansion,
it is thus reasonable to expect that the correlation functions of truly local operators are strongly peaked around 
$y \sim x$ in spite of the fact that those involving the Wigner operator are not. 

The off-equilibrium correction to the operator $:\wphi(x)\wphi(x):$ can be obtained by integrating the
correlation function of the Wigner operator and the stress-energy tensor in $\di^4 k$:
\begin{align*}
 \Delta \langle :\wphi(x)\wphi(x): \rangle &= \int \di^4 k \; \Delta W(x,k) = \frac{2}{(2\pi)^5} 
\int_{\Sigma_0} \!\!\! \di \Sigma_{\mu}(y) \; \Delta \beta_\nu(y,x) \int \di^4 k \; C^{\mu\nu}_{WT}(x-y,k) \nonumber \\
& \equiv \int_{\Sigma_0} \!\!\! \di \Sigma_{\mu}(y) \; \Delta \beta_\nu(y,x) C^{\mu\nu}(x-y)\;,
\end{align*}
plus a similar term involving $\Delta \zeta$ and $C^\mu_{Wj}(x-y)$.
This integral looks odd because the correlation function is peaked around $y\sim x$, while the point $y$ on the initial 
hypersurface $\Sigma_0$ is macroscopically distant from $x$. However, in a way similar to the deformation of a path
to calculate integrals over the complex plane, one can use the Gauss theorem to turn the integral above into:
\begin{align*}
 \Delta \langle :\wphi(x)\wphi(x): \rangle &= \int_{\Sigma_x} \!\!\! \di \Sigma_{\mu}(y) \; \Delta \beta_\nu(y,x) 
 C^{\mu\nu}(x-y) + \int_\Omega \di^4 y \; \Delta \beta_\nu(y,x) \frac{\partial}{\partial y^\mu} C^{\mu\nu}(y-x) \\
&+ \int_\Omega \di^4 y \;  C^{\mu\nu}(y-x) \frac{\partial}{\partial y^\mu}\Delta \beta_\nu(y,x)\;,
\end{align*}
where $\Sigma_x$ is a hypersurface passing through $x$. The first term on the right hand side corresponds to the 
local equilibrium correction to the global equilibrium, while the other two terms correspond to the dissipative
corrections. If the function $C^{\mu\nu}(x-y)$ is highly peaked 
around $y \sim x$ with a width governed by microscopic quantities, whereas $\Delta \beta_\nu(y-x)$ is slowly
varying, since $\Delta \beta_\nu (y=x) = 0$, the rightmost integral provides the largest contribution, so that:
$$
 \Delta \langle :\wphi(x)\wphi(x): \rangle \simeq \int_\Omega \di^4 y \;  C^{\mu\nu}(y-x) \partial_\mu \Delta \beta_\nu(y,x)
  \simeq \partial_\mu \beta_\nu(x) \int_\Omega \di^4 y \;  C^{\mu\nu}(y-x)\;,
$$
thus recovering the kind of familiar corrections proportional to the local gradient of $\beta$, with a coefficient given by 
the integral of the correlation function, much like the Kubo formulae of transport coefficients. 

\subsection{Zeroth order term: free vs interacting field}

As has been mentioned, the unexpected appearance of the zeroth order term \eqref{zerorder} indicates that memory effects are present in the 
full quantum statistical approach to the calculation of the Wigner function, hence of the momentum spectrum, at the decoupling. 
An insight about its nature can be gained by considering the free field Wigner operator. In this case, the full solution is 
known, see equation \eqref{fssol} and relevant discussion in Section \ref{density}:
$$
  W^+(x,k) = \Tr (\wrho\; \wW^+(X,k))    \;,      
$$
with $\wrho$ given by the equation \eqref{densop} and $X \in \Sigma_0$ being the intersection point between the wordline
drawn from $x$ with velocity $k/k^0$. The calculation of the right hand side can be carried out much the same way as for 
the Wigner function of the Dirac field at local equilibrium presented in ref. \cite{Sheng:2025cjk} and its leading term reads:
\be\label{leqsigma0}
   W^+_0(X,k) \simeq \frac{1}{(2\pi)^3\varepsilon_k} \delta\left(k^0-{\sqrt{\kb^2+m^2}}\right) n_B(\beta(X))\;.
\ee

Now, suppose we set out to calculate the leading order term of Wigner function of the free field in the point $x$ lying in the 
future of $\Sigma_0$ with the method presented in this work. For a free field the calculation is much easier than for an 
interacting field because the stress-energy tensor is known, and, chiefly, there is only one term in the series \eqref{deltawig3}, the one with $\delta(q\cdot k)$ (see ref. \cite{Sheng:2025cjk}). At the leading order the result is 
obtained from the equations \eqref{leading order W} and \eqref{zerorder} by using the free spectral function \eqref{freespectral}:
\begin{equation*}
W^+(x,k) \simeq \frac{1}{(2\pi)^3} \delta\left(k^0-{\sqrt{\kb^2+m^2}}\right) n_B(\beta(x)) \Big[ 
{\theta_k(x)}(1+n_{\rm B}(k))\left( k\cdot\beta(x)-k\cdot\beta\left(\overline{y}_k(x)\right)-
\zeta(x)+\zeta(\overline{y}_k(x))\right) \Big]\;.
\end{equation*}
The above equation should be an approximation of the \eqref{fssol} and, indeed, it can be obtained from \eqref{fssol}
and \eqref{leqsigma0} as the leading order expansion of the Bose-Einstein distribution function in 
$\Delta\beta = \beta(x)-\beta(X)$ and $\delta \zeta = \zeta(x)-\zeta(X)$ with $X = \bar y_k(x)$. In formula:
$$
 n_B(\beta(X))= n_B\left(\beta(x)-\Delta \beta\right) \simeq n_B(\beta(x)) + n_B(\beta(x))(1+n_B(\beta(x)) \Delta \beta\;.
$$
In the free field case, the zeroth order correction \eqref{zerorder} is thus justified by the obvious fact
that one should eventually reproduce, if all orders of the expansion in $\Delta \beta$ of the density operator
\eqref{densop2} were worked out, the simple free-streaming solution. 

Imagining to turn on the interaction coupling constants adiabatically, it is therefore reasonable to expect 
the zeroth order term to survive in the interacting field case and not to vanish abruptly. The somewhat 
surprising result is that, for the \eqref{zerorder}, the only difference with respect to the free case is that the 
particle mass is distributed according to the spectral function at finite temperature. We can thus attempt an 
interpretation of this term as the contribution of a free stream of virtual particles from the initial to the final 
hypersurface.

\subsection{Statistical quantum field theory vs classical relativistic kinetic theory}

An important question, which is related to the above discussion on the nature of the zeroth order term, is whether the 
expansion of the Wigner function in the initial gradients can be found in a similar form in relativistic kinetic theory 
as well. As has been discussed in Section \ref{density}, in principle one can obtain an expansion of the same function 
$W^+(x,k)$ in the final gradients by using the Taylor expansion of the gradients, for instance:
$$
\partial^{(k)} \beta (x)= \partial^{(k)} \beta(x_0) + \partial^{(k+1)} \beta(x_0) (x-x_0) + \ldots\;,
$$
and the issue is which expansion (in the initial or final gradients) provides the better approximation. It is worth noting 
that in the above Taylor expansion a long distance is introduced at each term (i.e. $(x-y)^k$, hence only the resummation of 
many terms may lead to a decent approximating formula.

In classical relativistic kinetic theory, it is well known that the distribution function $f(x,k)$ expanded about local 
equilibrium receives corrections proportional to the gradients of the thermo-hydrodynamic fields at the same point $x$
\cite{DeGroot:1980dk} under the assumption of separation of time scales (mean collision time $\ll$ hydrodynamic time scale) 
and factorization of the two-particle distribution in the collisional integral, i.e. molecular chaos hypothesis or correlation 
memory loss \cite{cercignani2002}:
$$
  \Delta f(x,k) \propto \partial (\beta, \zeta)\;.
$$
If either assumption is relaxed, one has memory effects and a dependence of $f(x,k)$ on the history of the system 
\cite{Danielewicz:1982kk,KODAMAKOIDE,KOIDE}, hence on the initial conditions. The obvious limiting example is the 
collision-less Boltzmann equation:
$$
  k^\mu \partial_\mu f(x,k) = 0\;,
$$
where the mean collision time is infinite, the solution is analogous to the equation \eqref{fssol} for the Wigner function, 
the memory of the initial distribution is fully retained, and an expansion in the gradients of $\beta(x)$, even if possible 
in principle if the particles leave in a fluid medium, does not provide a good approximation of $f(x,p)$. 
In modern formulations of relativistic kinetic theory \cite{DNMR} memory effects are limited to a finite relaxation time
but they do not involve convolution integrals in time (see also ref. \cite{Abbasi:2025teu}).

We can learn something more about the difference between the quantum statistical approach and the classical relativistic kinetic
wisdom by studying the zeroth order term \eqref{zerorder} in the special case where the hypersurfaces $\Sigma_0$ and $\Sigma_D$,
passing through $x$, are not too far from each other. In this case we can expand the gradients of the thermo-hydrodynamic fields
and retain only the leading order term. Notably, taking into account the equations \eqref{worldline} and \eqref{worldline2}:
\begin{align*}
 & \beta^\nu(x) - \beta^\nu(\bar y_k(x)) \simeq \partial_\lambda \beta^\nu(x) \frac{k^\lambda}{k^0} \Delta x^0(x,k)\;,
 \qquad \zeta(x) - \zeta(\bar y_k(x)) \simeq \partial_\lambda \zeta(x) \frac{k^\lambda}{k^0} \Delta x^0(x,k)\;, \\
 & \beta^\nu(x) - \beta^\nu(\bar y_\beta(x)) \simeq \partial_\lambda \beta^\nu(x) \frac{\beta^\lambda(x)}{\beta^0(x)} 
 \Delta x^0(x,\beta(x))\;,
 \qquad \zeta(x) - \zeta(\bar y_\beta(x)) \simeq \partial_\lambda \zeta(x) \frac{\beta^\lambda(x)}{\beta^0(x)} \Delta x^0(x,\beta(x))\;,\\
\end{align*}
where $\Delta x^0$ is the time difference between the point $x$ and the intersection between the world-lines \eqref{worldline} 
and \eqref{worldline2} starting from $x$ and the hypersurface $\Sigma_0$. The formula \eqref{zerorder} will come down to:
\begin{align}\label{leadorder}
\Delta^{(0)}W\left(x,k\right) &\simeq \frac{2\varrho(k)n_{\rm B}(k)}{(2\pi)^4} \Bigg\{ \theta_k(x) \Big[
\left(1+n_{\rm B}(k)\right)\Big( k^\lambda k^\nu \partial_\lambda \beta_\nu (x) \Big)
-\frac{\partial\log\varrho(k)}{\partial(k \cdot \beta)}\Big( k^\lambda k^\nu \partial_\lambda \beta_\nu (x) \Big)
\Big]  \frac{\Delta x^0(x,k)}{k^0} \nonumber \\
&\qquad\qquad\qquad\quad-2 \theta_\beta(x) \,\Big( \frac{\partial\log\varrho(k)}{\partial\beta^2}
\beta^\lambda(x)\beta^\nu(x)\partial_\lambda \beta_\nu(x) \Big) \frac{\Delta x^0(x,\beta(x))}{\beta^0} \Bigg\}\\
&-\frac{2\varrho^2(k)}{(2\pi)^2}\Bigg\{\theta_k(x)\,\Upsilon^k_1\left(k,0,\beta\right)\,k^\lambda
\partial_\lambda\zeta(x)+\theta_\beta(x)\,
\Upsilon^\beta_1\left(k,0,\beta\right)\,\beta^\lambda(x)\partial_\lambda\zeta(x)\Bigg\}\;,\nonumber
\end{align}
showing that the leading order correction is now proportional to the first order gradients of the fields in the point $x$,
like in the classical relativistic kinetic theory. Indeed, if we replace the ratio $\Delta x^0/k^0$ with a small relaxation
time $\tau_R$, we essentially retrieve a classical kinetic expression of the non-equilibrium correction to the distribution
function in the relaxation time approximation \cite{Romatschke:2017ejr}:
\be\label{leadorderkin}
 \Delta^{(0)}W\left(x,k\right) \propto  k^\lambda k^\nu \partial_\lambda \beta_\nu (x) \tau_R + \ldots\;.
\ee
This simple exercise shows that in the quantum statistical framework, the classical expressions are recovered provided that there 
is a {\em microscopically} small time distance between the current time hypersurface and the initial hypersurface where local 
equilibrium is previously achieved. On a macroscopic time scale, an expression such as \eqref{leadorderkin} applies if not just 
the fields evolve according to Heisenberg rules \eqref{heisenberg}, but also the quantum state varies on a relaxation time basis. 
In formula, if the quantum state collapses
\be\label{collapse}
 \wrho \longrightarrow \wrho_{\rm LE}\;,
\ee
every relaxation time step. This kind of decoherence in equation \eqref{collapse} makes entropy increase objective and 
not just an effect of restricting information to relevant observables (energy-momentum and charge density). The entropy 
$$
  S = -\Tr (\wrho \log \wrho)\;,
$$
unlike in the Heisenberg picture where $\wrho$ is fixed, does vary because of \eqref{collapse}. Similarly, the Boltzmann equation 
with molecular chaos hypothesis (i.e. factorization of the two-particle distribution function in the collisional integral) involves an objective increase of entropy through the $H$-theorem.

In conclusion, it should not be surprising that rigorous quantum statistical methods where system evolves according to the Heisenberg 
equation and do not include additional assumptions somehow equivalent to the equation \eqref{collapse}, provide off-equilibrium corrections
involving a memory of the initial state. Our method of calculating the Wigner function is completely equivalent to find a parametric 
solution of the differential equations of quantum kinetic theory with assigned initial conditions at the Cauchy hypersurface $\Sigma_0$
(notably, the so-called Kadanoff-Baym equations \cite{Berges:2004yj}). Indeed, the quantum kinetic equations in their original form 
are non-Markovian integro-differential equations whose solution must depend on the history of the system. 

\subsection{Global equilibrium and convergence to global equilibrium}

Another important point to address is the reduction to global equilibrium of the expressions found. If the density operator at $\Sigma_0$ 
is a global equilibrium one, the four-temperature $\beta$ is a Killing vector and the reduced chemical potential $\zeta$ is constant, i.e.:
\be\label{killing}
   \beta_\mu = b_\mu + \varpi_{\mu\nu} x^\nu\;, \qquad\qquad \zeta = {\rm const}\;,
\ee
with $b$ and $\varpi$, i.e. the thermal vorticity, constant. If $\varpi \ne 0$, the correction to the leading order expression 
\eqref{leading order W} of the Wigner function can be non-vanishing, as demonstrated in an exact calculation in ref. \cite{Becattini:2020qol}.
However, the leading order term \eqref{zerorder} vanishes at global equilibrium; plugging the equations \eqref{killing} into the 
correction term in the \eqref{zerorder} we have $\Delta \zeta = 0$ and:
\begin{equation*}
    \begin{split}
         k \cdot (\beta(x) - \beta(\bar y_k(x)) &= k^\mu \varpi_{\mu\nu} (x^\nu - \bar y_k(x)^\nu)=
   k^\mu \varpi_{\mu\nu} k^\nu \Delta \tau = 0\;,\\
   \beta\cdot\left(\beta(x)-\beta(\overline{y}_\beta(x))\right)&=\beta^\mu\varpi_{\mu\nu}\left(x^\nu-\overline{y}_\beta(x)^\nu\right)=\beta^\mu\varpi_{\mu\nu}\beta^\nu\Delta\tau=0\;,
    \end{split}
\end{equation*}
where we took advantage of the fact that $x$ and $\bar y$ are two events lying on the worldline whose tangent
vector is proportional to $k$ and $\beta$ respectively. 

Another crucial problem is whether, starting from a non-equilibrium density operator such as \eqref{densop}, the expected
form of the Wigner function at global equilibrium \eqref{leading order W} is achieved asymptotically in the limit 
$t \to +\infty$ because of dissipation. More specifically, if the system is confined within a finite region and if it 
evolves according to the laws of dissipative hydrodynamics, we expect the thermo-hydrodynamic fields to converge to a global 
equilibrium configuration where $\partial \beta,\, \partial \zeta \to 0$ (provided that the angular momentum vanishes, so thermal 
vorticity vanishes at equilibrium). In this case, one expects that in the same limit the off-equilibrium correction 
$\Delta W^+(x,k)$ vanishes thereby losing the memory of the initial state. 

\begin{figure}[h!]
    \centering
    \includegraphics[width=0.55\linewidth]{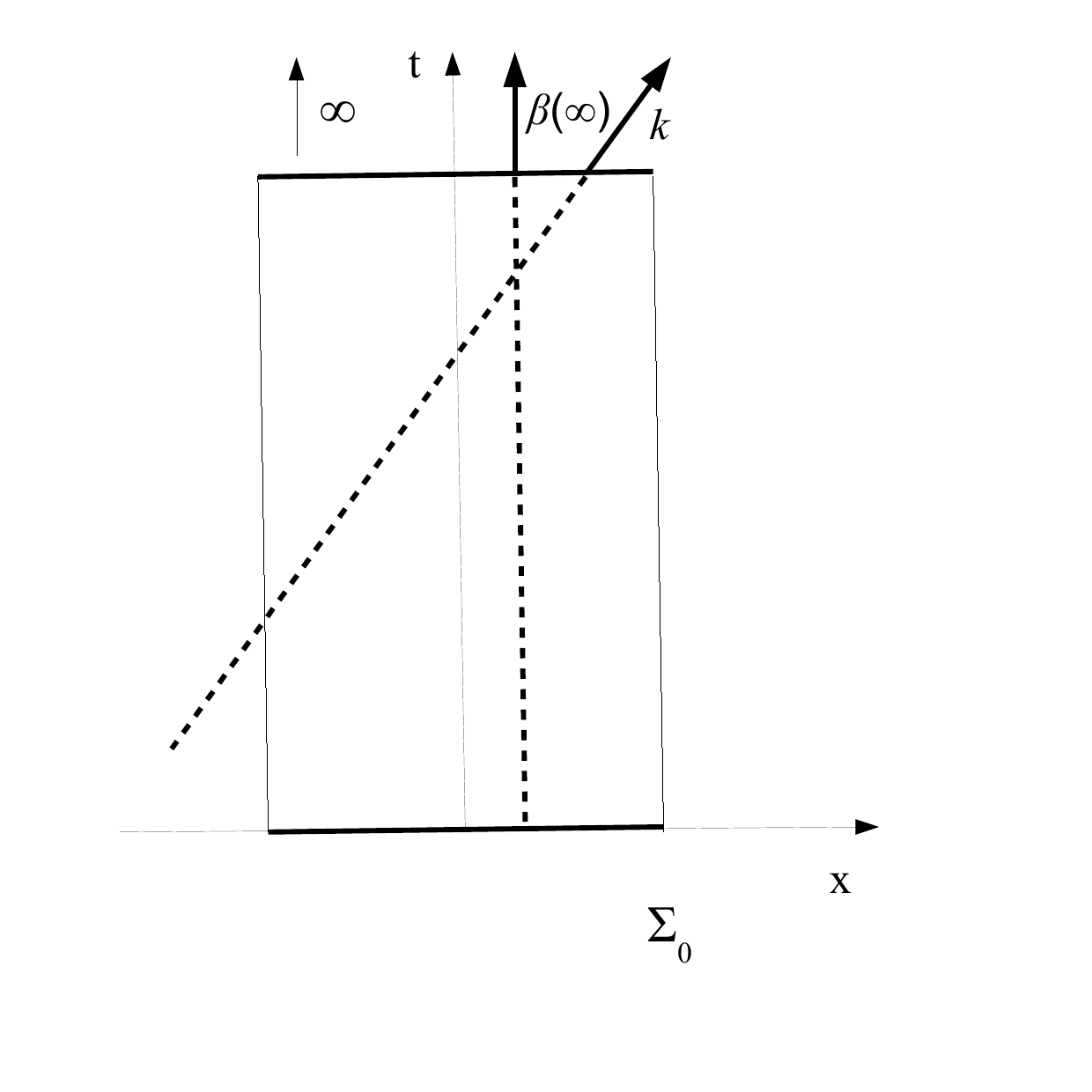}
    \caption{For a system with fixed volume (e.g. a gas within a vessel) in the limit $t \to +\infty$ all
    worldlines with a slope proportional to $k$ eventually have no intersection with the hypersurface $\Sigma_0$
    at $t=0$. Conversely, all worldlines whose tangent four-vector is $\lim_{t \to \infty} \beta(x) = \frac{1}{T_0} (1,{\bf 0})$
    worldlines intersect $\Sigma_0$.}
    \label{fig:longtime}
\end{figure}

According to the equation \eqref{deltawig-special}, \eqref{corrfunc}, as $\Delta \beta$ remains finite, this is the case 
if the correlation function vanishes for large $(x-y)^2$, which is not generally the case though, as we have discussed in this 
Section. Nevertheless, if $\Sigma_0$ a compact region, those terms of the correlation function \eqref{corrfunc} which are constant 
over the worldlines $(x-y)^\mu=k^\mu \tau$ and giving rise to the correction \eqref{deltawigk} do not survive in the limit 
$t \to +\infty$ because they not intersect $\Sigma_0$ except for $k^\mu \propto (1,{\bf 0})$ (see figure \ref{fig:longtime}). 
In fact, this does not necessarily apply to the worldlines $(x-y)^\mu = \beta^\mu \tau$ if the field $\beta^\mu$ has no spatial 
component in the limit $t \to +\infty$ and in this case the contribution \eqref{deltawigbeta} survives, particularly the 
terms proportional to $\theta_\beta(x)$ in the equation \eqref{zerorder}. It remains an open question whether this term
cancels out with other contributions ${\cal O}(\partial^2)$ to the Wigner function, that is the terms \eqref{deltawigjs3} 
as well as corrections beyond linear order response. 

\subsection{Dissipative vs non-dissipative}

Finally, another interesting question is whether the corrections found to the Wigner function are of dissipative or 
non-dissipative nature, according to the discussion in Section \ref{density}. As it was pointed out, the 
off-equilibrium correction can be split into a local equilibrium and a dissipative correction, see equation 
\eqref{gauss2}:
$$
 \Delta O(x) = \Delta O(x)_{\rm LE} + \Delta O(x)_{\rm diss}\;.
$$
Since the main term of the Wigner function is the one calculated at global equilibrium with four-temperature $\beta(x)$
and reduced chemical potential $\zeta(x)$, this question is very relevant because it is known that there are non-dissipative 
terms beyond the global equilibrium main term which vanish at global equilibrium, such as the shear-induced polarization 
\cite{Becattini:2021suc,Liu:2021uhn}. A relevant question is whether the leading order correction in $\Delta\beta$ in
equation \eqref{zerorder} is either a dissipative or a local equilibrium correction or both.

A point can be made that \eqref{zerorder} is essentially dissipative. Suppose we are to calculate the local equilibrium 
correction to the global equilibrium value of the Wigner operator, 
just like in eq. \eqref{lrtO}. We could repeat the same steps of the derivation presented in this work and obtain
the desired correction just by replacing $\Sigma_0$ with $\Sigma_{\rm D}$ in all expressions. This kind of 
calculation was indeed carried out in ref. \cite{Sheng:2025cjk} with the free field, but in fact the expression 
\eqref{deltawig3} with $\Sigma_D$ replacing $\Sigma_0$ would be an improvement of that calculation including the 
effect of interactions. For the leading order correction \eqref{zerorder}, we would then obtain the corresponding local 
equilibrium correction by replacing the intersection between the lines \eqref{worldline} and \eqref{worldline2} with the 
equilibrium hypersurface $\Sigma_0$ with those with the decoupling hypersurface $\Sigma_D$. If $x$ was the only one, then 
the correction would simply vanish because $\Delta \beta_\nu (x,x) = \Delta \zeta(x,x)=0$, whereas in case of multiple 
intersections - which 
requires $\Sigma_{\rm D}$ to have a time-like branches - we would get some non-vanishing contribution proportional to 
the difference between the values of the thermo-hydrodynamic fields at $x$ and at the other intersections. By no
means we would find, though, an expression involving the value of $\beta$ and $\zeta$ over the hypersurface $\Sigma_0$,
which points to a non-local equilibrium origin of the \eqref{zerorder}. A full calculation of the local equilibrium
and the dissipative contributions to the off-equilibrium correction of the Wigner function confirms that this is the 
case \cite{roselliphd}.

\section{The Wigner function and the momentum spectrum}
\label{sec:spectrum}

We finally come to the main phenomenological consequence of the off-equilibrium correction to the Wigner function. 
By using the Wigner function expression up to second order gradients in eqs. \eqref{leading order W} and and \eqref{zerorder} 
in the equation \eqref{spectrum}, we obtain an expression of the momentum spectrum of particles at the decoupling hypersurface, 
i.e. before collisional corrections:
\begin{equation}\label{spectrumfinal}
\begin{split}
\frac{\di N_k}{\di^3 \kk}(\kb) = & \frac{\di N_k}{\di^3 \kk}\Big|_0(\kb) + \frac{2}{(2\pi)^4}\int_0^{+\infty} 
\!\!\! \di k^0 \int_{\Sigma_D}\di\Sigma(x) \cdot k \;\varrho(k)\,\Big\{ \Big[ \theta_k(x)  n_{\rm B}(k) \left(1+n_{\rm B}(k)\right)\left( k\cdot\beta(x)-k\cdot\beta(\overline{y}_k(x))\right)
\\
&-\theta_k(x)\, n_{\rm B}(k) \frac{\partial\log\varrho(k)}
{\partial(k\cdot\beta)}\left(k\cdot\beta(x)-k\cdot\beta\left(\overline{y}_k(x)\right)\right)-2\theta_\beta(x)\frac{\partial\log\varrho(k)}{\partial\beta^2}\left(\beta^2(x)-\beta(x)\cdot 
\beta(\overline{y}_\beta(x))\right)\Big] \\
&-(2\pi)^2\varrho(k)\Big[\theta_k(x)\,\Upsilon^k_1\left(k,0,\beta\right)\left(\zeta(x)-\zeta\left(\bar{y}_k(x)\right)\right)+\theta_\beta(x)\Upsilon^\beta_1\left(k,0,\beta\right)\left(\zeta(x)-\zeta\left(\bar{y}_\beta(x)\right)\right)\Big]\Bigg\}\;,
\end{split}
\end{equation}
where:
\begin{equation*}
\frac{\di N_k}{\di^3 \kk}\Big|_0(\kb) = \frac{2}{(2\pi)^4}\int_0^{+\infty} \!\!\! \di k^0 \int_{\Sigma_D}
\di\Sigma(x) \cdot k \;\varrho(k)\, n_{\rm B}(k)\;,
\end{equation*}
and $\Upsilon^k_1\left(k,0,\beta\right)$ and $\Upsilon^\beta_1\left(k,0,\beta\right)$ given in \eqref{Upsilonkbeta q0}.
The relative weight of the correction in eq. \eqref{spectrumfinal} is thus given by:
\be\label{spectrumratio}
 R(\kb) = \frac{\frac{\di N_k}{\di^3 \kk}-\frac{\di N_k}{\di^3 \kk}|_0}{\frac{\di N_k}{\di^3 \kk}|_0}\;.
\ee
It should be reminded that the correction \eqref{zerorder} is just the leading order one in the expansion of the
density operator, hence it is a good approximation whenever:
$$
  k \cdot \Delta \beta \ll 1\;,  \qquad \qquad \Delta \zeta \ll 1\;,
$$
that is for small differences between initial and final four-temperature and reduced chemical potential.
In practice, if $R(\kb)$ defined above is not much smaller than 1, higher orders (quadratic response and beyond) 
should be considered. 

The spectrum \eqref{spectrumfinal} results from the convolution, in the variable $k^0$, of the spectral function 
calculated at the decoupling temperature with the familiar Bose-Einstein distribution and thermo-hydrodynamic fields. 
Indeed, it can be seen as the spectrum of particles with momentum $\kb$ and a mass distributed according to the spectral 
function. The space integration is carried out over a 3D hypersurface with the functions $\theta_k(x)$ and 
$\theta_\beta(x)$ which may cut off high momenta (see discussion below), reducing the ratio $R(\kb)$ possibly 
extending the range of applicability of the linear approximation. 

In the limit of a quasi-free spectral function \eqref{freespectral} the contributions from 
$\partial\log\varrho/\partial(k\cdot\beta)$, $\partial\log\varrho/\partial\beta^2$, $\partial\log\varrho/\partial\zeta$ and $\Lambda(k,\beta)$ in \eqref{Upsilonkbeta q0} are vanishing, and \eqref{spectrumfinal} simplifies to:
\begin{equation}\label{spectrumfinal2}
    \varepsilon \frac{\di N_k}{\di^3 \kk} = \frac{1}{(2\pi)^3} \int_{\Sigma_D}\di\Sigma(x)\cdot k \; n_B (k) 
   \Big[ 1+\theta_k(x) \left(1+n_B(k)\right)\left( k\cdot\beta(x)-k\cdot\beta(\overline{y}_k(x))
   -\zeta(x)+\zeta(\overline{y}_k(x))\right) \Big]\;,
\end{equation}
whose leading term is precisely the Cooper-Frye formula \eqref{cfrye}.

\begin{figure}[h!]
    \centering
    \includegraphics[width=0.5\linewidth]{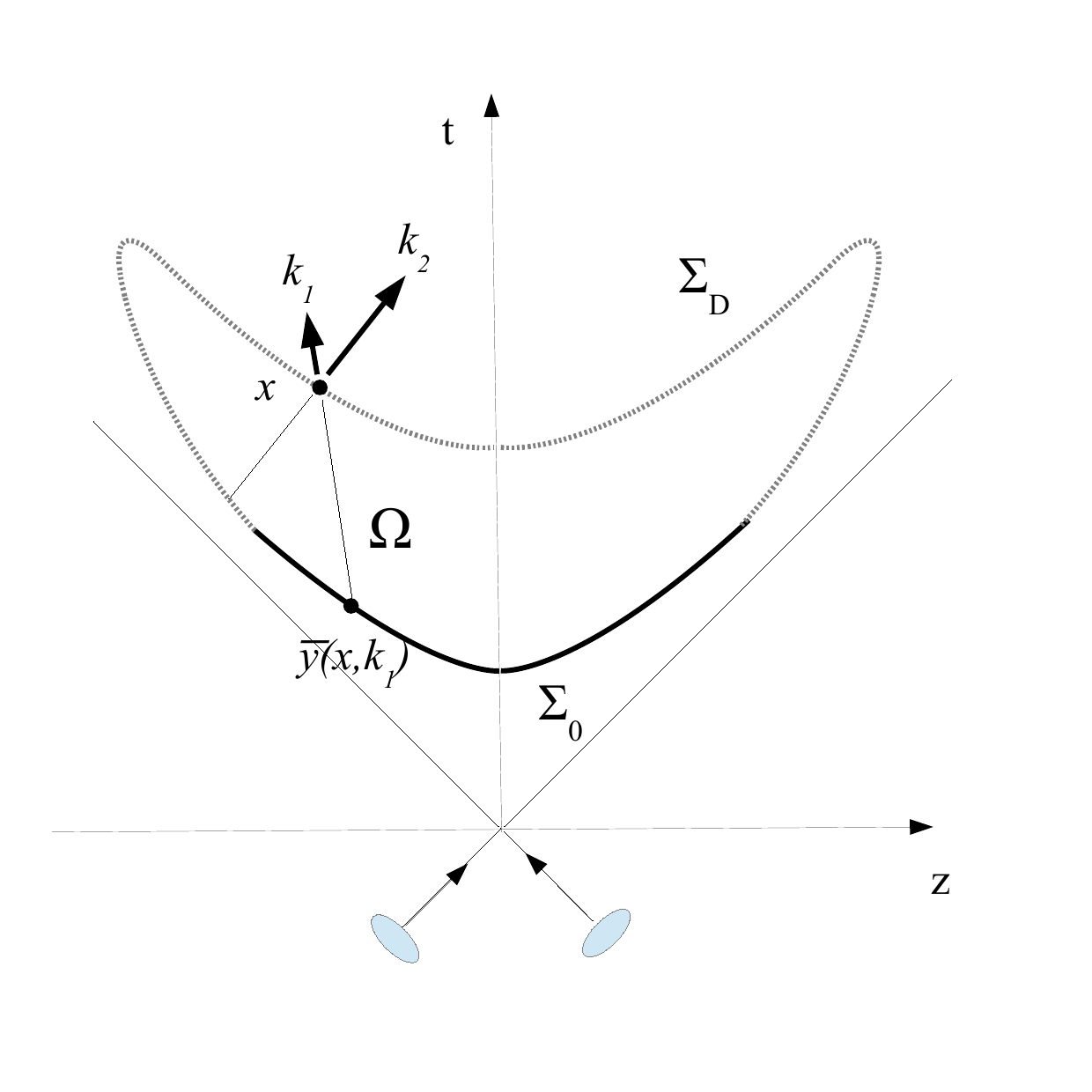}
    \caption{In a relativistic nuclear collision (see also figure \ref{fig:Gauss Theorem Region}) 
    particles with (off-shell) four-momentum $k_2$ do not receive zero-order correction from the Wigner function
    $W^+(x,k)$ at $x$ over the decoupling hypersurface because the world-line does not intersect the initial 
    hypersurface $\Sigma_0$; the converse is true for the momentum $k_1$.}
    \label{qgpexpa}
\end{figure}

An important feature of the formulae \eqref{spectrumfinal} or its simplified version \eqref{spectrumfinal2} is the function
$\theta_k(x)$ which cuts off momenta whose corresponding worldline does not intersect the hypersurface $\Sigma_0$; this is
shown in the figure \ref{qgpexpa} for the typical longitudinal projection of the initial hypersurface and the decoupling
hypersurface. To quantify the effect of this cutoff on the momentum spectrum we can use a simple argument, which applies at 
vanishing momentum component along the beam line, that is $p_z =0$, and in the transverse projection of the expansion. 
Setting $\zeta=0$ for simplicity, the region weighing the most for the integration in both the numerator and denominator of 
the equation \eqref{spectrumratio}, is the one where $\beta \cdot k$ is the smallest because of the exponential factor 
$\exp[-\beta \cdot k]$. Therefore, taking into account that the transverse component of the initial flow velocity ${\bf v}_{Ti}$ 
vanishes, for $p_z=0$ one has, for the correction in the numerator of \eqref{spectrumratio}: 
$$
   k \cdot \Delta \beta = k \cdot \beta_f - k \cdot \beta_i = \varepsilon \left(\frac{\gamma_f}{T_f} 
 - \frac{\gamma_i}{T_i} \right) - \frac{{\bf p}_T\cdot \gamma_f {\bf v}_{Tf}}{T_f} \;,
$$
and, according to the above argument, one can take ${\bf p}_T$ collinear to ${\bf v}_T$, so as to get:
\be\label{kdeltab}
 k \cdot \Delta \beta \longrightarrow \varepsilon \left(\frac{\gamma_f}{T_f}  - \frac{\gamma_i}{T_i} \right) - 
 \frac{p_T \gamma_f v_{Tf}}{T_f}\;.
\ee
For $p_T = 0$ this quantity is positive:
\begin{equation*}
 k \cdot \Delta \beta =  m \left(\frac{\gamma_f}{T_f}  - \frac{\gamma_i}{T_i} \right) > 0\;,
\end{equation*}
because, in general we have $\gamma_f \ge \gamma_i$ and $T_i > T_f$. At $p_T = 0$ the derivative of the function 
on the right hand side of eq. \eqref{kdeltab} is negative, so that the ratio $R$ presumably decreases until $p_T$
reaches a critical value, which may well be beyond the natural geometric cut-off. Therefore, in relativistic heavy ion 
collision, an enhancement of the transverse momentum spectrum at low transverse momenta is expected due to the correction 
term in eq. \eqref{spectrumfinal}. Indeed, an excess of pions at low $p_T$ \cite{ALICE:2013mez,ALICE:2019hno} at very high 
energy is a long-standing phenomenological issue which has been addressed in several papers in literature \cite{Lu:2024shm,McNelis:2021acu,Begun:2015ifa}; it is still premature to say that this correction can account for this phenomenon, 
however this is an effect going into the right direction.

\section{Summary and conclusions}
\label{summary}

In summary, we have derived the dissipative corrections, up to linear order in the gradients of the thermo-hydrodynamic fields, 
to the Wigner function and to the single-particle momentum spectrum of scalar particles emitted from an expanding decoupling 
fluid that is initially in local thermodynamic equilibrium .

We have performed an \emph{ab initio} calculation of the Wigner function within the framework of statistical quantum field theory 
by employing the appropriate density operator and a novel approximation scheme introduced in our recent work~\cite{Sheng:2025cjk}. 
The calculation has been carried out for a generally interacting scalar quantum field. Moreover, no specific assumptions have 
been made regarding the microscopic structure of the stress-energy tensor and four-current operators which include the
contribution of all the remaining interacting fields. Retaining the full generality led us to parametrize correlators of
the field Fourier transforms with thermal-gravitational and thermal-charged form factors, for which the only assumption made has 
been the requirement of analyticity in a four-momentum variable. 
We have shown that the leading-order expansion of the non-equilibrium contribution to the density operator within linear response 
theory naturally gives rise to a series involving gradients of the hydrodynamic fields evaluated on the initial local-equilibrium 
hypersurface, rather than on the final decoupling hypersurface, as it is customary in classical kinetic theory. The emergence 
of gradients at the initial hypersurface is a direct consequence of the long-distance persistence of the correlation function 
between the Wigner operator and the stress-energy tensor and current operators, that we have discussed in detail, entailing a
memory of the initial state.

The leading contribution in the resulting expansion is a zeroth-order correction proportional to the difference between the hydrodynamic 
fields evaluated at the decoupling point $x$ and at the intersection of the initial hypersurface with the worldline passing through 
$x$ and having a tangent four-vector proportional to the four-momentum argument of the Wigner function. On the other hand, the 
first-order gradient contribution identically vanishes. The zeroth-order term provides a clear manifestation of the memory 
of the initial state and, at least for the main contribution, has a clear counterpart in the free field limit, where its interpretation is straightforward in the free-streaming solution of the Wigner function. Its survival in the interacting case (weighted by the spectral 
function and that can be interpreted as a term related to the free propagation of virtual particles) should not be surprising in
the limit of a weakly interacting theory; in a strongly interacting theory its relative importance for the off-equilibrium correction 
to the Wigner function depends on how large the coefficients of all higher-order gradients as well as all the contributions of the 
terms beyond linear response.

The zeroth order term reduces to a first-order gradient correction of the familiar form encountered in relativistic kinetic theory only 
when the initial and decoupling hypersurfaces are microscopically close. More precisely, it is necessary that their separation is of the 
order of the classical relaxation time within a kinetic description. This observation suggests that, within a fully quantum-mechanical 
statistical framework, the suppression of initial-state memory effects needs an internal process of quantum decoherence, 
whereby the density operator undergoes a continuous reduction toward a local-equilibrium form and the time evolution becomes effectively 
non-unitary. However, such a mechanism is not expected to take place in an isolated system. Our results therefore represent the proper 
quantum-mechanical prediction for the physical situation under consideration.

Finally, the resulting momentum spectrum acquires corrections proportional to integrals over the decoupling hypersurface of the 
aforementioned field differences. These corrections induce a distortion of the spectrum, which is expected to be most pronounced at 
low momenta. A quantitative assessment of the magnitude of these effects, as well as an evaluation of the validity of the 
linear-gradient approximation in such systems, calls for a dedicated numerical investigation. We finally note that the derived 
expressions hold at finite chemical potential and do not rely on any approximation for the geometry of the decoupling 
hypersurface. Consequently, the formalism applies to collisions over a broad range of energies.

\section*{Acknowledgements}
We are grateful to M. Buzzegoli and E. Grossi for very useful discussions. D. R. would like to express sincere 
gratitude to Professor Huang Xu-Guang for his warm hospitality during his research work, as well 
as for valuable suggestions and insightful discussions. This work is supported in part by the Italian 
Ministry of University and Research, project PRIN2022 “Advanced probes of the Quark Gluon Plasma”, Next 
Generation EU, Mission 4 Component 1.


\bibliographystyle{apsrev4-1}
\bibliography{main}

@article{Sheng:2025cjk,
    author = "Sheng, Xin-Li and Becattini, Francesco and Roselli, Daniele",
    title = "{An improved formula for Wigner function and spin polarization in a decoupling relativistic fluid at local thermodynamic equilibrium}",
    eprint = "2509.14301",
    archivePrefix = "arXiv",
    primaryClass = "nucl-th",
    month = "9",
    year = "2025"
}

@article{Polyakov:2018zvc,
    author = "Polyakov, Maxim V. and Schweitzer, Peter",
    title = "{Forces inside hadrons: pressure, surface tension, mechanical radius, and all that}",
    eprint = "1805.06596",
    archivePrefix = "arXiv",
    primaryClass = "hep-ph",
    doi = "10.1142/S0217751X18300259",
    journal = "Int. J. Mod. Phys. A",
    volume = "33",
    number = "26",
    pages = "1830025",
    year = "2018"
}

@article{Zub2,
    author = "Zubarev, D. N. and Prozorkevich, A. V. and Smolyanskii, S. A.",
    title = "{Derivation of nonlinear generalized equations of quantum relativistic hydrodynamics}",
    doi = "10.1007/BF01032069",
    journal = "Theor. Math. Phys.",
    volume = "40",
    number = "3",
    pages = "821--831",
    year = "1979"
}

@article{VANWEERT1982133,
title = {Maximum entropy principle and relativistic hydrodynamics},
journal = {Annals of Physics},
volume = {140},
number = {1},
pages = {133-162},
year = {1982},
issn = {0003-4916},
doi = {https://doi.org/10.1016/0003-4916(82)90338-4},
url = {https://www.sciencedirect.com/science/article/pii/0003491682903384},
author = {Ch.G {van Weert}}
}

@article{Zhang:2024mhs,
    author = "Zhang, Zhong-Hua and Huang, Xu-Guang and Becattini, Francesco and Sheng, Xin-Li",
    title = "{Vector and tensor spin polarization for vector bosons at local equilibrium}",
    eprint = "2412.19416",
    archivePrefix = "arXiv",
    primaryClass = "hep-ph",
    doi = "10.1007/JHEP07(2025)224",
    journal = "JHEP",
    volume = "07",
    pages = "224",
    year = "2025"
}

@article{Donoghue:1991qv,
    author = "Donoghue, John F. and Leutwyler, H.",
    title = "{Energy and momentum in chiral theories}",
    reportNumber = "UMHEP-321, NSF-ITP-91-119I",
    doi = "10.1007/BF01560453",
    journal = "Z. Phys. C",
    volume = "52",
    pages = "343--351",
    year = "1991"
}

@article{Pagels:1966zza,
    author = "Pagels, Heinz",
    title = "{Energy-Momentum Structure Form Factors of Particles}",
    doi = "10.1103/PhysRev.144.1250",
    journal = "Phys. Rev.",
    volume = "144",
    pages = "1250--1260",
    year = "1966"
}

@article{Becattini:2019dxo,
    author = "Becattini, F. and Buzzegoli, M. and Grossi, E.",
    title = "{Reworking the Zubarev's approach to non-equilibrium quantum statistical mechanics}",
    eprint = "1902.01089",
    archivePrefix = "arXiv",
    primaryClass = "cond-mat.stat-mech",
    doi = "10.3390/particles2020014",
    journal = "Particles",
    volume = "2",
    number = "2",
    pages = "197--207",
    year = "2019"
}

@article{Hosoya:1983id,
    author = "Hosoya, Akio and Sakagami, Masa-aki and Takao, Masaru",
    title = "{Nonequilibrium Thermodynamics in Field Theory: Transport Coefficients}",
    reportNumber = "OU-HET-53",
    doi = "10.1016/0003-4916(84)90144-1",
    journal = "Annals Phys.",
    volume = "154",
    pages = "229",
    year = "1984"
}

@book{DeGroot:1980dk,
    author = "De Groot, S. R.",
    editor = "Van Leeuwen, W. A. and Van Weert, C. G.",
    title = "{Relativistic Kinetic Theory. Principles and Applications}",
    year = "1980"
}

@book{Romatschke:2017ejr,
    author = "Romatschke, Paul and Romatschke, Ulrike",
    title = "{Relativistic Fluid Dynamics In and Out of Equilibrium}",
    eprint = "1712.05815",
    archivePrefix = "arXiv",
    primaryClass = "nucl-th",
    doi = "10.1017/9781108651998",
    isbn = "978-1-108-48368-1, 978-1-108-75002-8",
    publisher = "Cambridge University Press",
    series = "Cambridge Monographs on Mathematical Physics",
    month = "5",
    year = "2019"
}

@article{Denicol:2012cn,
    author = "Denicol, G. S. and Niemi, H. and Molnar, E. and Rischke, D. H.",
    title = "{Derivation of transient relativistic fluid dynamics from the Boltzmann equation}",
    eprint = "1202.4551",
    archivePrefix = "arXiv",
    primaryClass = "nucl-th",
    doi = "10.1103/PhysRevD.85.114047",
    journal = "Phys. Rev. D",
    volume = "85",
    pages = "114047",
    year = "2012",
    note = "[Erratum: Phys.Rev.D 91, 039902 (2015)]"
}

@article{Becattini:2021suc,
    author = "Becattini, F. and Buzzegoli, M. and Palermo, A.",
    title = "{Spin-thermal shear coupling in a relativistic fluid}",
    eprint = "2103.10917",
    archivePrefix = "arXiv",
    primaryClass = "nucl-th",
    doi = "10.1016/j.physletb.2021.136519",
    journal = "Phys. Lett. B",
    volume = "820",
    pages = "136519",
    year = "2021"
}

@article{Fu:2021pok,
    author = "Fu, Baochi and Liu, Shuai Y. F. and Pang, Longgang and Song, Huichao and Yin, Yi",
    title = "{Shear-Induced Spin Polarization in Heavy-Ion Collisions}",
    eprint = "2103.10403",
    archivePrefix = "arXiv",
    primaryClass = "hep-ph",
    doi = "10.1103/PhysRevLett.127.142301",
    journal = "Phys. Rev. Lett.",
    volume = "127",
    number = "14",
    pages = "142301",
    year = "2021"
}

@article{Liu:2021uhn,
    author = "Liu, Shuai Y. F. and Yin, Yi",
    title = "{Spin polarization induced by the hydrodynamic gradients}",
    eprint = "2103.09200",
    archivePrefix = "arXiv",
    primaryClass = "hep-ph",
    doi = "10.1007/JHEP07(2021)188",
    journal = "JHEP",
    volume = "07",
    pages = "188",
    year = "2021"
}

@article{Li:2025pef,
    author = "Li, Youyu and Liu, Shuai Y. F.",
    title = "{Zubarev response approach to polarization phenomena in local equilibrium}",
    eprint = "2501.17861",
    archivePrefix = "arXiv",
    primaryClass = "nucl-th",
    month = "1",
    year = "2025"
}

@article{Buzzegoli:2025zud,
    author = "Buzzegoli, Matteo",
    title = "{Kubo formulas for spin polarization in dissipative relativistic spin hydrodynamics: a first-order gradient expansion approach}",
    eprint = "2502.15520",
    archivePrefix = "arXiv",
    primaryClass = "nucl-th",
    month = "2",
    year = "2025"
}

@article{Becattini:2020sww,
    author = "Becattini, Francesco",
    title = "{Polarization in Relativistic Fluids: A Quantum Field Theoretical Derivation}",
    eprint = "2004.04050",
    archivePrefix = "arXiv",
    primaryClass = "hep-th",
    doi = "10.1007/978-3-030-71427-7_2",
    journal = "Lect. Notes Phys.",
    volume = "987",
    pages = "15--52",
    year = "2021"
}

@article{Sheng:2024pbw,
    author = "Sheng, Xin-Li and Becattini, Francesco and Huang, Xu-Guang and Zhang, Zhong-Hua",
    title = "{Spin polarization of fermions at local equilibrium: Second-order gradient expansion}",
    eprint = "2407.12130",
    archivePrefix = "arXiv",
    primaryClass = "hep-th",
    doi = "10.1103/PhysRevC.110.064908",
    journal = "Phys. Rev. C",
    volume = "110",
    number = "6",
    pages = "064908",
    year = "2024"
}

@article{Cooper:1974mv,
    author = "Cooper, Fred and Frye, Graham",
    title = "{Comment on the Single Particle Distribution in the Hydrodynamic and Statistical Thermodynamic Models of Multiparticle Production}",
    reportNumber = "Print-74-0742 (YESHIVA)",
    doi = "10.1103/PhysRevD.10.186",
    journal = "Phys. Rev. D",
    volume = "10",
    pages = "186",
    year = "1974"
}

@article{Harutyunyan:2021rmb,
    author = "Harutyunyan, Arus and Sedrakian, Armen and Rischke, Dirk H.",
    title = "{Relativistic second-order dissipative hydrodynamics from Zubarev{\textquoteright}s non-equilibrium statistical operator}",
    eprint = "2110.04595",
    archivePrefix = "arXiv",
    primaryClass = "nucl-th",
    doi = "10.1016/j.aop.2022.168755",
    journal = "Annals Phys.",
    volume = "438",
    pages = "168755",
    year = "2022"
}

@article{Bhadury:2020puc,
    author = "Bhadury, Samapan and Florkowski, Wojciech and Jaiswal, Amaresh and Kumar, Avdhesh and Ryblewski, Radoslaw",
    title = "{Relativistic dissipative spin dynamics in the relaxation time approximation}",
    eprint = "2002.03937",
    archivePrefix = "arXiv",
    primaryClass = "hep-ph",
    doi = "10.1016/j.physletb.2021.136096",
    journal = "Phys. Lett. B",
    volume = "814",
    pages = "136096",
    year = "2021"
}

@article{Yang:2020hri,
    author = "Yang, Di-Lun and Hattori, Koichi and Hidaka, Yoshimasa",
    title = "{Effective quantum kinetic theory for spin transport of fermions with collsional effects}",
    eprint = "2002.02612",
    archivePrefix = "arXiv",
    primaryClass = "hep-ph",
    reportNumber = "RIKEN-QHP-441, YITP-20-17, RIKEN-iTHEMS-Report-20",
    doi = "10.1007/JHEP07(2020)070",
    journal = "JHEP",
    volume = "07",
    pages = "070",
    year = "2020"
}

@article{Weickgenannt:2022qvh,
    author = "Weickgenannt, Nora and Wagner, David and Speranza, Enrico and Rischke, Dirk H.",
    title = "{Relativistic dissipative spin hydrodynamics from kinetic theory with a nonlocal collision term}",
    eprint = "2208.01955",
    archivePrefix = "arXiv",
    primaryClass = "nucl-th",
    doi = "10.1103/PhysRevD.106.L091901",
    journal = "Phys. Rev. D",
    volume = "106",
    number = "9",
    pages = "L091901",
    year = "2022"
}

@article{Wang:2025mfz,
    author = "Wang, Jia-Rong and Fang, Shuo and Yang, Di-Lun and Pu, Shi",
    title = "{Is the shear induced spin polarization non-dissipative?}",
    eprint = "2507.15238",
    archivePrefix = "arXiv",
    primaryClass = "hep-ph",
    month = "7",
    year = "2025"
}

@article{Lu:2024shm,
    author = "Lu, Pengzhong and Kavak, Rafet and Dubla, Andrea and Masciocchi, Silvia and Selyuzhenkov, Ilya",
    title = "{Quantification of the low-$p_\textrm{T}$ pion excess in heavy-ion collisions at the LHC and top RHIC energy}",
    eprint = "2407.09207",
    archivePrefix = "arXiv",
    primaryClass = "hep-ph",
    doi = "10.1007/s41365-025-01718-z",
    journal = "Nucl. Sci. Tech.",
    volume = "36",
    number = "8",
    pages = "142",
    year = "2025"
}

@article{McNelis:2021acu,
    author = "McNelis, M. and Heinz, U.",
    title = "{Modified equilibrium distributions for Cooper--Frye particlization}",
    eprint = "2103.03401",
    archivePrefix = "arXiv",
    primaryClass = "nucl-th",
    doi = "10.1103/PhysRevC.103.064903",
    journal = "Phys. Rev. C",
    volume = "103",
    number = "6",
    pages = "064903",
    year = "2021"
}

@article{ALICE:2019hno,
    author = "Acharya, Shreyasi and others",
    collaboration = "ALICE",
    title = "{Production of charged pions, kaons, and (anti-)protons in Pb-Pb and inelastic $pp$ collisions at $\sqrt {s_{NN}}$ = 5.02 TeV}",
    eprint = "1910.07678",
    archivePrefix = "arXiv",
    primaryClass = "nucl-ex",
    reportNumber = "CERN-EP-2019-208",
    doi = "10.1103/PhysRevC.101.044907",
    journal = "Phys. Rev. C",
    volume = "101",
    number = "4",
    pages = "044907",
    year = "2020"
}

@article{ALICE:2013mez,
    author = "Abelev, Betty and others",
    collaboration = "ALICE",
    title = "{Centrality dependence of $\pi$, K, p production in Pb-Pb collisions at $\sqrt{s_{NN}}$ = 2.76 TeV}",
    eprint = "1303.0737",
    archivePrefix = "arXiv",
    primaryClass = "hep-ex",
    reportNumber = "CERN-PH-EP-2013-019",
    doi = "10.1103/PhysRevC.88.044910",
    journal = "Phys. Rev. C",
    volume = "88",
    pages = "044910",
    year = "2013"
}

@article{Begun:2015ifa,
    author = "Begun, Viktor and Florkowski, Wojciech",
    title = "{Bose-Einstein condensation of pions in heavy-ion collisions at the CERN Large Hadron Collider (LHC) energies}",
    eprint = "1503.04040",
    archivePrefix = "arXiv",
    primaryClass = "nucl-th",
    doi = "10.1103/PhysRevC.91.054909",
    journal = "Phys. Rev. C",
    volume = "91",
    pages = "054909",
    year = "2015"
}

@article{Jeon:1995zm,
    author = "Jeon, Sangyong and Yaffe, Laurence G.",
    title = "{From quantum field theory to hydrodynamics: Transport coefficients and effective kinetic theory}",
    eprint = "hep-ph/9512263",
    archivePrefix = "arXiv",
    reportNumber = "UW-PT-95-09",
    doi = "10.1103/PhysRevD.53.5799",
    journal = "Phys. Rev. D",
    volume = "53",
    pages = "5799--5809",
    year = "1996"
}

@book{LeBellac_1996, 
 place= {Cambridge},
 series={Cambridge Monographs on Mathematical Physics},
 title={Thermal Field Theory}, publisher={Cambridge University Press}, 
author={Le Bellac, Michel}, year={1996}, collection={Cambridge Monographs on Mathematical Physics}
}

@article{Berges:2004yj,
    author = "Berges, Juergen",
    editor = "Bracco, Mirian and Chiapparini, Marcelo and Ferreira, Erasmo and Kodama, Takeshi",
    title = "{Introduction to nonequilibrium quantum field theory}",
    eprint = "hep-ph/0409233",
    archivePrefix = "arXiv",
    doi = "10.1063/1.1843591",
    journal = "AIP Conf. Proc.",
    volume = "739",
    number = "1",
    pages = "3--62",
    year = "2004"
}

@article{PhysRev.79.972,
  title = {The S-Matrix in the Heisenberg Representation},
  author = {Yang, C. N. and Feldman, David},
  journal = {Phys. Rev.},
  volume = {79},
  issue = {6},
  pages = {972--978},
  numpages = {0},
  year = {1950},
  month = {Sep},
  publisher = {American Physical Society},
  doi = {10.1103/PhysRev.79.972},
  url = {https://link.aps.org/doi/10.1103/PhysRev.79.972}
}

@article{Jeon:1994if,
    author = "Jeon, Sangyong",
    title = "{Hydrodynamic transport coefficients in relativistic scalar field theory}",
    eprint = "hep-ph/9409250",
    archivePrefix = "arXiv",
    reportNumber = "UW-PT-94-09",
    doi = "10.1103/PhysRevD.52.3591",
    journal = "Phys. Rev. D",
    volume = "52",
    pages = "3591--3642",
    year = "1995"
}

@article{Hidaka:2016yjf,
    author = "Hidaka, Yoshimasa and Pu, Shi and Yang, Di-Lun",
    title = "{Relativistic Chiral Kinetic Theory from Quantum Field Theories}",
    eprint = "1612.04630",
    archivePrefix = "arXiv",
    primaryClass = "hep-th",
    doi = "10.1103/PhysRevD.95.091901",
    journal = "Phys. Rev. D",
    volume = "95",
    number = "9",
    pages = "091901",
    year = "2017"
}

@article{Zhang:2019xya,
    author = "Zhang, Jun-jie and Fang, Ren-hong and Wang, Qun and Wang, Xin-Nian",
    title = "{A microscopic description for polarization in particle scatterings}",
    eprint = "1904.09152",
    archivePrefix = "arXiv",
    primaryClass = "nucl-th",
    doi = "10.1103/PhysRevC.100.064904",
    journal = "Phys. Rev. C",
    volume = "100",
    number = "6",
    pages = "064904",
    year = "2019"
}

@article{Kovtun:2012rj,
    author = "Kovtun, Pavel",
    title = "{Lectures on hydrodynamic fluctuations in relativistic theories}",
    eprint = "1205.5040",
    archivePrefix = "arXiv",
    primaryClass = "hep-th",
    doi = "10.1088/1751-8113/45/47/473001",
    journal = "J. Phys. A",
    volume = "45",
    pages = "473001",
    year = "2012"
}

@article{Jaiswal:2014isa,
    author = "Jaiswal, Amaresh and Ryblewski, Radoslaw and Strickland, Michael",
    title = "{Transport coefficients for bulk viscous evolution in the relaxation time approximation}",
    eprint = "1407.7231",
    archivePrefix = "arXiv",
    primaryClass = "hep-ph",
    doi = "10.1103/PhysRevC.90.044908",
    journal = "Phys. Rev. C",
    volume = "90",
    number = "4",
    pages = "044908",
    year = "2014"
}

@book{Kubo1991,
  author       = {Ryogo Kubo and Masao Toda and Norio Hashitsume},
  title        = {Statistical Physics II: Nonequilibrium Statistical Mechanics},
  series       = {Springer Series in Solid-State Sciences},
  volume       = {31},
  publisher    = {Springer},
  edition      = {2},
  year         = {1991},
  doi          = {10.1007/978-3-642-58244-5},
}

@article{Zhang:2025vlk,
    author = "Zhang, Zhong-Hua and Huang, Xu-Guang",
    title = "{Tensor Spin Polarization Induced by Curved Freeze-Out Hypersurface}",
    eprint = "2509.20200",
    archivePrefix = "arXiv",
    primaryClass = "hep-ph",
    month = "9",
    year = "2025"
}

@article{Danielewicz:1982kk,
    author = "Danielewicz, P.",
    title = "{Quantum Theory of Nonequilibrium Processes. 1.}",
    doi = "10.1016/0003-4916(84)90092-7",
    journal = "Annals Phys.",
    volume = "152",
    pages = "239--304",
    year = "1984"
}

@book{cercignani2002,
  author    = {C. Cercignani and G. M. Kremer},
  title     = {The Relativistic Boltzmann Equation: Theory and Applications},
  publisher = {Birkh\"auser},
  year      = {2002}
}

@article{Becattini:2020qol,
    author = "Becattini, F. and Buzzegoli, M. and Palermo, A.",
    title = "{Exact equilibrium distributions in statistical quantum field theory with rotation and acceleration: scalar field}",
    eprint = "2007.08249",
    archivePrefix = "arXiv",
    primaryClass = "hep-th",
    doi = "10.1007/JHEP02(2021)101",
    journal = "JHEP",
    volume = "02",
    pages = "101",
    year = "2021"
}

@phdthesis{RoselliPhD,
  author       = {Roselli, Daniele},
  title        = {Off-Equilibrium Corrections to the Covariant Wigner Function in Relativistic Hydrodynamics},
  school       = {University of Florence},
  year         = {2026}
}

@article{Teaney:2003kp,
    author = "Teaney, Derek",
    title = "{The Effects of viscosity on spectra, elliptic flow, and HBT radii}",
    eprint = "nucl-th/0301099",
    archivePrefix = "arXiv",
    doi = "10.1103/PhysRevC.68.034913",
    journal = "Phys. Rev. C",
    volume = "68",
    pages = "034913",
    year = "2003"
}

@article{Dusling:2007gi,
    author = "Dusling, K. and Teaney, D.",
    title = "{Simulating elliptic flow with viscous hydrodynamics}",
    eprint = "0710.5932",
    archivePrefix = "arXiv",
    primaryClass = "nucl-th",
    doi = "10.1103/PhysRevC.77.034905",
    journal = "Phys. Rev. C",
    volume = "77",
    pages = "034905",
    year = "2008"
}

@article{Denicol:2009am,
    author = "Denicol, G. S. and Kodama, T. and Koide, T. and Mota, Ph.",
    title = "{Effect of bulk viscosity on Elliptic Flow near QCD phase transition}",
    eprint = "0903.3595",
    archivePrefix = "arXiv",
    primaryClass = "hep-ph",
    doi = "10.1103/PhysRevC.80.064901",
    journal = "Phys. Rev. C",
    volume = "80",
    pages = "064901",
    year = "2009"
}

@article{Pratt:2010jt,
    author = "Pratt, Scott and Torrieri, Giorgio",
    title = "{Coupling Relativistic Viscous Hydrodynamics to Boltzmann Descriptions}",
    eprint = "1003.0413",
    archivePrefix = "arXiv",
    primaryClass = "nucl-th",
    doi = "10.1103/PhysRevC.82.044901",
    journal = "Phys. Rev. C",
    volume = "82",
    pages = "044901",
    year = "2010"
}

@article{Luzum:2010ad,
    author = "Luzum, Matthew and Ollitrault, Jean-Yves",
    title = "{Constraining the viscous freeze-out distribution function with data obtained at the BNL Relativistic Heavy Ion Collider (RHIC)}",
    eprint = "1004.2023",
    archivePrefix = "arXiv",
    primaryClass = "nucl-th",
    doi = "10.1103/PhysRevC.82.014906",
    journal = "Phys. Rev. C",
    volume = "82",
    pages = "014906",
    year = "2010"
}

@article{Teaney:2013gca,
    author = "Teaney, Derek and Yan, Li",
    title = "{Second order viscous corrections to the harmonic spectrum in heavy ion collisions}",
    eprint = "1304.3753",
    archivePrefix = "arXiv",
    primaryClass = "nucl-th",
    doi = "10.1103/PhysRevC.89.014901",
    journal = "Phys. Rev. C",
    volume = "89",
    number = "1",
    pages = "014901",
    year = "2014"
}

@article{ShenEtAl2014,
  author = {Shen, Chun and Qiu, Zhi and Song, Huichao and Bernhard, Jonah and Bass, Steffen and Heinz, Ulrich},
  title = {The iEBE-VISHNU code package for relativistic heavy-ion collisions},
  journal = {arXiv:1409.8164},
  year = {2014}
}

@article{McNelis:2019auj,
    author = "McNelis, Mike and Everett, Derek and Heinz, Ulrich",
    title = "{Particlization in fluid dynamical simulations of heavy-ion collisions: The i S3D module}",
    eprint = "1912.08271",
    archivePrefix = "arXiv",
    primaryClass = "nucl-th",
    doi = "10.1016/j.cpc.2020.107604",
    journal = "Comput. Phys. Commun.",
    volume = "258",
    pages = "107604",
    year = "2021"
}

@article{KOIDE,
    author = "Koide, T.",
    title = "{Microscopic derivation of causal diffusion equation using projection operator method}",
    eprint = "cond-mat/0501696",
    archivePrefix = "arXiv",
    doi = "10.1103/PhysRevE.72.026135",
    journal = "Phys. Rev. E",
    volume = "72",
    pages = "026135",
    year = "2005"
}

@article{KODAMAKOIDE,
    author = "Kodama, T. and Koide, T.",
    editor = "Liu, Feng and Xiao, Zhigang and Zhuang, Pengfei",
    title = "{Memory Effects and Transport Coefficients for Non-Newtonian Fluids}",
    eprint = "0812.4138",
    archivePrefix = "arXiv",
    primaryClass = "hep-ph",
    doi = "10.1088/0954-3899/36/6/064063",
    journal = "J. Phys. G",
    volume = "36",
    pages = "064063",
    year = "2009"
}

@article{DNMR,
    author = "Denicol, G. S. and Niemi, H. and Molnar, E. and Rischke, D. H.",
    title = "{Derivation of transient relativistic fluid dynamics from the Boltzmann equation}",
    eprint = "1202.4551",
    archivePrefix = "arXiv",
    primaryClass = "nucl-th",
    doi = "10.1103/PhysRevD.85.114047",
    journal = "Phys. Rev. D",
    volume = "85",
    pages = "114047",
    year = "2012",
    note = "[Erratum: Phys.Rev.D 91, 039902 (2015)]"
}

@article{Abbasi:2025teu,
    author = "Abbasi, Navid and Kaminski, Matthias and Rischke, Dirk H.",
    title = "{Comparison between Causal and Acausal Diffusion: a Schwinger-Keldysh Effective Field Theory Perspective}",
    eprint = "2506.20500",
    archivePrefix = "arXiv",
    primaryClass = "hep-th",
    month = "6",
    year = "2025"
}

\appendix

\section{Decomposition of $a^{\mu}a^{\nu}$}\label{SECTION: aa is not independent}

We show that the tensor $a^{\mu}a^{\nu}$, where $a^{\mu}\equiv\epsilon^{\mu\rho\sigma\tau}k_{\rho}q_{\sigma}\beta_{\tau}$, 
can be expressed in terms of other symmetric tensors built with $k,q$ and $\beta$. To make notation compact, we introduce 
the four-vectors $\bar{q}^{\mu}$and $\bar{\beta}^{\mu}$ such that $k^{\mu},\bar{q}^{\mu},\bar{\beta}^{\mu}$ are perpendicular 
to each other:
as:
\begin{eqnarray}\label{qbetabar}
\bar{q}^{\mu} & = & \left(g^{\mu\nu}-\frac{k^{\mu}k^{\nu}}{k\cdot k}\right)q_{\nu}\nonumber\;, \nonumber \\
\bar{\beta}^{\mu} & = & \left(g^{\mu\nu}-\frac{\bar{q}^{\mu}\bar{q}^{\nu}}{\bar{q}\cdot\bar{q}}\right)
\left(g_{\nu\rho}-\frac{k_{\nu}k_{\rho}}{k\cdot k}\right)\beta^{\rho}\;.
\end{eqnarray}
Thereby $a^{\mu}$ can be written as:
\begin{equation*}
a^{\mu}=\epsilon^{\mu\rho\sigma\tau}k_{\rho}\bar{q}_{\sigma}\bar{\beta}_{\tau}\; ;
\end{equation*}
thus:
\be\label{amu1}
a^{\mu}a^{\nu} = \epsilon^{\mu\rho\sigma\tau}k_{\rho}\bar{q}_{\sigma}\bar{\beta}_{\tau}
\epsilon^{\nu\alpha\lambda\xi}k_{\alpha}\bar{q}_{\lambda}\bar{\beta}_{\xi} = 
\epsilon^{\mu\rho\sigma\tau}k_{\rho}\tilde{q}_{\sigma}\bar{\beta}_{\tau}
\epsilon_{\ \alpha}^{\nu\ \lambda\xi}k^{\alpha}\bar{q}_{\lambda}\bar{\beta}_{\xi}\;.
\ee
With the help of the Schouten identity, we can write:
\begin{equation*}
\epsilon^{\mu\rho\sigma\tau}k^{\alpha}=-\left(\epsilon^{\rho\sigma\tau\alpha}k^{\mu}+
\epsilon^{\sigma\tau\alpha\mu}k^{\rho}+
\epsilon^{\tau\alpha\mu\rho}k^{\sigma}+\epsilon^{\alpha\mu\rho\sigma}k^{\tau}\right)\;,
\end{equation*}
hence the \eqref{amu1} can be rewritten as:
\begin{equation*}
a^{\mu}a^{\nu}=-\left(\epsilon^{\rho\sigma\tau\alpha}k^{\mu}+\epsilon^{\sigma\tau\alpha\mu}k^{\rho}+
\epsilon^{\tau\alpha\mu\rho}k^{\sigma}+\epsilon^{\alpha\mu\rho\sigma}k^{\tau}\right)
\epsilon_{\ \alpha}^{\nu\ \lambda\xi}k_{\rho}\bar{q}_{\sigma}\bar{\beta}_{\tau}\bar{q}_{\lambda}\bar{\beta}_{\xi}\;.
\end{equation*}
The contraction of two Levi-Civita symbols in the above equation can be expanded as:
\begin{eqnarray*}
\epsilon^{\rho\sigma\tau\alpha}\epsilon_{\ \alpha}^{\nu\ \lambda\xi} & = & 
\epsilon^{\alpha\rho\sigma\tau}\epsilon_{\alpha}^{\ \nu\lambda\xi}\nonumber \\
 & = & -g^{\rho\nu}(g^{\sigma\lambda}g^{\tau\xi}-g^{\sigma\xi}g^{\tau\lambda})
 -g^{\rho\lambda}(g^{\sigma\xi}g^{\tau\nu}-g^{\sigma\nu}g^{\tau\xi}) -g^{\rho\xi}
 (g^{\sigma\nu}g^{\tau\lambda}-g^{\sigma\lambda}g^{\tau\nu})\;.
\end{eqnarray*}
Since $k\cdot\bar{q}=k\cdot\bar{\beta}=\bar{q}\cdot\bar{\beta}=0$, the \eqref{amu1} can be finally cast
in the following form:
\begin{eqnarray*}
a^{\mu}a^{\nu} & = & -\left(\epsilon^{\alpha\rho\sigma\tau}\epsilon_{\alpha}^{\ \nu\lambda\xi}k^{\mu}
-\epsilon^{\alpha\sigma\tau\mu}\epsilon_{\alpha}^{\ \nu\lambda\xi}k^{\rho}+\epsilon^{\alpha\tau\mu\rho}
\epsilon_{\alpha}^{\ \nu\lambda\xi}k^{\sigma}-\epsilon^{\alpha\mu\rho\sigma}
\epsilon_{\alpha}^{ \nu\lambda\xi}k^{\tau}\right)k_{\rho}\bar{q}_{\sigma}\bar{\beta}_{\tau}\bar{q}_{\lambda}
\bar{\beta}_{\xi}\nonumber \\
 & = & \bar{q}^{2}\bar{\beta}^{2}k^{\mu}k^{\nu}+k^{2}\bar{\beta}^{2}\bar{q}^{\mu}\bar{q}^{\nu}+
 k^{2}\bar{q}^{2}\bar{\beta}^{\mu}\bar{\beta}^{\nu}-k^{2}\bar{q}^{2}\bar{\beta}^{2}g^{\mu\nu}\;.
\end{eqnarray*}
Since $\bar{q}^{\mu}$ and $\bar{\beta}^{\mu}$ are defined as linear combinations of $k^{\mu}$, $q^{\mu}$, and $\beta^{\mu}$
in equation \eqref{qbetabar}, we can further expand $\bar{q}^{\mu}\bar{q}^{\nu}$ and $\bar{\beta}^{\mu}\bar{\beta}^{\nu}$
in terms of $k^{\mu}k^{\nu}$, $q^{\mu}q^{\nu}$, $\beta^{\mu}\beta^{\nu}$, $k^{\mu}q^{\nu}+k^{\nu}q^{\mu}$, 
$k^{\mu}\beta^{\nu}+k^{\nu}\beta^{\mu}$ and $q^{\mu}\beta^{\nu}+q^{\nu}\beta^{\mu}$. As a consequence, $a^{\mu}a^{\nu}$
turns out to be a linear combination of these symmetric tensors and $g^{\mu\nu}$. 

\section{Complex conjugation, time-reversal and parity}
\label{SECTION: symmetries}

The correlators in the equation \eqref{thff} are constrained by the properties of the density operator, creation/annihilation operators and stress-energy tensor operator under discrete transformations: complex conjugation, 
time-reversal and parity. The correlator is defined as:
\begin{equation}\label{App: Definizione Exp Val}
 \begin{split}
     \langle\wAd(k_+)\wA(k_-),\wT^{\mu\nu}(0)\rangle_{c,\mathrm{GE}}  &=\frac{1}{Z}\Tr\left(\e^{-\beta(x)\cdot\wP+\zeta(x)\wQ}\wAd(k_+)\wA(k_-)\wT^{\mu\nu}(0)\right)\\
 &-
\frac{1}{Z}\Tr\left(\e^{-\beta(x)\cdot\wP+\zeta(x)\wQ}\wAd(k_+)\wA(k_-)\right) 
\frac{1}{Z}\Tr\left(\e^{-\beta(x)\cdot\wP+\zeta(x)\wQ}\,\wT^{\mu\nu}(0)\right)\;.
 \end{split}
\end{equation}
Taking the complex conjugate of both sides, using $\Tr(\wO)^*=\Tr(\wO^\dagger)$ and the relations \eqref{abtransl}, \eqref{abtransl2}
one obtains:
\begin{equation*}
 \langle\wAd(k_+)\wA(k_-),\wT^{\mu\nu}(0)\rangle_{c,\mathrm{GE}}^* = 
 \langle\wAd(k_-)\wA(k_+),\wT^{\mu\nu}(0)\rangle_{c,\mathrm{GE}} \, \e^{-\beta(x) \cdot q}
\end{equation*}
where $q=k_+-k_-$. Hence, according to the definition \eqref{thff}:
\begin{equation}\label{app parity:ThetaKMS}
    \Theta^{\mu\nu}\left(k,q,\beta\right)^*=\e^{-\beta(x)\cdot q}\,\Theta^{\mu\nu}\left(k,-q,\beta\right)\;.
\end{equation}
We now come to the time-reversal and parity transformations. At operator level time-reversal and parity are described by an 
involutive anti-unitary $\wmT$ and a unitary operator $\wPi$ respectively:
\begin{subequations}\label{App1: TR operator}
    \begin{align}
        \wmT^{\dagger}&=\wmT^{-1},\quad\wmT^2=\mathrm{I},\quad \wmT\left(\alpha\left|\mathcal{H}\right.\rangle\right)=\alpha^*\wmT\left|\mathcal{H}\right.\rangle\;,\\
        \wPi^\dagger&=\wPi^{-1},\quad \wPi^2=\mathrm{I},\quad \wPi\left(\alpha\left|\mathcal{H}\right.\rangle\right)=\alpha\wPi\left|\mathcal{H}\right.\rangle\;,
    \end{align}
\end{subequations}
with $\left|\mathcal{H}\right.\rangle$ is a complex vector on an Hilbert space and $\alpha\in\mathbb{C}$ is a complex 
number.
The field transforms under time-reversal and parity as follows:
$$
  \wmT \wphi(x) \wmT = \eta_T \wphi(-x^0,\x) \qquad \wPi \wphi(x) \wPi = \eta_\Pi \wphi(x^0,-\x)
$$
where $\eta_T$ and $\eta_\Pi$ are phase factors $= \pm 1$. Hence, from \eqref{intfield}, it follows:
\begin{subequations}
    \begin{align}
        \wmT\,\widehat{A}(k)\,\wmT&=\eta_{T}\widehat{A}(\widetilde{k}),\;\qquad\wmT\,\widehat{A}^\dagger(k)\,\wmT=\eta^*_T\widehat{A}^\dagger(\widetilde{k})\;,\\
        \wPi\,\widehat{A}(k)\,\wPi&=\eta_{\Pi}\widehat{A}(\widetilde{k}),\;\qquad\wPi\,\widehat{A}^\dagger(k)\,\wPi=\eta^*_\Pi\widehat{A}^\dagger(\widetilde{k})\;,
    \end{align}
\end{subequations}
where $\widetilde{k}$ is the time-reversal/parity transformed of the four-momentum $k$:
\begin{equation}\label{tr k}
    k=\left(k^0,{\bf k}\right)\mapsto\widetilde{k}=\left(k^0,-{\bf k}\right)\;.
\end{equation}
The stress-energy tensor operator in $x=0$ transforms under time reversal as:
$$
   \wmT \wT^{\mu\nu}(0) \wmT = \theta^{\mu}_\alpha \theta^\nu_\beta \wT^{\alpha\beta}(0) 
$$
and likewise for parity, with $\theta^\mu_\alpha=\text{diag}\left(1,-1,-1,-1\right)$. In turn, the density
operator at global equilibrium is such that:
\begin{equation}\label{trhot}
    \wmT \wrho_{\rm GE}(\beta,\zeta) \wmT = \wrho_{\rm GE}(\widetilde \beta,\zeta) 
\end{equation}
and likewise for parity, where $\widetilde\beta$ is defined the same way as $\widetilde k$ in eq. \eqref{tr k}. 
From the \eqref{trhot} and the general relation:
$$
  \Tr (\wO^\dagger) = \Tr (\wO)^* = \Tr (\wmT \wO \wmT) 
$$
with $\wO$ any operator, the following relation can be obtained for the correlator in eq. \eqref{thff}:
\begin{equation}\label{App1: TR relation per Gamma finale}
    \Theta^{\mu\nu}\left(k,q,\beta\right)=\e^{-\beta(x)\cdot q}\theta^\mu_\alpha\theta^\nu_\beta 
    \:\Theta^{\alpha\beta}\left(\widetilde{k},-\widetilde{q},\tilde{\beta}\right)\;.
\end{equation}
Likewise, for parity, being $\wPi$ linear, one has:
$$
  \Tr (\wO) = \Tr (\wPi\, \wO\, \wPi)\;, 
$$
and correspondingly:
\begin{equation}\label{App1: Parity relation per Gamma finale}
\Theta^{\mu\nu}\left(k,q,\beta\right)=\theta^\mu_\alpha\theta^\nu_\beta\;
\Theta^{\alpha\beta}\left(\widetilde{k},\widetilde{q},\widetilde{\beta}\right)\;.
\end{equation}
The extension of the relations~\eqref{app parity:ThetaKMS}, \eqref{App1: TR relation per Gamma finale}, 
and \eqref{App1: Parity relation per Gamma finale} to the tensor coefficients $\Gamma^{\mu\nu}_j$ appearing 
in Eq.~\eqref{Theta deltas2} is not straightforward. The scalar arguments $S$ defined in Eq.~\eqref{scalars} 
are invariant under the replacement of all four-vectors by their tilde-transformed counterparts, corresponding 
to parity or time-reversal transformations. However, they are not, in general, invariant under the transformation 
$q \mapsto -q$. As a consequence, while the constraints imposed by a $\delta$ function and by the same $\delta$ 
function with tilde-transformed arguments coincide, this is not generally the case when the sign of $q$ is 
also sign-reversed.

For the parity transformations, the invariance of arguments of the $\delta$ functions allows one to extend the 
relation~\eqref{App1: Parity relation per Gamma finale} to each tensor $\Gamma^{\mu\nu}_j$ independently, yielding:
\begin{equation*}
    \Gamma^{\mu\nu}_{j}\left(k,q,\beta\right) =
    \theta^\mu_{\alpha}\,\theta^\nu_{\beta}\;
    \Gamma^{\alpha\beta}_{j}\left(\widetilde{k},\widetilde{q},\widetilde{\beta}\right)\, .
\end{equation*}
On the other hand, upon plugging the expansion~\eqref{Theta deltas2} into the relations~\eqref{app parity:ThetaKMS} and 
\eqref{App1: TR relation per Gamma finale}, it is found that the transformation $q \mapsto -q$ changes, in general, the argument of the delta distribution, mapping the tensor $\Gamma_j$ into a different term. In symbols:
\begin{equation*}
    \begin{split}
        \Gamma^{\mu\nu}_j\left(k,q,\beta\right)
        &= \e^{-\beta(x)\cdot q}\,
        \Gamma^{\mu\nu}_{\underline{j}}\left(k,-q,\beta\right)\, , \\
        \Gamma^{\mu\nu}_j\left(k,q,\beta\right)
        &= \e^{-\beta(x)\cdot q}\,
        \theta^\mu_{\alpha}\,\theta^\nu_{\beta}\,
        \Gamma^{\alpha\beta}_{\underline{j}}\left(\widetilde{k},-\widetilde{q},\widetilde{\beta}\right)\, .
    \end{split}
\end{equation*}
where $\Gamma_{\underline{j}}$ denotes the tensor in the expansion~\eqref{Theta deltas2} associated with the $\delta$ function fulfilling:
\begin{equation*}
    \delta\!\left(s_{\underline{j}} - f_{\underline{j}}(S)\right)\Big|_{q \mapsto -q}
    =\delta\!\left(s_j - f_j(S)\right)\, .
\end{equation*}
For the two specific cases $s = q \cdot k$ and $s = q \cdot \beta$, with $f_j(S) = 0$, denoted by $j = k$ and $j = \beta$ respectively, 
the transformation $q \mapsto -q$ leaves the $\delta$-constraint unchanged. Thus we get:
\begin{subequations}\label{app: Gamma relations k}
    \begin{align}
        \Gamma^{\mu\nu}_{k}\left(k,q,\beta\right)^*
        &= \e^{-\beta(x)\cdot q}\,
        \Gamma^{\mu\nu}_{k}\left(k,-q,\beta\right)\, , \\
        \Gamma^{\mu\nu}_{k}\left(k,q,\beta\right)
        &= \e^{-\beta(x)\cdot q}\,
        \theta^\mu_{\alpha}\,\theta^\nu_{\beta}\,
        \Gamma^{\alpha\beta}_{k}\left(\widetilde{k},-\widetilde{q},\widetilde{\beta}\right)\, , \\
        \Gamma^{\mu\nu}_{k}\left(k,q,\beta\right)
        &= \theta^\mu_{\alpha}\,\theta^\nu_{\beta}\,
        \Gamma^{\alpha\beta}_{k}\left(\widetilde{k},\widetilde{q},\widetilde{\beta}\right)\, .
    \end{align}
\end{subequations}
Similarly, for the case $j = \beta$, one finds:
\begin{subequations}\label{app: Gamma relations beta}
    \begin{align}
        \Gamma^{\mu\nu}_{\beta}\left(k,q,\beta\right)^*
        &= \Gamma^{\mu\nu}_{\beta}\left(k,-q,\beta\right)\, , \\
        \Gamma^{\mu\nu}_{\beta}\left(k,q,\beta\right)
        &= \theta^\mu_{\alpha}\,\theta^\nu_{\beta}\,
        \Gamma^{\alpha\beta}_{\beta}\left(\widetilde{k},-\widetilde{q},\widetilde{\beta}\right)\, , \\
        \Gamma^{\mu\nu}_{\beta}\left(k,q,\beta\right)
        &= \theta^\mu_{\alpha}\,\theta^\nu_{\beta}\,
        \Gamma^{\alpha\beta}_{\beta}\left(\widetilde{k},\widetilde{q},\widetilde{\beta}\right)\, .
    \end{align}
\end{subequations}

Finally, since the four-current operator $\widehat{j}^\mu(0)$ is Hermitian and transforms under parity and time 
reversal as:
\begin{equation*}
    \widehat{j}^\mu(0) \mapsto \theta^\mu_{\alpha}\,\widehat{j}^\alpha(0)\, ,
\end{equation*}
the relations~\eqref{app parity:ThetaKMS}, \eqref{App1: TR relation per Gamma finale}, and \eqref{App1: Parity relation per Gamma finale} can be straightforwardly extended to the current expectation values, yielding
\begin{subequations}\label{app: relations Upsilon*}
    \begin{align}
        Y^\mu\left(k,q,\beta\right)^*
        &= \e^{-\beta(x)\cdot q}\,
        Y^\mu\left(k,-q,\beta\right)\, , \\
        Y^\mu\left(k,q,\beta\right)
        &= \e^{-\beta(x)\cdot q}\,
        \theta^\mu_{\alpha}\,
        Y^\alpha\left(\widetilde{k},-\widetilde{q},\widetilde{\beta}\right)\, , \\
        Y^\mu\left(k,q,\beta\right)
        &= \theta^\mu_{\alpha}\,
        Y^\alpha\left(\widetilde{k},\widetilde{q},\widetilde{\beta}\right)\, .
    \end{align}
\end{subequations}
%

\section{Calculation of the Wigner function}
\label{SECTION: Wigner computation}

The calculation of the correction to the Wigner function for the branches proportional to $\delta(q\cdot k)$ and $\delta(q \cdot \beta)$ 
in the equation \eqref{deltawig3} essentially proceeds through the same steps of the calculation in ref. \cite{Sheng:2025cjk}, the 
main difference being that the integration is done over the initial equilibrium hypersurface and not the decoupling one. We show the 
full calculation for the branch proportional to $\delta(q\cdot k)$, the computation for the branch proportional to $\delta(q\cdot\beta)$ 
being very similar.

We start from the equation \eqref{deltawig5} and we assume the hypersurface $\Sigma_0$ to be space-like and future-oriented 
and parametrized as a single function $y^0 = f({\bf y})$ in Cartesian coordinates. This simplifies the derivation, which can be 
anyhow extended to more complicated topologies \cite{Sheng:2025cjk}. The vector perpendicular to the hypersurface and future-oriented 
is the four-gradient of $(y^0 - f({\bf y}))$:
\begin{equation*}
\sigma_{\mu}(y)=\left(1,-\frac{\partial f}{\partial{\bf y}}\right)\;.
\end{equation*}
Denoting:
$$
 \Theta(y,x) = \partial^{\nu_{1}}_{x}\cdots\partial^{\nu_{M}}_{x} \Delta\beta_{\nu}(y,x)\;,
 \qquad {\rm or} \qquad \Theta(y,x) = \partial^{\nu_{1}}_{x}\cdots\partial^{\nu_{M}}_{x} \Delta\zeta(y,x)\;,
$$
the integral in \eqref{deltawig5} gives rise to:
\begin{eqnarray}\label{deltas}
&& \int_{\Sigma_0} \di \Sigma_{\mu}(y) \; \delta^{3}\left({\bf y}-{\bf x}-\frac{\bf k}{k^{0}}
(y^{0}-x^{0})\right) \Theta(y,x) = \int_{\Sigma_0} \di^3 {\rm y} \; \sigma_{\mu}(y)\;  
\delta^{3}\left({\bf y}-{\bf x}-\frac{\bf k}{k^{0}}(y^{0}-x^{0})\right) \Theta(y,x) \nonumber \\
&& = \sum_{i} \int_{\Sigma_0} \di^3 {\rm y} \; \sigma_{\mu}(y)\; \frac{|k^0|}{|k\cdot\sigma(y)|}
\delta^{3}({\bf y}-\bar{\bf y}_{i}) \Theta(y,x) = \sum_{i} \sigma_{\mu}(\bar y_{i})\; \frac{|k^0|}{|k\cdot\sigma(\bar y_{i})|} \Theta(\bar y_{i}(x),x)\;,
\end{eqnarray} 
where $\bar{\bf y}_{i}$ is the $i$-th solution of the equation ${\bf y} = {\bf x}-({\bf k}/k^0)(f({\bf y})-
x^{0})$, yielding the intersection of the hypersurface $\Sigma_0$ with the world-line of an off-mass-shell particle moving 
with velocity ${\bf k}/k^0$:
\begin{equation*}
    {\bf y} = {\bf x}-\frac{\bf k}{k^{0}}(y^0-x^0)\;,
\end{equation*}
and it is obviously a function of $x$ and $k$. Indeed, if $\Sigma_0$ is space-like, at it supposedly is, there is at 
most one intersection, denoted by $\bar y_k(x)$. Note that in the equation \eqref{deltas} the pre-factor $k^0/ k\cdot \sigma$ 
is the inverse of the determinant of the matrix:
$$
\frac{\partial}{\partial y^j}\left[{\bf y}-{\bf x}- \frac{\bf k}{k^0}(f({\bf y})-x^{0})\right]^i =
\delta^i_j - \frac{k^i}{k^0} \frac{\partial f({\bf y})}{\partial y^j} = \delta^i_j - \frac{k^i}{k^0} 
\sigma^j \;. 
$$
Using the relations \eqref{deltas}, and replacing $\sigma$ with the unit vector normal to the hypersurface:
$$
  n_\mu = \frac{\sigma_\mu}{\sqrt{|\sigma \cdot \sigma|}}\;,
$$
the equation \eqref{deltawig5} is converted to:
\begin{eqnarray}\label{deltawig-y}
\Delta W^{+}(x,k) & = & \frac{2}{k^0\left(2\pi\right)^2}\sum_{N=0}^{\infty}\frac{(-1)^{N}}{N!}
\left.\left[\partial^{q}_{\nu_{1}}\ldots\partial^{q}_{\nu_{N}}G^{\mu\nu}_k(q)\right]\right|_{q=0}
 \sum_{M=0}^{N}\frac{N!(-1)^{M}}{M!(N-M)!} \nonumber \\
 && \times \di^{\nu_{M+1}}_{x}\ldots\di^{\nu_{N}}_{x}
 \left[\left.\frac{n_{\mu}}{|k\cdot n|}\partial^{\nu_{1}}_{x}\ldots\partial^{\nu_{M}}_{x}
 \Delta\beta_{\nu}(y,x)\right] \right|_{y=\bar{y}(x,k)} +\; {\rm analogous \; term \; for \; ,
 \Delta\zeta}
\end{eqnarray}
where we have introduced the total derivative:
$$
\di^\mu_x = \frac{\di}{\di x^\mu}\;,
$$
to emphasize the difference between the derivative acting on the function {\em before} setting 
$y=\bar y_k(x)$ (that is $\partial_x$) and the derivative acting on the function {\em after}
setting $y=\bar y_k(x)$. 
Now, the derivative of the function $y^\nu = \bar y_{k}^\nu(x)$ is obtained by taking into account
that:
\begin{equation*}
\bar{{\bf y}}_{k}-\frac{{\bf k}}{k^{0}}\bar{y}_{k}^{0}={\bf x}-\frac{{\bf k}}{k^{0}}x^{0}\;,
\end{equation*}
with $\bar{y}_{k}^{0} =f_{k}(\bar{{\bf y}}_{k})$. Taking partial derivatives with respect to $x^\mu$
of the above equation:
\begin{equation*}
\frac{\partial\bar{y}_{k}^{j}}{\partial x^{\mu}}-\frac{k^{j}}{k^{0}}\frac{\partial\bar{y}_{k}^{0}}
{\partial\bar{y}_{k}^{l}}\frac{\partial\bar{y}_{k,i}^{l}}{\partial x^{\mu}} = 
\left[ \delta^j_l + \frac{k^j}{k^0} \sigma_l \right] \frac{\partial\bar{y}_{k}^{l}}{\partial x^{\mu}} 
= \delta_{\mu}^{j}-\frac{k^{j}}{k^{0}}\delta_{\mu}^{0}\;,
\end{equation*}
where $j,l=1,2,3$ and where $\sigma_{\mu}(\bar{y}_{k})$ is the normal vector of $\Sigma_{\rm D}$ at 
the spacetime point $\bar{y}_{k}$,
\begin{equation*}
\sigma_{\mu}(\bar{y}_{k})=\left(1,-\frac{\partial f_{k}(\bar{{\bf y}}_{k})}
{\partial\bar{{\bf y}}_{k}}\right)\;.
\end{equation*}
The $3 \times 3$ matrix:
$$
  A^j_l = \left[ \delta^j_l + \frac{k^j}{k^0} \sigma_l \right]\;,
$$
can be inverted and one obtains:
\be\label{deriv1}
\frac{\partial\bar{y}_{k,i}^{j}}{\partial x^{\mu}} = \left(A^{-1}\right)^{j}_{\;\; l} \left(\delta_{\mu}^{l}-
\frac{k^{l}}{k^{0}}\delta_{\mu}^{0}\right) = \left( \delta^j_l - \frac{k^j \sigma_l}{k \cdot \sigma} \right)  
\left(\delta_{\mu}^{l}-\frac{k^{l}}{k^{0}}g_{\mu}^{0}\right)\;.
\ee
Similarly, one can calculate the derivative of $y^0$ with respect to $x$:
$$
\frac{\partial\bar{y}_{k}^{0}}{\partial x^{\mu}} = \frac{\partial\bar{y}_{k}^{0}}
{\partial\bar{y}_{k}^{j}}\frac{\partial\bar{y}_{k}^{j}}{\partial x^{\mu}} = 
-\sigma_j \left( \delta^j_\mu - \frac{k^j \sigma_\mu}{k \cdot \sigma} \right)\;,
$$
whence we obtain:
\be\label{deriv2}
 \frac{\partial\bar{y}_{k}^{0}}{\partial x^{0}} = \sigma_0 - \frac{k^0}{k \cdot \sigma}\;, \qquad\qquad
 \frac{\partial\bar{y}_{k}^{0}}{\partial x^{m}} = - \frac{k^0 \sigma_m}{ k \cdot \sigma}\;. 
\ee
The equations \eqref{deriv1} and \eqref{deriv2} can be written in a compact form as:
\begin{equation*}
 \frac{\partial\bar{y}_{k}^{\nu}}{\partial x^{\mu}}=\delta_{\mu}^{\nu}-\frac{k^{\nu}
 \sigma_{\mu}(\bar{y}_{k})}{k\cdot\sigma(\bar{y}_{k})} \equiv \Delta_{\mu}^{\nu}(\bar{y}_{k})\;.
\end{equation*}
Finally, using the chain rule for the derivative of an implicit function:
\begin{equation*}
\frac{\di}{\di x^\mu} g(x,y(x))= \frac{\partial}{\partial x^\mu} g(x,y(x))+ 
\Delta^{\; \nu}_\mu \frac{\partial g(x,y)}{\partial y^\nu}\Bigg|_{y=y(x)},
\end{equation*}
the equation \eqref{deltawig5} can be rewritten as:
\begin{eqnarray*}
&& \Delta W^{+}(x,k) =\frac{2}{\left(2\pi\right)^2}
\sum_{N=0}^{\infty}\frac{(-\ii)^{N}}{N!}
\left.\left[\partial^{q}_{\nu_{1}}\ldots\partial^{q}_{\nu_{N}}G^{\mu\nu}_k(q)\right]\right|_{q=0}
\sum_{M=0}^{N}\frac{N!(-1)^{M}}{M!(N-M)!} \left[\partial^{\nu_{M+1}}_{x}+
\Delta^{\alpha_{M+1}\rho_{M+1}}(\bar{y}_{k})\partial_{\rho_{M+1}}^{y}\right]\nonumber \\
 && \times\left.\ldots\left[\partial^{\nu_{n}}_{x}+\Delta^{\nu_n\rho_{n}}(\bar{y}_{k})
 \partial_{\rho_{n}}^{y}\right]\partial^{\nu_{1}}_{x}\cdots\partial^{\nu_{M}}_{x}
 \frac{n_{\mu}(y)}{|k\cdot\ n(y)|}\Delta\beta_{\nu}(y,x)\right|_{y=\bar{y}_k(x)}
 +\; {\rm analogous \; term \; for \; \Delta\zeta}\;.
\end{eqnarray*}
It is convenient to introduce the following differential operator:
\begin{equation}\label{D}
D_y(\bar y_k) \equiv  \Delta^{\nu\rho}(\bar y_k)
\partial_{\rho}^{y}\partial_{\nu}^{q}\;,
\end{equation}
By using the binomial theorem, the last expression can be finally recast as:
\begin{equation*}
\Delta W^{+}(x,k)=\frac{2}{\left(2\pi\right)^2}\sum_{N=0}^{\infty} \frac{(-\ii)^{N}}{N!}
\left(D_y\left(\bar{y}_k(x)\right)\right)^{N}
\left[ G^{\mu\nu}_k(q)\frac{n_{\mu}(y)}{|k\cdot n(y)|} \Delta\beta_\nu(y,x) \right] 
\Bigg|_{q=0,y=\bar{y}_k(x)} \hspace{-1.5cm} +\; {\rm analogous \; term \; for \; \Delta\zeta}\;,
\end{equation*}
which coincides with eq. \eqref{deltawigk} in the main text. Note that the differential operator \eqref{D} acts only on the square bracket and not on the projector $\Delta$ which is already computed on the intersection $\bar{y}_k(x)$ hence \eqref{D} never acts upon itself.

\end{document}